\documentclass[aps,prd,twocolumn,groupedaddress,floatfix,longbibliography,superscriptaddress]{revtex4-2}

\usepackage{amsmath}
\usepackage{amssymb}
\usepackage{amsfonts}
\usepackage{bbold}
\usepackage{bm}
\usepackage{cancel}
\usepackage{times,float}
\usepackage{graphicx}
\graphicspath{{Figures/}}
\usepackage[usenames,dvipsnames,svgnames]{xcolor}
\usepackage{hyperref}
\hypersetup{colorlinks=true, linkcolor=Blue, citecolor=Blue,urlcolor=Blue}
\usepackage{multirow}
\usepackage{ulem}
\usepackage{color,epsfig}
\usepackage{slashed}
\usepackage{feynmf}
\usepackage{array}
\usepackage{physics}
\usepackage{subcaption}
\usepackage{soul}

\begin{document}
\sloppy 

\title{Non-Hermitian thermoelectric transport in graphene: Tunable anomalous transmission through complex barriers}

\author{Daniel A. Bonilla}
\email{daniel.bonillam@correo.nucleares.unam.mx}
\affiliation{Instituto de Ciencias Nucleares, Universidad Nacional Aut\'{o}noma de M\'{e}xico, 04510 Ciudad de M\'{e}xico, M\'{e}xico}
\author{Juan A. Ca\~nas}
\email{juan.canas@correo.nucleares.unam.mx}
\affiliation{Instituto de Ciencias Nucleares, Universidad Nacional Aut\'{o}noma de M\'{e}xico, 04510 Ciudad de M\'{e}xico, M\'{e}xico}
\author{J. C. Pérez-Pedraza}
\email{julio.perez@correo.nucleares.unam.mx}
\affiliation{Instituto de Ciencias Nucleares, Universidad Nacional Aut\'{o}noma de M\'{e}xico, 04510 Ciudad de M\'{e}xico, M\'{e}xico}
\author{A. Mart\'{i}n-Ruiz}
\email{alberto.martin@nucleares.unam.mx}
\affiliation{Instituto de Ciencias Nucleares, Universidad Nacional Aut\'{o}noma de M\'{e}xico, 04510 Ciudad de M\'{e}xico, M\'{e}xico}

\begin{abstract}
We investigate thermoelectric transport in monolayer graphene across a finite complex barrier within a Landauer scattering framework. Solving the Dirac-Weyl problem exactly, we show that the imaginary part of the barrier renders the scattering matrix nonunitary and replaces the usual Hermitian flux conservation by a generalized flux-balance relation determined by the net gain or loss inside the barrier. In the Hermitian limit, the standard graphene $n$-$p$-$n$ barrier behavior is recovered, including perfect transmission at normal incidence and Fabry-Perot-type resonances. For a finite imaginary part, however, the same resonant channels are selectively attenuated or amplified, which significantly modifies both the angular response and the conductance profile. We further show that the lead-resolved conductances become dependent on the bias partition, providing a direct signature of the breakdown of gauge invariance in the effective two-terminal response. At finite temperature, the exact linear-response coefficients reveal a clear trade-off controlled by the imaginary part of the barrier: gain enhances both the electrical and thermal conductances, whereas loss suppresses the thermal conductance more efficiently and yields the largest thermoelectric figure of merit within the parameter range considered. These results demonstrate that complex barriers extend the range of transport behaviors accessible in graphene beyond the usual Hermitian $n$-$p$-$n$ junction. They also suggest a practical interpretation of the imaginary potential as an effective reduced description of unresolved source-sink channels or additional probes coupled to the device, particularly when a fully microscopic model of the environment is not available.
\end{abstract}

\maketitle

\section{Introduction}

Graphene has become a paradigmatic platform for the study of quantum transport in low-dimensional systems. Its low-energy quasiparticles are described by a massless Dirac-Weyl equation, leading to a linear dispersion relation, a chiral structure of the wave functions, and transport phenomena such as Klein tunnelling and minimal conductivity at the charge neutrality point \cite{CastroNeto2009,DasSarma2011,Katsnelson2012,Geim_Klein}. In high-mobility devices, phase-coherent transport over mesoscopic length scales is routinely achieved, so that the current is governed by transmission amplitudes rather than by a local Ohmic response. This makes graphene an ideal system to explore how fundamental concepts of quantum scattering and open-system dynamics manifest in electronic transport.

In the mesoscopic regime, the Landauer-B\"uttiker formalism provides the standard framework to relate the conductance to transmission probabilities between ideal leads \cite{Landauer1970,Landauer1981,Buttiker1985,Buttiker1986,Datta1995,Nazarov}. For non-interacting particles, the conductance of a multi-terminal device is expressed in terms of a unitary scattering matrix that connects incoming and outgoing modes in each lead. Unitarity encodes particle-number conservation in the scattering region and ensures several key properties: the balance between transmission and reflection, the equality of currents measured at different leads, and gauge invariance of the conductance under uniform shifts of the electrochemical potential. In graphene, this formalism has been extensively applied to understand ballistic transport across electrostatic and mass-like barriers, disordered regions, and constrictions, both at zero and finite temperatures \cite{CastroNeto2009,DasSarma2011,Katsnelson2012}.

Related analytical strategies, based on exact scattering states, phase shifts, transmission functions, disorder-averaged Green functions, and semiclassical transport equations, have also been used in graphene and in other Dirac materials. In graphene, such methods have been developed for magneto-strain transport through nanobubbles, for electromagnetic-coupling problems in graphene heterostructures, for strain-induced impurity models, and for smooth finite-range disorder \cite{MunozSotoGarrido2017,BonillaCastanoYepesMartinRuizMunoz2023,nanolett3c04703,CanasBonillaMartinRuiz2025,CanasBonillaPerezPedrazaMartinRuiz2026}. In Weyl semimetals, methods of the same spirit were used for torsional strain and dislocation defects, including ballistic Landauer calculations, thermoelectric extensions, and linear-response treatments based on T-matrix, Kubo, and Onsager formalisms \cite{SotoGarridoMunoz2018,MunozSotoGarrido2019,BonillaMunozSotoGarrido2021,BonillaMunoz2022,BonillaMunoz2024}.

In many physical situations, however, the effective Hamiltonian governing transport is non-Hermitian. These kind of Hamiltonians emerge naturally as reduced descriptions of open quantum systems coupled to continua or reservoirs, and as effective models with gain and loss in photonics, cold atoms, and electronic devices \cite{Moiseyev2011,Rotter2009}. In this context, complex energy eigenvalues encode finite lifetimes, while the imaginary part of an effective potential describes absorption or amplification. The development of non-Hermitian quantum mechanics and of $\mathcal{PT}$-symmetric Hamiltonians has shown that such systems can display real spectra, phase transitions at exceptional points, and a rich variety of unconventional scattering and transport phenomena \cite{Bender2007,ElGanainy2018}. At the same time, non-Hermitian topology has provided a broad language to describe skin effects, anomalous edge states, and the interplay between topology, interactions, and gain and loss \cite{FoaTorres2020,Kawabata105165137,ElGanainy2018}.

From the scattering perspective, non-Hermitian systems are characterized by nonunitary $S$ matrices, which break the balance between incoming and outgoing fluxes. This allows for phenomena such as coherent perfect absorption, unidirectional reflectionlessness, and non-reciprocal transport \cite{Schomerus2013,Muga2004,GhaemiDizicheh2023}. Recent works have explored transport in non-Hermitian tight-binding lattices with complex on-site potentials, where randomness in the imaginary part can strongly modify the spectral and transport properties, and lead to features such as anomalous diffusion and unusual conductance statistics \cite{Tzortzakakis2021}. More recently, a general transport theory for non-Hermitian systems has been formulated in terms of Green functions and continuity equations \cite{Yan2024}, and the question of gauge invariance in non-Hermitian quantum transport has been analyzed explicitly in the Landauer setting \cite{Wei2025}. Very recently, an extended Landauer-B\"uttiker formula for open quantum systems with local gain or loss has been derived within a Lindblad-Keldysh approach \cite{Yang2026}. These developments make the problem particularly timely.

Graphene and related Dirac materials offer a particularly attractive arena in which to study non-Hermitian physics. On the one hand, they provide a continuum Dirac description with a well-defined microscopic origin. On the other hand, realistic devices are unavoidably open and may exhibit effective loss or gain channels, for example due to inelastic processes or coupling to additional degrees of freedom. Several works have introduced non-Hermitian deformations of graphene in different settings. A $\mathcal{PT}$-symmetric modification of the tight-binding model under a magnetic field has been shown to lead to complex Landau levels and modified transport channels \cite{Bagarello2016}. Complex magnetic fields in the Dirac-Weyl Hamiltonian of graphene generate non-Hermitian spectra and unusual current patterns that can be treated analytically using supersymmetric quantum mechanics \cite{Fernandez2022}. In parallel, non-Hermitian and $\mathcal{PT}$-symmetric photonic analogues of graphene have been engineered, and phenomena such as unidirectional invisibility and reflection control in graphene-based structures have been demonstrated theoretically \cite{Sarisaman2018,ElGanainy2018}.

More recently, non-Hermitian topology has been directly connected to transport measurements in graphene devices operating in the quantum Hall regime. By driving multi-terminal graphene heterostructures and analyzing the full conductance matrix, it has been shown that non-Hermitian topological invariants can be encoded into robust current-voltage characteristics, even though the underlying microscopic Hamiltonian is Hermitian and the effective non-Hermiticity arises at the level of the transport problem \cite{Chaturvedi2025,Ozer2024}. These developments position graphene as a promising solid-state platform to study non-Hermitian effects in a controlled way, thereby bridging the gap between photonic realizations and electronic transport experiments.

At a more direct level, the present setup is close to a well-established family of ballistic Dirac-barrier problems in graphene. Resonant transmission through graphene-based double barriers and wells, and its impact on the ballistic conductance, was analyzed in early work such as Ref.~\cite{Pereira2007}. Fabry-P\'erot resonances in single and double graphene barriers, including the role of magnetic fields and barrier asymmetry, were studied in Ref.~\cite{RamezaniMasir2010}. Exact analytical expressions for the transmission through one and two rectangular electrostatic barriers, together with Landauer conductance formulas, were derived in Ref.~\cite{Lejarreta2013}. Related results for angle-dependent conductance in graphene FET geometries showed that non-normal incidence can generate effective transport gaps in gated barrier structures \cite{Fuentevilla2015}. A broader comparison with piecewise velocity and potential barriers, as well as superlattice effects, was presented in Ref.~\cite{Krstajic2011}. More recently, transfer-matrix methods were extended to locally periodic and superperiodic rectangular electrostatic barriers in graphene, where transmission, conductance, and the Fano factor acquire a richer resonance structure as the number and ordering of barriers increase \cite{Shekhar2025}. These works are particularly relevant for the present study because they share the same basic scattering geometry of a rectangular barrier region between ballistic graphene leads. Our problem differs from them in a single key aspect: the barrier potential is complex, which allows us to ask how gain and loss modify the familiar Klein, Fabry-P\'erot, and conductance features known from the Hermitian case.

Despite this progress, much less attention has been paid to the simplest continuum realization of non-Hermitian transport in graphene: a local complex scalar potential acting as a gain/loss region. In the Schr\"odinger setting, complex square wells and barriers have long been used as phenomenological optical potentials that model absorption, and the associated continuity equation contains explicit source or sink terms related to the imaginary part of the potential \cite{Muga2004}. However, a systematic analysis of how a rectangular complex barrier in the Dirac-Weyl equation for graphene modifies the scattering matrix, the associated probability balance, and the Landauer conductance of a two-terminal device is still lacking. In particular, it is unclear how the non-unitarity of the $S$ matrix translates into different currents in the left and right leads, and how this affects fundamental properties such as the gauge invariance of the conductance.

This question is further reinforced in view of the recent transport literature. The works of Refs.~\cite{Yan2024,Wei2025,Yang2026} make clear that gain and loss can modify the current formula itself, the gauge properties of the conductance, and the relation between transport coefficients in open systems. At the same time, studies based on simple non-Hermitian scattering and lattice models have shown that unusual transport effects can already appear at the one-body level \cite{Tzortzakakis2021,GhaemiDizicheh2023}. These results motivate a closer examination of simple continuum models where the role of the imaginary potential can be tracked analytically, and where the link between non-unitary scattering, current imbalance, and gauge dependence can be established in a transparent and controlled manner. In this sense, the present problem is also attractive because it remains sufficiently simple to allow the full scattering analysis to be carried out analytically.

In this work we study non-Hermitian transport through a rectangular barrier in monolayer graphene, modeled by a spatially localized complex potential $V(x)=U+iW$. We analyze the two-dimensional Dirac equation appropriate for graphene, construct the scattering states for arbitrary incidence angles, and obtain the full $2\times 2$ scattering matrix connecting the left and right leads. This framework allows us to quantify deviations from unitarity  in terms of a gain/loss coefficient that is directly controlled by the imaginary part of the potential and by the probability density inside the barrier.

Building on these results, we formulate a Landauer-type description of transport through the non-Hermitian barrier. Because probability is not conserved inside the scattering region, the currents in the left and right leads are no longer equal and depend differently on the Fermi-Dirac distributions in the reservoirs. As a consequence, the conductances defined from the left and right currents differ, and they acquire an explicit dependence on how the voltage bias is distributed between the two leads. We show that, in contrast to the Hermitian case, this leads to a breakdown of gauge invariance at the level of the effective two-terminal conductance, and we relate this effect quantitatively to the gain/loss coefficient associated with the non-Hermitian barrier. Our analysis thus provides a minimal continuum model connecting non-unitary scattering in graphene to modified Landauer transport, and complements more general approaches based on nonequilibrium Green functions and Lindblad formalisms \cite{Datta1995,Nazarov,Yan2024,Wei2025,Yang2026}.

The paper is organized as follows. In Sec.~\ref{formalism_section} we develop the scattering formalism for graphene with a complex barrier: we first obtain the Dirac scattering states, then derive the $S$ matrix and the reflection and transmission amplitudes, and finally establish the continuity equation and the corresponding gain-loss flux balance. In Sec.~\ref{linear_response_section} we formulate the Landauer-type linear-response transport theory, including the charge and heat currents and the finite-temperature thermoelectric coefficients. In Sec.~\ref{results_discussion_section} we present and discuss the numerical results, focusing on the angular behavior of the scattering coefficients, the non-Hermitian flux balance, the zero-temperature conductance, and the finite-temperature thermoelectric response. Finally, in Sec.~\ref{conclusions} we summarize the main conclusions of the work.

\section{Scattering formalism for graphene with a complex barrier} \label{formalism_section}

In this section we set up the scattering formalism for Dirac quasiparticles in graphene across a finite complex barrier. We first obtain the spinor scattering states from the Dirac equation, then we extract the reflection and transmission amplitudes via interface matching and constructing the corresponding $S$ matrix. Finally, we discuss the continuity equation in the presence of gain/loss and the resulting flux-balance relation that replaces the usual Hermitian flux conservation.

\subsection{Model and Dirac equation for a complex barrier}

\begin{figure}[t]
    \centering
    \includegraphics[width=\linewidth]{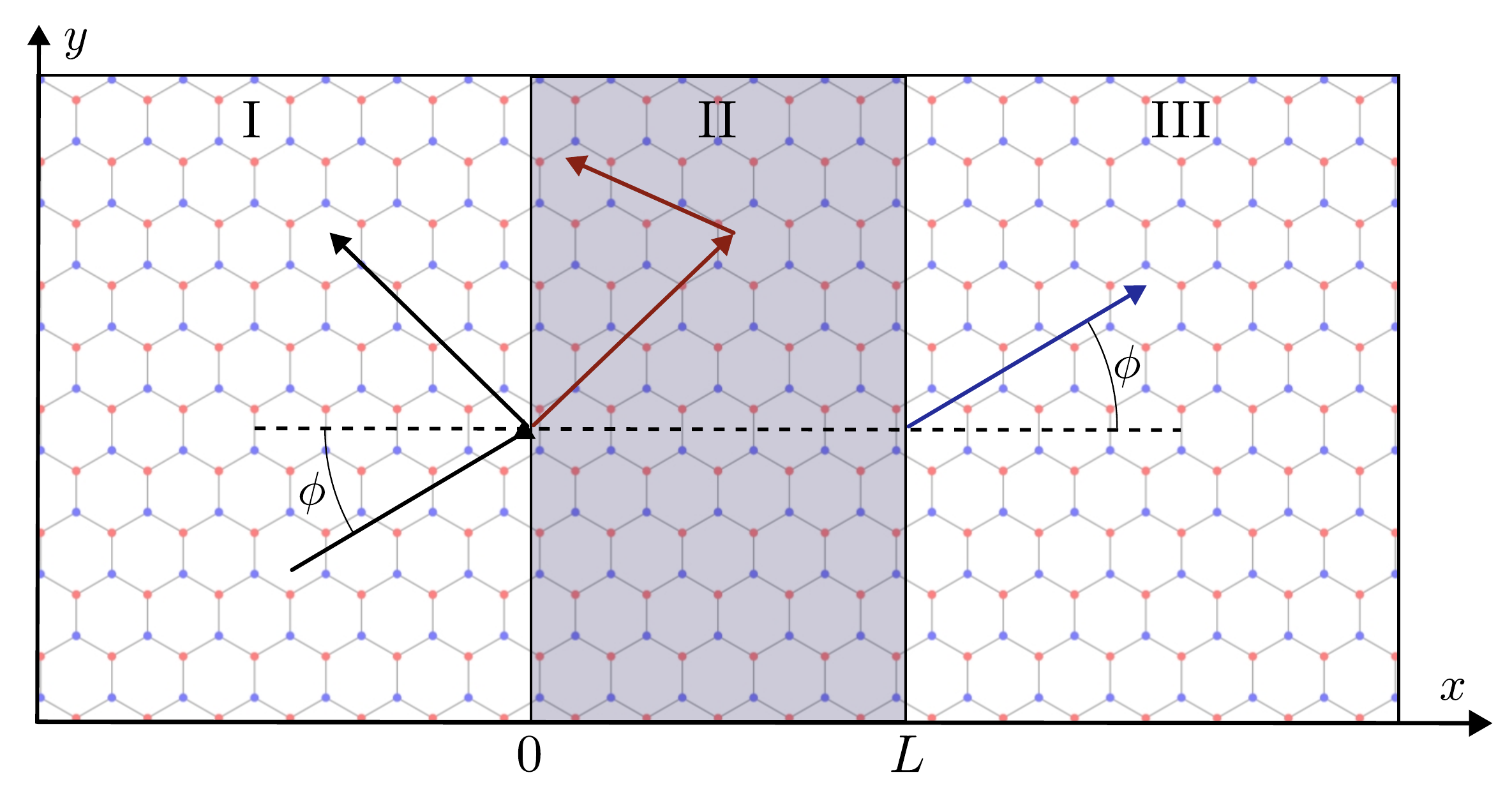}
    \caption{Schematic setup. A monolayer graphene sheet is subject to a finite complex barrier $V(x)=U+iW$ in the strip $0<x<L$ (region II), while the potential vanishes outside (regions I and III). Scattering states are injected from the left and right reservoirs, with incidence angle $\phi$ measured with respect to the $x$ axis.}
    \label{fig:setup}
\end{figure}

We consider a monolayer graphene sheet and study ballistic scattering of Dirac quasiparticles by a finite non-Hermitian (complex) barrier. The setup is illustrated in FIG.~\ref{fig:setup}. The barrier is modeled as a piecewise constant complex potential,
\begin{align}
V(x) =
\begin{cases}
U + i W, & \text{for } 0<x<L \\
0, & \text{otherwise }
\end{cases}
\label{eq:nonHermitian_square_well}
\end{align}
where $U$ and $W$ are real constants and the barrier length is $L$. This profile divides the system into three regions: region I ($x<0$), region II ($0<x<L$), and region III ($x>L$). Within the Landauer framework, carriers are injected from both reservoirs. Accordingly, we consider scattering states incident from the left and from the right, with reflection and transmission at the interfaces $x=0$ and $x=L$.

The complex barrier in Eq.~\eqref{eq:nonHermitian_square_well} should be understood as an effective open-system model. In quantum transport, coupling a finite system to external continua, sinks, or leads naturally gives rise to an effective non-Hermitian Hamiltonian after the external degrees of freedom are eliminated, and phenomenological non-Hermitian terms are often used to summarize that coupling in a reduced description \cite{Celardo2009,Giusteri2015}. A closely related viewpoint is provided by complex absorbing potentials, where an imaginary potential is added to finite lead regions in order to reproduce the effect of semi-infinite contacts and the corresponding self-energies; this strategy has been used successfully in graphene-based transport calculations \cite{Cook2011,Xie2014}. There is also a graphene-specific motivation for such a description, since coupling to a substrate can induce dissipative effects that may be modeled by a non-Hermitian graphene Hamiltonian \cite{Wu2022}. In the same spirit, recent mesoscopic transport work shows that reservoir-induced non-Hermitian self-energies can lead to measurable transport signatures \cite{Geng2023}. For this reason, in the present work the term $iW$ is taken as a simple effective way to model gain, loss or additional source--probe channels attached to the barrier region, especially in situations where those external channels are not known or controlled microscopically in detail. This is therefore an effective model, with clear limitations, but it has the practical advantage of keeping the scattering problem simple and transparent while still capturing the main open-system physics relevant to transport, especially when compared with more sophisticated microscopic approaches.

For monolayer graphene, low-energy carriers near each valley are described by the two-dimensional Dirac equation,
\begin{align}
    \chi\Big[ v_F\boldsymbol{\sigma}\cdot \mathbf{p}+\sigma_0 V(x)\Big]\Psi(x,y)&=E\,\Psi(x,y) \label{eq:Dirac_eqn_V}
\end{align}
where $\chi=\pm$ is the valley index, $\boldsymbol{\sigma}=(\sigma_x,\sigma_y)$ are Pauli matrices, $\sigma_0$ is the $2\times 2$ identity matrix, and $\mathbf{p}=-i\hbar\nabla$ is the momentum operator. The potential $V(x)$ is given in Eq.~\eqref{eq:nonHermitian_square_well}. Since the potential depends only on $x$, the system is translationally invariant along $y$, and the transverse momentum $k_2$ is conserved. We therefore use the plane-wave form
\begin{align}
    \Psi(x,y)=e^{ik_2 y}\psi(x),
\end{align}
which reduces the Dirac equation to an effective one-dimensional problem,
\begin{align}
 -i \chi \hbar v_F\Big(\sigma_x \partial_x+\sigma_yik_2 \Big)\psi(x)&=\Big[E-V(x)\Big]\psi(x).
\end{align}

Following the standard procedure \cite{Geim_Klein,CastroNeto2009,Katsnelson2012}, we build scattering states by solving the Dirac equation in each region and matching the two-component spinor at the interfaces. In the outer regions the potential vanishes, so the solutions are free Dirac spinors. We write
\begin{align}
\Psi_\text{I}(x,y) =\,&A
\begin{pmatrix}
1 \\
s\,e^{i\phi}
\end{pmatrix}
e^{i(k_1 x + k_2 y)} 
\notag\\
&\qquad+ B
\begin{pmatrix}
1 \\
-s\,e^{-i\phi}
\end{pmatrix}
e^{i(-k_1 x + k_2 y)},\label{eq:psi_I}\\
\Psi_\text{III}(x,y) =\,& F
\begin{pmatrix}
1 \\
s\,e^{i\phi}
\end{pmatrix}
e^{i(k_1 x + k_2 y)} 
\notag\\
&\qquad + G
\begin{pmatrix}
1 \\
-s\,e^{-i\phi}
\end{pmatrix}
e^{i(-k_1 x + k_2 y)},\label{eq:psi_III}
\end{align}
where $\phi$ is the incidence angle, defined by $\tan\phi=k_2/k_1$. The reflected component corresponds to $k_1\to -k_1$ (equivalently, the propagation direction changes from $\phi$ to $\pi-\phi$). Here $s=\text{sign}\,(E)$ is the band index. The coefficients $A$, $B$, $F$, and $G$ are determined by the scattering boundary conditions.

Inside the barrier (region II), the Dirac equation has the same plane-wave form, but the longitudinal wave vector is generally complex because $V=U+iW$. This is captured by the replacement
\begin{align}
    k_1\to q=\frac{1}{\hbar v_F}\sqrt{(E-U-iW)^2-\hbar^2v_F^2k_2^2}.
\end{align}
We write $q=q_R+iq_I$ and define 
\begin{align}
    \alpha=\frac{1}{\hbar^2v_F^2},\quad \Delta=E-U,
\end{align}
which gives
\begin{align}
    q_R^2-q_I^2&=\alpha\Delta^2-\alpha W^2-k_2^2,\\
    q_Rq_I&=-\alpha \Delta W.
\end{align}
Solving for $q_R$ and $q_I$, and using that $k_2=(|E|/\hbar v_F)\sin\phi$, we express the result in terms of $E$, $U$, $W$, and $\phi$ as
\begin{widetext}
\begin{align}
    q_R&=\frac{1}{\hbar v_F\sqrt{2}}\sqrt{\sqrt{\left[\left(E-U\right)^2-W^2-E^2\sin^2\phi\right]^2 +4 (E-U)^2W^2}+\left(E-U\right)^2-W^2-E^2\sin^2\phi},\label{eq:q_R}\\
    q_I&=-\frac{\mathrm{sign}\Big((E-U)W\Big)}{\hbar v_F\sqrt{2}}\sqrt{\sqrt{\left[\left(E-U\right)^2-W^2-E^2\sin^2\phi\right]^2 +4 (E-U)^2W^2}-\left(E-U\right)^2+W^2+E^2\sin^2\phi}.\label{eq:q_I}
\end{align}
\end{widetext}
Then, in region II we have
\begin{align}
\Psi_\text{II}(x,y) =\, & C 
\begin{pmatrix}
1 \\
\frac{(E-V)/\hbar v_F}{q-ik_2}
\end{pmatrix}
e^{i(q x + k_2 y)} 
 \notag\\
&\qquad +D
\begin{pmatrix}
1 \\
-\frac{(E-V)/\hbar v_F}{q+ik_2}
\end{pmatrix}
e^{i(-q x + k_2 y)}.\label{eq:psi_II}
\end{align}
The coefficients $C$ and $D$ are determined by the same boundary-matching conditions at the interfaces. Note that we write the spinor directly in terms of $q$ and $E - V$, so it is not necessary to introduce $s' = \text{sgn}(E - V)$ as is often done \cite{Geim_Klein}. Here $E - V$ is complex, so introducing $s'$ is not useful.

We determine all coefficients by imposing continuity of the two-component spinor at the interfaces,
\begin{align}
\Psi_\text{I}(0,y) &= \Psi_\text{II}(0,y), \\
\Psi_\text{II}(L,y) &= \Psi_\text{III}(L,y). 
\end{align}
Replacing the wavefunctions with their general expressions, the condition at $x=0$ yields
\begin{subequations}\label{eq:interface_0}
\begin{align}
A+B &= C + D, \label{eq:interface_0a}\\
sAe^{i\phi}-sBe^{-i\phi} &= C\frac{(E-V)/\hbar v_F}{q-ik_2}-D\frac{(E-V)/\hbar v_F}{q+ik_2},\label{eq:interface_0b}
\end{align}
\end{subequations}
while the condition at $x=L$ gives
\begin{subequations}\label{eq:interface_L}
\begin{align}
&C e^{i q L} + D e^{-i q L} = F e^{i k_1 L}+G e^{-i k_1 L}, \label{eq:interface_La}
\end{align}
\begin{align}
 &C\,e^{i q L}\frac{(E-V)/\hbar v_F}{q-ik_2}-D\,e^{-i q L}\frac{(E-V)/\hbar v_F}{q+ik_2} \notag\\
 &\quad= sFe^{i k_1 L} e^{i\phi}-sGe^{-i k_1 L} e^{-i\phi}.\label{eq:interface_Lb}
\end{align}
\end{subequations}

\subsection{S matrix and scattering amplitudes}

From the boundary conditions in Eqs.~\eqref{eq:interface_0}-\eqref{eq:interface_L} we solve for the unknown amplitudes and express $B$, $C$, $D$, and $F$ in terms of the incoming ones $A$ and $G$. Here $A$ and $G$ are the amplitudes of the waves incident from the left and right reservoirs, respectively, while $B$ and $F$ are the outgoing amplitudes in the left and right leads. The coefficients $C$ and $D$ correspond to the right- and left-moving components inside the barrier. For completeness, the coefficients in region II are
\begin{widetext}
\begin{align}
    C=\,& 2se^{i\phi}\cos\phi\Bigr[\hbar^2v_F^2q^2+E^2\sin^2\phi\Bigl]\Bigr[e^{i\phi}(s\hbar v_Fq+i|E|\sin\phi)+E-V\Bigl]\frac{A}
    {\Delta_C}\notag\\
    &\quad +2se^{iqL+i\phi}e^{-i|E|\cos\phi/\hbar v_F}\cos\phi\Bigr[\hbar^2v_F^2q^2+E^2\sin^2\phi\Bigl]\Bigr[E-V-e^{-i\phi}(s\hbar v_Fq+i|E|\sin\phi)\Bigl]\frac{G}
    {\Delta_C},\label{eq:coeff_C}
\end{align}
where the denominator is given by
\begin{align}
\Delta_C=\,&\left(\hbar v_F q + i |E|\sin\phi\right)
\Biggl\{
e^{2iqL}\Bigl[se^{i\phi}\bigl(\hbar v_F q - i |E|\sin\phi\bigr)-E+V\Bigr]
\Bigl[e^{i\phi}(E-V)-s\hbar v_F q-i E\sin\phi\Bigr]
\notag\\
&\quad+\Bigl[e^{i\phi}(E-V)+s\hbar v_F q-i E\sin\phi\Bigr]
\Bigl[se^{i\phi}\bigl(\hbar v_F q+i |E|\sin\phi\bigr)+E-V\Bigr]
\Biggr\},
\end{align}
and
\begin{align}
    D=\,&2se^{2iqL+i\phi}\cos\phi\Bigl(\hbar v_F q+i|E|\sin\phi\Bigr)\Bigl[E-V-se^{i\phi}(\hbar v_F q-i|E|\sin\phi)\Bigr]\frac{A}{\Delta_D}\notag\\
    &\quad +2e^{iqL}e^{-i|E|\cos\phi/\hbar v_F}\cos\phi(\hbar v_F q+i|E|\sin\phi)\Bigl[\hbar v_F q-i|E|\sin\phi+se^{i\phi}(E-V)\Bigr]\frac{G}{\Delta_D},\label{eq:coeff_D}
\end{align}
with the corresponding denominator
\begin{align}
    \Delta_D=\,e^{iqL+i\phi}\Bigl\lbrace4(E-V)s\hbar v_Fq \cos\phi\cos qL-2i\sin qL\Bigr[(E-V)^2+\hbar^2 v_F^2 q^2+E(2V-E)\sin^2\phi\Bigl]\Bigr\rbrace.
\end{align}
Using the explicit solution for the outgoing amplitudes $B$ and $F$ in terms of the incoming amplitudes $A$ and $G$, we define the scattering matrix $\hat{S}$ as
\begin{align}
    \begin{pmatrix}
        B \\ F
    \end{pmatrix}=\hat{S}\begin{pmatrix}
        A \\ G
    \end{pmatrix}.\label{eq:S_matrix_definition}
\end{align}
In the usual notation \cite{Nazarov}, the $S$ matrix is written as
\begin{align}
    \hat{S}=\begin{pmatrix}
        r_L & t_R \\
        t_L & r_R
    \end{pmatrix}\label{eq:S_matrix_elements}
\end{align}
where $r_{L(R)}$ and $t_{L(R)}$ are the reflection and transmission amplitudes for incidence from the left (right). Solving the matching equations yields
\begin{subequations}\label{eq:scattering_amplitudes}
\begin{align}
    r_L&=\frac{e^{2i\phi}\left[E^2\sin^2\phi+\hbar^2 v_F^2q^2-(E-V)(E-e^{-2i\phi}V)\right]}{2i\,s\,\hbar v_Fq(E-V)\cot(qL)\cos\phi+\left[(E-V)^2+\hbar^2v_F^2q^2-E(E-2V)\sin^2\phi\right]},\label{eq:r_L}\\
    r_R&=\frac{e^{-2i\phi}e^{2i|E|\cos\phi/\hbar v_F}\left[E^2\sin^2\phi+\hbar^2 v_F^2q^2-(E-V)(E-e^{-2i\phi}V)\right]}{2i\,s\,\hbar v_Fq(E-V)\cot(qL)\cos\phi+\left[(E-V)^2+\hbar^2v_F^2q^2-E(E-2V)\sin^2\phi\right]},\label{eq:r_R}\\
    t_L&=\frac{2i\,s\,\hbar v_F q(E-V)e^{i|E|\cos\phi/\hbar v_F}\csc(qL)\cos\phi}{2i\,s\,\hbar v_Fq(E-V)\cot(qL)\cos\phi+\left[(E-V)^2+\hbar^2v_F^2q^2-E(E-2V)\sin^2\phi\right]},\label{eq:t_L}\\
    t_R&=\frac{2i\,s\,\hbar v_F q(E-V)e^{i|E|\cos\phi/\hbar v_F}\csc(qL)\cos\phi}{2i\,s\,\hbar v_Fq(E-V)\cot(qL)\cos\phi+\left[(E-V)^2+\hbar^2v_F^2q^2-E(E-2V)\sin^2\phi\right]},\label{eq:t_R}
\end{align}
\end{subequations}
\end{widetext}
where we use $V=U+iW$ and $q=q_R+iq_I$. From Eqs.~\eqref{eq:t_L} and \eqref{eq:t_R} we immediately find
\begin{align}
    t_L=t_R.\label{eq:t_R_t_L}
\end{align}
Although $r_L$ and $r_R$ differ by a phase as can be seen from Eqs.~\eqref{eq:r_L} and \eqref{eq:r_R}, their reflection coefficients coincide, so that
\begin{align}
    |r_L|^2= |r_R|^2.\label{eq:r_R_r_L}
\end{align}

\subsection{Continuity equation and gain-loss flux}

Starting from Eq.~\eqref{eq:Dirac_eqn_V}, one can derive the continuity equation
\begin{align}
	\frac{\partial  \rho}{\partial t}+\nabla\cdot\mathbf{j}=-\frac{i}{\hbar}\Psi^{\dagger}\left(V-V^{*}\right)\Psi, \label{eq:prob_conservation}
\end{align}
where the probability density is
\begin{align}
	\rho(\mathbf{x})=\Psi^{\dagger}(\mathbf{x})\Psi(\mathbf{x}),\label{eq:rho}
\end{align}
and the probability (particle) current is
\begin{align}
	\mathbf{j}(\mathbf{x})= v_F\Psi^{\dagger} (\mathbf{x})\boldsymbol{\sigma}\Psi(\mathbf{x}).\label{eq:j_current}
\end{align}
Equation~\eqref{eq:prob_conservation} shows that probability is conserved when the potential is Hermitian. For a complex potential, the right-hand side is a source or sink term, which means that particles can be injected or absorbed inside the barrier. In the present time-independent case this reduces to
\begin{align}
    \nabla\cdot\mathbf{j}=\frac{2}{\hbar}\,W\,\rho(\mathbf{x}).\label{eq:prob_conservation_2}
\end{align}
Using the spinors in Eqs.~\eqref{eq:psi_I}, \eqref{eq:psi_III}, and \eqref{eq:psi_II}, and noting that $k_2$ is real, both $\mathbf{j}$ and $\rho$ are independent of $y$. Equation~\eqref{eq:prob_conservation_2} then becomes
\begin{align}
\frac{d}{dx}j_x(x)=\frac{2W}{\hbar}\rho(x).
\end{align}
Integrating across the barrier we obtain
\begin{align}
    j_x(L)-j_x(0)=\frac{2W}{\hbar}\int_0^L\rho(x)\,dx=j_\text{NH}.\label{eq:prob_conservation_3}
\end{align}
The current to the right of the barrier is constant and, from Eq.~\eqref{eq:psi_III}, reads
\begin{align}
    j_x(L)=2\,s\,v_F\,\cos\phi\left(|F|^2-|G|^2\right).\label{eq:j_x_L}
\end{align}
Likewise, the current to the left of the barrier is constant and, from Eq.~\eqref{eq:psi_I}, is
\begin{align}
    j_x(0)=2\,s\,v_F\,\cos\phi\left(|A|^2-|B|^2\right).\label{eq:j_x_0}
\end{align}
With the definition of $j_\text{NH}$ in Eq.~\eqref{eq:prob_conservation_3}, we obtain the flux-balance relation
\begin{align}
   2\,s\,v_F\,\cos\phi\left(|B|^2+|F|^2-|A|^2-|G|^2\right)=j_\text{NH}.
\end{align}
Using the $S$-matrix definition in Eq.~\eqref{eq:S_matrix_definition}, this can be written compactly as
\begin{align}
   2\,s\,v_F\,\cos\phi\,\begin{pmatrix}
       A^* & G^*
   \end{pmatrix}\left(\hat{S}^{\dagger}\hat{S}-\mathbb{1}\right)\begin{pmatrix}
       A \\ G
   \end{pmatrix}=j_\text{NH}.\label{eq:S_mat_Gamma}
\end{align}
In the Hermitian limit $W=0$ we have $j_\text{NH}=0$ and $\hat{S}$ is unitary. For $W\neq 0$, $\hat{S}$ is not unitary and the deviation is fixed by the net gain or loss inside the barrier. To make this explicit, consider incidence from left to right and set $G=0$. After flux normalization we take $A=1$, and Eq.~\eqref{eq:S_mat_Gamma} gives
\begin{align}
    2\,s\,v_F\,\cos\phi\left(|r_L|^2+|t_L|^2-1\right)=\frac{2W}{\hbar}\int_0^L\rho_\text{LR}(x)\,dx,\label{eq:prob_conservation_LR}
\end{align}
where $\rho_\text{LR}(x)$ is the density inside the barrier computed using the solution for region II in Eq.~\eqref{eq:psi_II} with the coefficients $C$ and $D$ given in Eqs.~\eqref{eq:coeff_C} and \eqref{eq:coeff_D} for $A=1$ and $G=0$. This leads to
\begin{align}
   |r_L|^2+|t_L|^2 = 1 + \frac{W}{s\hbar v_F \cos\phi}\int_0^L\rho_\text{LR}(x)\,dx.\label{eq:r_L_t_L_Gamma}
\end{align}
We define the gain or loss coefficient as
\begin{align}
    \Gamma_\text{LR}=\frac{W}{s\hbar v_F \cos\phi}\int_0^L\rho_\text{LR}(x)\,dx.\label{eq:Gamma}
\end{align}
For incidence from right to left we set $A=0$ and $G=1$, which gives
\begin{align}
   |r_R|^2+|t_R|^2 = 1 + \Gamma_\text{RL}.\label{eq:r_R_t_R_Gamma}
\end{align}
with the corresponding definition of $\Gamma_\text{RL}$. Equations~\eqref{eq:r_L_t_L_Gamma} and \eqref{eq:r_R_t_R_Gamma} are the generalized current-balance relations for a non-Hermitian barrier.

The equality $\Gamma_\text{LR}=\Gamma_\text{RL}$ follows from Eq.~\eqref{eq:Gamma} and a mirror symmetry of the barrier problem. Since $V(x)=U+iW$ is uniform in the strip $0<x<L$, reversing the incidence side is equivalent, inside region II, to the mirror transformation $x\to L-x$. Under this mapping, the region-II spinor in Eq.~\eqref{eq:psi_II} keeps the same form, while the two counter-propagating components exchange their roles (the coefficients $C$ and $D$ are interchanged up to an overall phase fixed by matching). Therefore, we simply write the gain-loss factor as $\Gamma(E,\phi)$.

For convenience, we define the transmission function as
\begin{align}
    \mathcal{T}(E,\phi)=|t_L(E,\phi)|^2=|t_\mathcal{R}(E,\phi)|^2,
    \label{eq:T_def}
\end{align}
and the reflection function as
\begin{align}
    \mathcal{R}(E,\phi)=|r_L(E,\phi)|^2=|r_\mathcal{R}(E,\phi)|^2,
    \label{eq:R_def}
\end{align}
which are consistent with Eqs.~\eqref{eq:t_R_t_L} and \eqref{eq:r_R_r_L}. In the present non-Hermitian case $\mathcal{T}$ and $\mathcal{R}$ should be read as asymptotic \textit{flux ratios} in the Hermitian leads; in particular, they may exceed unity when the barrier provides gain. Using Eqs.~\eqref{eq:r_L_t_L_Gamma} and \eqref{eq:r_R_t_R_Gamma}, the flux balance can be written as
\begin{align}
    \mathcal{T}+\mathcal{R}=1+\Gamma.\label{eq:1-R}
\end{align}
In the Hermitian limit $W=0$ we have $\Gamma=0$, and Eq.~\eqref{eq:1-R} reduces to the standard result $\mathcal{T}+\mathcal{R}=1$.

\section{Linear-response thermoelectric transport: Landauer approach} \label{linear_response_section}

In this section we compute thermoelectric transport in the mesoscopic regime within the Landauer approach \cite{Landauer1970,Landauer1981,Buttiker1985,Buttiker1986,Nazarov,Kloeckner2017_PRB96_205405}. The left and right reservoirs are large contacts that remain in equilibrium at fixed chemical potential and temperature. Two identical leads connect the reservoirs to the transport region. We write the charge and heat currents in terms of the reservoir Fermi functions, linearize them with respect to a small voltage bias and a small temperature difference, identify the linear-response coefficients, and obtain explicit formulas for the electrical conductance, thermal conductance, Seebeck coefficient, and the figure of merit $ZT$.

\begin{figure}[t]
    \centering
    \includegraphics[width=\linewidth]{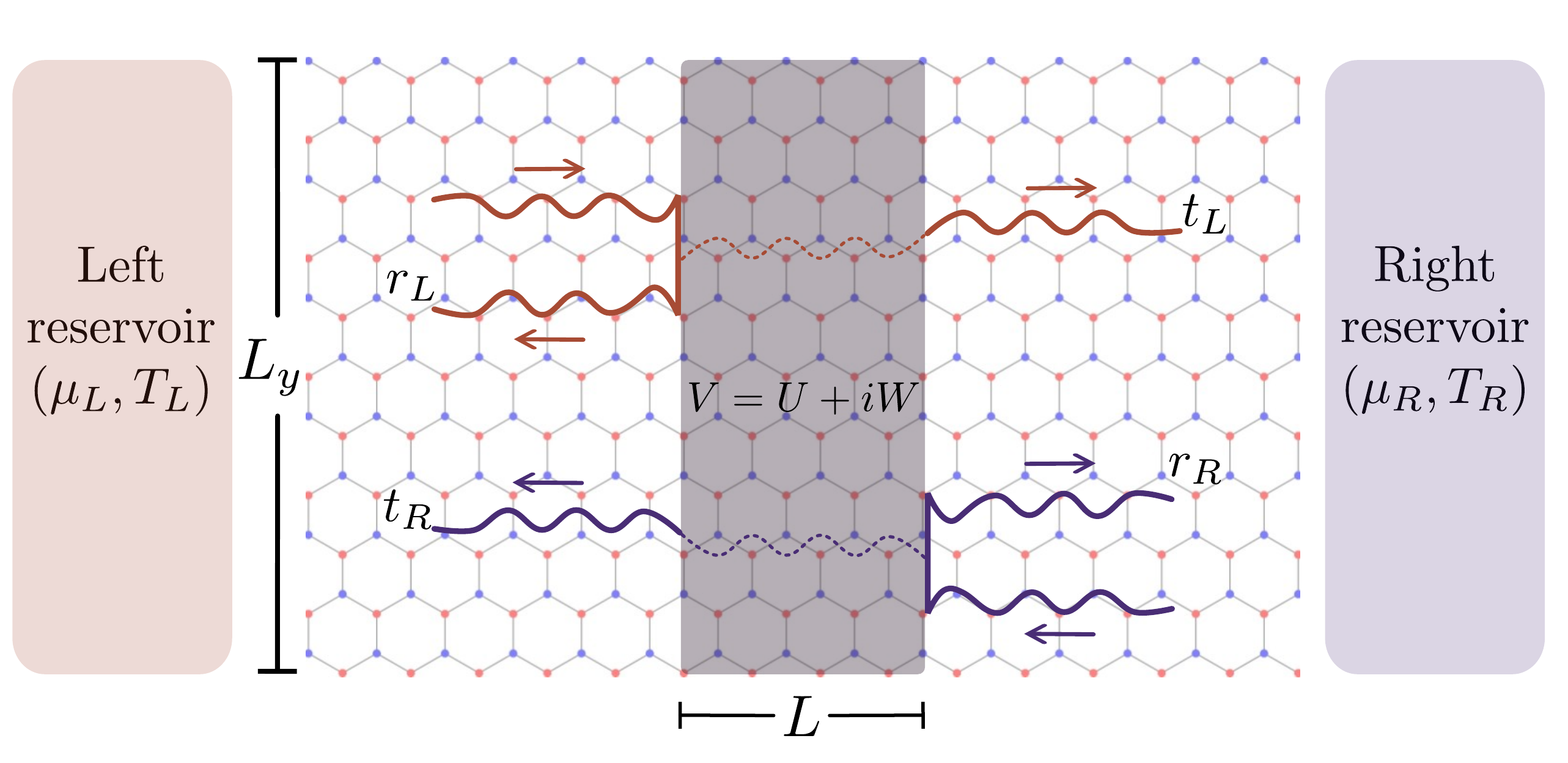}
    \caption{Setup for a monolayer graphene sheet with a finite complex barrier. Red trajectories indicate scattering states injected from the left reservoir, which is in equilibrium at chemical potential $\mu_L$ and temperature $T_L$; using flux normalization the reflected and transmitted amplitudes are $r_L$ and $t_L$. Violet trajectories indicate states injected from the right reservoir, which is in equilibrium at $\mu_R$ and $T_R$ with amplitudes $r_R$ and $t_R$.}
    \label{fig:landauer_setup}
\end{figure}

\subsection{Charge and heat currents}
 
The main difference from the Hermitian case is that, due to the complex potential in region II, the probability current, and therefore the charge and heat currents, are no longer independent of $x$ throughout the system. As a result, the currents evaluated in the left and right leads generally differ, leading to different conductance values.

To begin, we consider a monolayer graphene sheet connected to two graphene reservoirs on the left and right of a non-Hermitian barrier, as shown in FIG.~\ref{fig:landauer_setup}. Two identical graphene leads connect the reservoirs to the barrier region. A small voltage bias and a small temperature difference are applied between the reservoirs, so that the net transport is from left to right along the $x$ axis. The left reservoir is in equilibrium at chemical potential $\mu_L$ and temperature $T_L$, with Fermi-Dirac distribution $f_0(E,\mu_L,T_L)\equiv f_L(E)$, while the right reservoir is in equilibrium at chemical potential $\mu_R$ and temperature $T_R$, with $f_0(E,\mu_R,T_R)\equiv f_R(E)$. These reservoir distributions set the occupations of the incoming scattering states in the corresponding leads. Because the barrier is non-Hermitian, particle flux is not conserved inside region II, and the current evaluated in the left lead generally differs from the current evaluated in the right lead; we therefore compute both currents separately.

We first compute the charge current in the left lead. Following the standard heuristic procedure used to derive the Landauer formula \cite{Nazarov,Landauer1970}, we write the charge current as the net flux along the transport direction, weighted by the reservoir occupations and by the scattering probabilities. Carriers incident at an angle $\phi$ have longitudinal velocity $v_F\cos\phi$ along $x$, and the number of available states at energy $E$ is accounted for by the graphene density of states $D(E)$. With these ingredients, the current in the left lead can be written as
\begin{widetext}
 \begin{align}
    I_{L}=eg_sg_v\,\frac{L_y}{2\pi}\int_{-\infty}^{\infty}dE\int_{-\pi/2}^{\pi/2}d\phi \,v_F\cos\phi\,D(E)
    \Big\lbrace\big[1-\mathcal{R}(E,\phi)\big]f_L(E)-\mathcal{T}(E,\phi)f_R(E)\Big\rbrace.\label{eq:I_L}
\end{align}   
\end{widetext}
where $e>0$ denotes the elementary charge. For convenience, we define the charge current to be positive when the net electron flow is from left to right. The factors $g_s$ and $g_v$ are the spin and valley degeneracies, and $L_y$ is the sample width in the transverse $y$ direction. The term proportional to $f_L(E)$ accounts for particles injected from the left reservoir: right-moving states in the left lead, including the incident and reflected components, are populated according to $f_L(E)$, and their net contribution is reduced by the reflection probability $\mathcal{R}(E,\phi)$. The term proportional to $f_R(E)$ accounts for particles injected from the right reservoir that are transmitted into the left lead with probability $\mathcal{T}(E,\phi)$. The overall sign reflects the electron charge. Here $D(E)$ denotes the usual two-dimensional density of states per unit area, per unit energy, per spin and valley for monolayer graphene,
\begin{align}
    D(E)=\frac{|E|}{2\pi(\hbar v_F)^2}.
\end{align}
By the same reasoning, we obtain that the charge current in the right lead must be
\begin{widetext}
 \begin{align}
    I_{R}=eg_sg_v\,\frac{L_y}{2\pi}\int_{-\infty}^{\infty}dE\int_{-\pi/2}^{\pi/2}d\phi \,v_F\cos\phi\,D(E)
    \Big\lbrace \mathcal{T}(E,\phi)f_L(E)-\big[1-\mathcal{R}(E,\phi)\big]f_R(E)\Big\rbrace.\label{eq:I_R}
\end{align}   
Introducing the sign factors $s_R=+1$ and $s_L=-1$, we can write the charge currents in Eqs.~\eqref{eq:I_L} and \eqref{eq:I_R} in the unified form
\begin{equation}
I_{R/L}=eg_sg_v\,\frac{L_y}{2\pi}\int_{-\infty}^{\infty}dE\int_{-\pi/2}^{\pi/2}d\phi \,
v_F\cos\phi\,D(E)\,s_{R/L}
\Big\{
\mathcal{T}(E,\phi)f_{L/R}(E)-\big[1-\mathcal{R}(E,\phi)\big]f_{R/L}(E)
\Big\}.
\label{eq:I_R/L}
\end{equation}
\end{widetext}
\noindent
Now we turn to heat transport. To proceed, we follow Refs.~\cite{Imry1986,Butcher1990,Cuevas2010,Kloeckner2017_PRB96_205405} and first introduce the \textit{energy} current $I_{E,\alpha}$, defined as the net energy flux carried by the scattering states out of reservoir $\alpha\in\{L,R\}$ and into the conductor. With this convention, $I_{E,\alpha}$ is positive when it leaves the reservoir. Then, the energy currents in the two leads can be written in the unified form
\begin{widetext}
\begin{equation}
I_{E,R/L}=g_sg_v\,\frac{L_y}{2\pi}\int_{-\infty}^{\infty}dE\int_{-\pi/2}^{\pi/2}d\phi \,v_F\cos\phi\,D(E)\,E\,
\Big\{
\big[1-\mathcal{R}(E,\phi)\big]f_{R/L}(E)-\mathcal{T}(E,\phi)f_{L/R}(E)
\Big\}.
\label{eq:I_E_R/L}
\end{equation}
\end{widetext}

\noindent
In complete analogy, we define the heat current in lead $\alpha=L,R$ as the energy flux leaving reservoir $\alpha$, measured relative to the chemical potential of the same reservoir. We write $\dot{Q}_{\alpha}=I_{E,\alpha}-\mu_\alpha I_{N,\alpha}$, where $I_{N,\alpha}$ is the particle current leaving reservoir $\alpha$. The occupation factors are still set by the reservoir that injects each incoming state: in the left lead, right-moving states are populated by $f_L(E)$, while left-moving states come from reflection of left-injected waves and from waves transmitted from the right reservoir; the same physics applies in the right lead. Then, the heat currents take the unified form
\begin{widetext}
\begin{equation}
\dot{Q}_{R/L}=g_sg_v\,\frac{L_y}{2\pi}\int_{-\infty}^{\infty}dE\int_{-\pi/2}^{\pi/2}d\phi \,v_F\cos\phi\,D(E)\,\big(E-\mu_{R/L}\big)
\Big\{
\big[1-\mathcal{R}(E,\phi)\big]f_{R/L}(E)-\mathcal{T}(E,\phi)f_{L/R}(E)
\Big\}.
\label{eq:I_Q_R/L}
\end{equation}
\end{widetext}

In the Hermitian limit, where the barrier is real, i.e., $W=0$, and $\mathcal{R}+\mathcal{T}=1$, our expressions reduce to the standard two-terminal result \cite{Cuevas2010}. In the absence of a voltage bias, $\mu_L=\mu_R$, the heat currents defined as leaving the reservoirs are equal in magnitude and opposite in sign, $\dot{Q}_L=-\dot{Q}_R$. At finite bias, the magnitude of the sum $\dot{Q}_L+\dot{Q}_R$ equals $IV$, which is the total Joule power dissipated in the graphene reservoirs \cite{Cuevas2010}.

\subsection{Linear-response thermoelectric coefficients}

Now we are in position to compute the thermoelectric transport coefficients \cite{Imry1986,Butcher1990,Cuevas2010,Kloeckner2017_PRB96_205405}. In this work we focus on electronic heat transport and neglect the contribution from graphene lattice vibrations.

In the general non-Hermitian case, particle and energy flux are not conserved inside the barrier region, so the currents evaluated in the left and right graphene leads generally differ. We therefore work with lead-resolved charge and heat currents, $I_{\alpha}$ and $\dot{Q}_{\alpha}$. In linear response, these currents are driven by a small chemical-potential difference $\Delta\mu=\mu_L-\mu_R$ and a small temperature difference $\Delta T=T_L-T_R$ between the reservoirs.

An important point in the non-Hermitian case is how the voltage and temperature biases are applied, since the lead-resolved currents can depend on this choice, unlike in the Hermitian limit. We therefore allow for an asymmetric bias drop and write
\begin{align}
\mu_{R/L}&=\mu\mp \eta_{R/L}\,eV,
&
T_{R/L}&=T\mp \xi_{R/L}\,\Delta T,
\label{eq:bias_partition}
\end{align}
where
\begin{align}
\eta_L=\eta,\qquad \eta_R=1-\eta,\qquad
\xi_L=\xi,\qquad \xi_R=1-\xi,\label{eq:eta_xi}
\end{align}
with $0\le\eta,\xi\le 1$ and $V,\Delta T>0$. The symmetric choice corresponds to $\eta=\xi=1/2$. By construction, these definitions give $\mu_L-\mu_R=eV$ and $T_L-T_R=\Delta T$.

We now expand the expressions for the charge and heat currents obtained in the previous subsection to first order in the applied biases. Specifically, we linearize $I_{\alpha}$ and $\dot{Q}_{\alpha}$ with respect to the voltage bias $V$ and the temperature difference $\Delta T$ between the reservoirs, keeping only terms up to ${\cal O}(V,\Delta T)$. Then, using Eq.~\eqref{eq:I_R/L}, the charge currents in the two leads are, to linear order in $V$ and $\Delta T$,
\begin{widetext}
\begin{align}
I_{R/L}\simeq&\frac{2eL_y}{\pi}\int_{-\infty}^{\infty} dE\int_{-\pi/2}^{\pi/2} d\phi \,v_F\cos\phi\,D(E)
\Bigg\{
s_{R/L}\Big[\mathcal{T}(E,\phi)+\mathcal{R}(E,\phi)-1\Big]f_0(E)\notag
\\
&+\Big(-\frac{\partial f_0}{\partial E}\Big)\Bigg[
\Big(\eta_{R/L}\big[1-\mathcal{R}(E,\phi)\big]+\eta_{L/R}\mathcal{T}(E,\phi)\Big)eV
+\frac{E-\mu}{T}\Big(\xi_{R/L}\big[1-\mathcal{R}(E,\phi)\big]+\xi_{L/R}\mathcal{T}(E,\phi)\Big)\Delta T
\Bigg]
\Bigg\},
\label{eq:I_R/L_linear}
\end{align}

In the same way, using Eq.~\eqref{eq:I_Q_R/L}, the heat currents in the two leads can be expanded to first order in $V$ and $\Delta T$, yielding
\begin{align}
\dot{Q}_{R/L}\simeq&\frac{2L_y}{\pi}\int_{-\infty}^{\infty}dE\int_{-\pi/2}^{\pi/2}d\phi \,v_F\cos\phi\,D(E)
\Bigg\{
\big(E-\mu\big)\big[1-\mathcal{R}(E,\phi)-\mathcal{T}(E,\phi)\big]f_0(E)\notag
\\
&+s_{R/L}\eta_{R/L}\big[1-\mathcal{R}(E,\phi)-\mathcal{T}(E,\phi)\big]f_0(E)\,eV\notag
\\
&-s_{R/L}\Big(-\frac{\partial f_0}{\partial E}\Big)\big(E-\mu\big)\Bigg[
\Big(\eta_{R/L}\big[1-\mathcal{R}(E,\phi)\big]+\eta_{L/R}\mathcal{T}(E,\phi)\Big)eV\notag
\\
&+\frac{E-\mu}{T}\Big(\xi_{R/L}\big[1-\mathcal{R}(E,\phi)\big]+\xi_{L/R}\mathcal{T}(E,\phi)\Big)\Delta T
\Bigg]
\Bigg\}.
\label{eq:Q_R/L_linear}
\end{align}
\end{widetext}

Eqs.~\eqref{eq:I_R/L_linear} and \eqref{eq:Q_R/L_linear} show that, besides the terms linear in $V$ and $\Delta T$, there is a contribution that remains finite at $V=0$ and $\Delta T=0$. This contribution is proportional to $1-\mathcal{R}(E,\phi)-\mathcal{T}(E,\phi)$ and to the equilibrium distribution $f_0(E)$. We refer to it as a \textit{zero-bias current} and the corresponding \textit{zero-bias heat flow}. In our setup this term is entirely generated by the non-Hermitian part of the barrier, that is, the imaginary potential $W$, which breaks flux conservation in region II. In the Hermitian limit $W=0$ one has $\mathcal{R}+\mathcal{T}=1$, so these equilibrium terms cancel identically and the currents vanish when the two reservoirs share the same $\mu$ and $T$.

At $T=0$ the contrast between the Hermitian and non-Hermitian cases becomes even sharper. In the Hermitian case, the equilibrium terms in Eq.~\eqref{eq:Q_R/L_linear} vanish, and the remaining linear-response contributions are proportional to $(E-\mu)\,(-\partial f_0/\partial E)$, which gives zero because $f_0(E)=\Theta(\mu-E)$ and $(E-\mu)\delta(E-\mu)=0$. Therefore, $\dot{Q}_\alpha$ is zero at $T=0$ to first order in $V$ and $\Delta T$, and finite heat flow only appears beyond linear response. In contrast, for $W\neq 0$ there are extra terms proportional to $f_0(E)$, rather than to $-\partial f_0/\partial E$, multiplying $1-\mathcal{R}(E,\phi)-\mathcal{T}(E,\phi)$. These terms, including the contribution $(E-\mu)\,[1-\mathcal{R}(E,\phi)-\mathcal{T}(E,\phi)]\,f_0(E)$ in Eq.~\eqref{eq:Q_R/L_linear}, generally survive at $T=0$ and yield a nonzero \textit{zero-bias heat flow}, which is entirely controlled by the imaginary potential.

In order to obtain the thermoelectric transport coefficients, we note that for each lead, Eqs.~\eqref{eq:I_R/L_linear} and \eqref{eq:Q_R/L_linear} can be cast in the form
\begin{align}
\begin{pmatrix}
I_{\alpha}\\
\dot{Q}_{\alpha}
\end{pmatrix}
=\begin{pmatrix}
I^0_{\alpha}\\
\dot{Q}^0_{\alpha}
\end{pmatrix}+
\begin{pmatrix}
G_{\alpha} & L_{\alpha}\\
M_{\alpha} & K_{\alpha}
\end{pmatrix}
\begin{pmatrix}
V\\
\Delta T
\end{pmatrix},
\label{eq:current_vs_grads}
\end{align}
where the first term in the right-hand side corresponds to the zero-bias current and heat flow mentioned earlier, and the coefficients $G_{\alpha},L_{\alpha},M_{\alpha},K_{\alpha}$ are defined from the corresponding Landauer expressions for $I_{\alpha}$ and $\dot{Q}_{\alpha}$.

From Eq.~\eqref{eq:I_R/L_linear}, the electrical conductance and the thermoelectric coefficient in the two leads are
\begin{widetext}
\begin{align}
G_{R/L}(\mu,T)=&\frac{2e^2L_y}{\pi}\int_{-\infty}^{\infty} dE\int_{-\pi/2}^{\pi/2} d\phi \,v_F\cos\phi\,D(E)\,
\Big(-\frac{\partial f_0}{\partial E}\Big)
\Big[\eta_{R/L}\big(1-\mathcal{R}(E,\phi)\big)+\eta_{L/R}\mathcal{T}(E,\phi)\Big],
\label{eq:G_R/L}
\\
L_{R/L}(\mu,T)=&\frac{2eL_y}{\pi}\int_{-\infty}^{\infty} dE\int_{-\pi/2}^{\pi/2} d\phi \,v_F\cos\phi\,D(E)
\Big(-\frac{\partial f_0}{\partial E}\Big)\frac{E-\mu}{T}
\Big[\xi_{R/L}\big(1-\mathcal{R}(E,\phi)\big)+\xi_{L/R}\mathcal{T}(E,\phi)\Big],
\label{eq:L_R/L}
\end{align}
Now, from Eq.~\eqref{eq:Q_R/L_linear}, the coefficients $K_{R/L}$ and $M_{R/L}$ are
\begin{align}
K_{R/L}(\mu,T)=&-s_{R/L}\frac{2L_y}{\pi}\int_{-\infty}^{\infty} dE\int_{-\pi/2}^{\pi/2} d\phi \,v_F\cos\phi\,D(E)\,
\Big(-\frac{\partial f_0}{\partial E}\Big)\,
\frac{(E-\mu)^2}{T}\,
\Big[\xi_{R/L}\big(1-\mathcal{R}(E,\phi)\big)+\xi_{L/R}\mathcal{T}(E,\phi)\Big],
\label{eq:K_R/L}
\\
M_{R/L}(\mu,T)=&\frac{2eL_y}{\pi}\int_{-\infty}^{\infty} dE\int_{-\pi/2}^{\pi/2} d\phi \,v_F\cos\phi\,D(E)
\Bigg\{
s_{R/L}\eta_{R/L}\big[1-\mathcal{R}(E,\phi)-\mathcal{T}(E,\phi)\big]f_0(E)
\nonumber\\
&\qquad\qquad
-s_{R/L}(E-\mu)\Big(-\frac{\partial f_0}{\partial E}\Big)
\Big[\eta_{R/L}\big(1-\mathcal{R}(E,\phi)\big)+\eta_{L/R}\mathcal{T}(E,\phi)\Big]
\Bigg\}.
\label{eq:M_R/L}
\end{align}
\end{widetext}
Because the non-Hermitian barrier produces a finite \textit{zero-bias} contribution, $I_\alpha^0$ and $\dot{Q}_\alpha^0$ in Eq.~\eqref{eq:current_vs_grads}, the usual ratio definitions of thermoelectric coefficients must be understood as \textit{linear-response slopes} rather than as ratios through the origin \cite{Yang2026}. In practice, one considers the response of the currents to small changes in $V$ and $\Delta T$ around the reference state, keeping the zero-bias offsets fixed.

In general, the usual Onsager relation does not hold in the present non-Hermitian problem. This can be seen directly by comparing Eqs.~\eqref{eq:L_R/L} and \eqref{eq:M_R/L}. The coefficient $L_{\alpha}$ is generated only by the thermal derivative term proportional to $(E-\mu)(-\partial f_0/\partial E)$, whereas $M_{\alpha}$ contains, in addition, a genuine non-Hermitian contribution proportional to $f_0(E)\,[1-\mathcal{R}(E,\phi)-\mathcal{T}(E,\phi)]$. Since this extra term is nonzero whenever the scattering is nonunitary, one does not generally have $M_{\alpha}=-T L_{\alpha}$. Physically, this reflects the fact that the complex barrier acts as a source or sink of flux and produces finite zero-bias charge and heat currents, so the linear response is built around a shifted reference state rather than around the standard Hermitian equilibrium state. Only in the Hermitian limit, where $\mathcal{R}+\mathcal{T}=1$ and the zero-bias offsets vanish, does the usual Onsager structure reemerge.

At zero temperature, one has $f_0(E)=\Theta(\mu-E)$ and $-\partial f_0/\partial E=\delta(E-\mu)$. Therefore, the energy integrals in Eq.~\eqref{eq:G_R/L} collapse to the chemical potential, while all terms in Eq.~\eqref{eq:L_R/L} and Eq.~\eqref{eq:K_R/L} vanish because they are proportional to $(E-\mu)\delta(E-\mu)$ or $(E-\mu)^2\delta(E-\mu)$. In the same way, the second term in Eq.~\eqref{eq:M_R/L} also vanishes, and only the zero-bias non-Hermitian contribution proportional to $f_0(E)$ remains. Hence, the zero-temperature coefficients are $L_R(\mu)=\,L_L(\mu)=K_R(\mu)=K_L(\mu)=0$, while
\begin{widetext}
\begin{align}
G_{R/L}(\mu)=&\frac{2e^2L_y}{\pi}\int_{-\pi/2}^{\pi/2} d\phi\, v_F\cos\phi\, D(\mu)\,
\Big[\eta_{R/L}\big(1-\mathcal{R}(\mu,\phi)\big)+\eta_{L/R}\mathcal{T}(\mu,\phi)\Big],
\label{eq:G_R/L_T0}
\\
M_{R/L}(\mu)=&\frac{2eL_y}{\pi}\,s_{R/L}\eta_{R/L}\int_{-\infty}^{\mu} dE\int_{-\pi/2}^{\pi/2} d\phi\, v_F\cos\phi\, D(E)\,
\big[1-\mathcal{R}(E,\phi)-\mathcal{T}(E,\phi)\big].
\label{eq:M_R/L_T0}
\end{align}
\end{widetext}
The Seebeck coefficient (thermopower) $S_{\alpha}$ is obtained under open-circuit conditions in lead $\alpha$, namely $I_{\alpha}=0$. Physically, $S_{\alpha}$ measures how much voltage must develop to cancel the charge flow induced by a temperature difference. Since $I_\alpha$ contains the offset $I_\alpha^0$, open circuit implies a nonzero baseline voltage $V_{0,\alpha}$ even when $\Delta T=0$. We therefore define $S_\alpha$ from the slope of the open-circuit voltage,
\begin{align}
S_{\alpha}=-\left.\frac{dV}{d(\Delta T)}\right|_{I_{\alpha}=0},
\label{eq:seebeck_def}
\end{align}
which, using Eq.~\eqref{eq:current_vs_grads}, gives
\begin{align}
S_{\alpha}=\frac{L_{\alpha}}{G_{\alpha}}.
\label{eq:seebeck_ratio}
\end{align}

The thermal conductance $\kappa_{\alpha}$ is defined as the heat flow driven by a temperature difference under the same open-circuit condition, so that charge transport is blocked. In the presence of $\dot{Q}_\alpha^0$, the relevant quantity is again the linear-response slope, and we define
\begin{align}
\kappa_{\alpha}=\left.\frac{d\dot{Q}_{\alpha}}{d(\Delta T)}\right|_{I_{\alpha}=0}.
\label{eq:kappa_def}
\end{align}
Combining Eq.~\eqref{eq:kappa_def} with Eq.~\eqref{eq:current_vs_grads} yields
\begin{align}
\kappa_{\alpha}=K_{\alpha}-\frac{M_{\alpha}L_{\alpha}}{G_{\alpha}},
\label{eq:kappa_MLK}
\end{align}
which reduces to the usual expression $\kappa_\alpha=-K_\alpha-TL_\alpha^2/G_\alpha$ when the Onsager relation $M_\alpha=-TL_\alpha$ applies in the Hermitian case. We do not write the explicit integral forms of $S_{\alpha}$ and $\kappa_{\alpha}$ since they are straightforward combinations of the coefficients $G_{\alpha},L_{\alpha},M_{\alpha},K_{\alpha}$ given above, but their closed expressions become lengthy.

Finally, the efficiency of thermoelectric conversion is commonly characterized by the dimensionless \textit{figure of merit}
\begin{align}
ZT_{\alpha}=\frac{S_{\alpha}^2G_{\alpha}T}{\kappa_{\alpha}}.
\label{eq:ZT_def}
\end{align}
It is worth noting that, in our treatment of the figure of merit, we use the electronic thermal conductance $\kappa_{\alpha}$ only, and we do not include the lattice contribution from graphene phonons that would enter the total thermal conductance. A larger $ZT_{\alpha}$ indicates a better thermoelectric performance, since it requires a large thermopower $S_{\alpha}$ and electrical conductance $G_{\alpha}$ together with a small thermal conductance $\kappa_{\alpha}$.

\section{Results and Discussion} \label{results_discussion_section}

In this section we present the main results for transport across a finite complex barrier in monolayer graphene. We discuss the angular scattering pattern, the non-Hermitian flux balance, the zero-temperature conductance, and the finite-temperature thermoelectric response. Throughout this section we use standard graphene parameters. In particular, we take $v_F\sim10^{6}\,\mathrm{m/s}$, consider electrons as the carriers, and work with chemical potentials of the order of a few hundred meV, as in typical doped graphene devices. Spin and valley degeneracies are included throughout.

\subsection{Angular behavior of the scattering coefficients}

\begin{figure}[t]
    \centering
    \begin{subfigure}{0.32\linewidth}
        \centering
        \includegraphics[width=\linewidth]{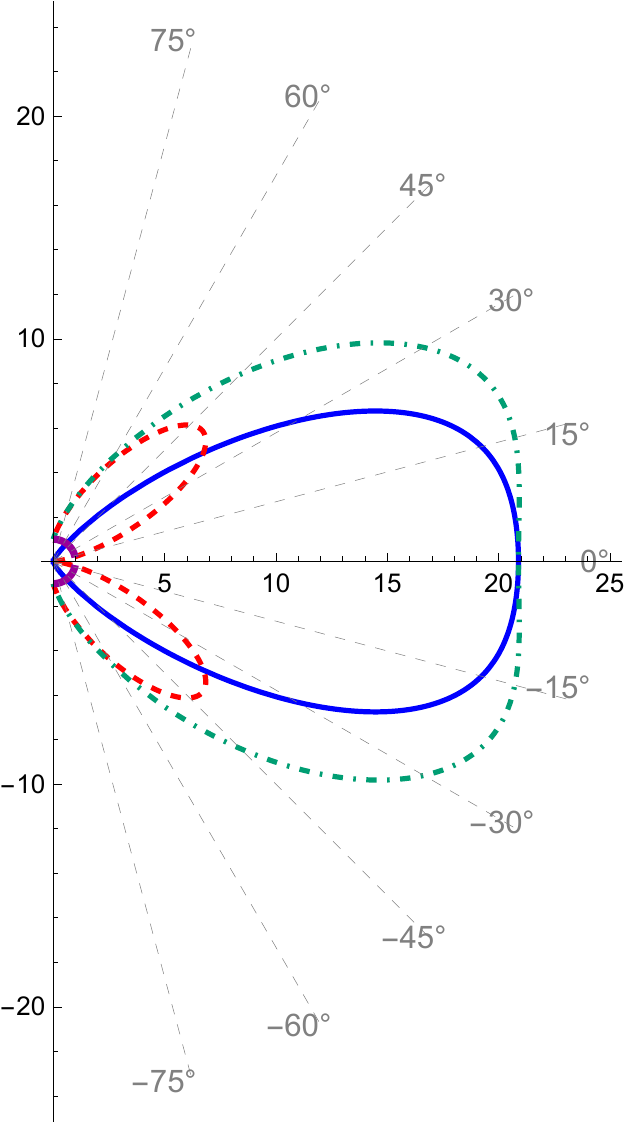}
        \caption{$\mu=20$ meV}
    \end{subfigure}
    \begin{subfigure}{0.32\linewidth}
        \centering
        \includegraphics[width=\linewidth]{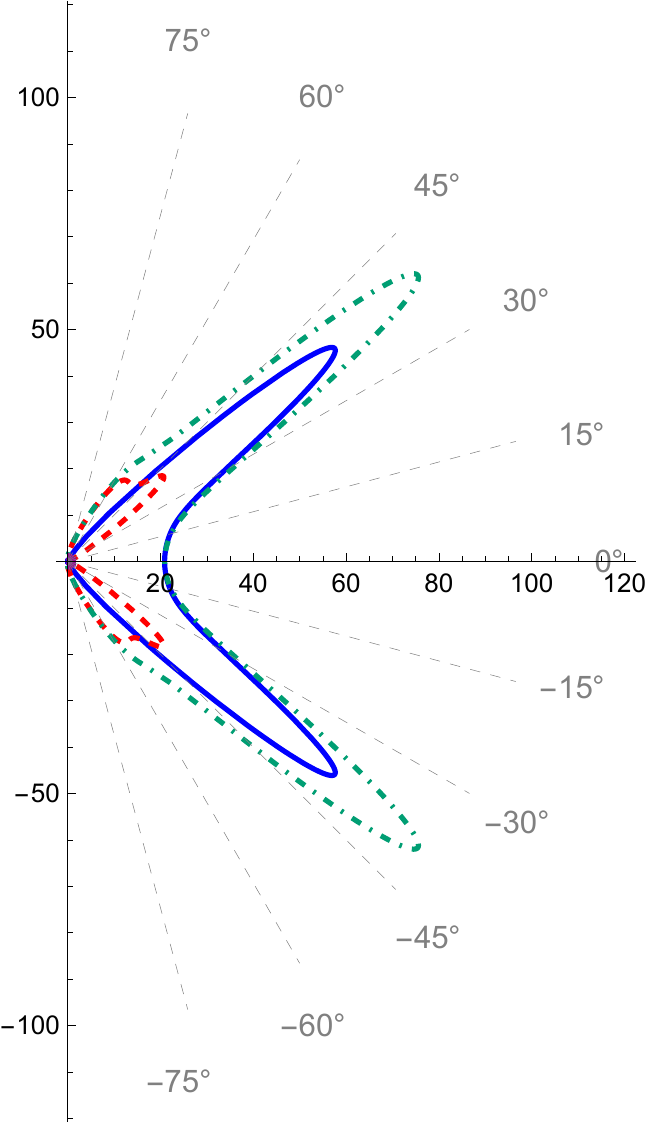}
        \caption{$\mu=60$ meV}
    \end{subfigure}
    \begin{subfigure}{0.32\linewidth}
        \centering
        \includegraphics[width=\linewidth]{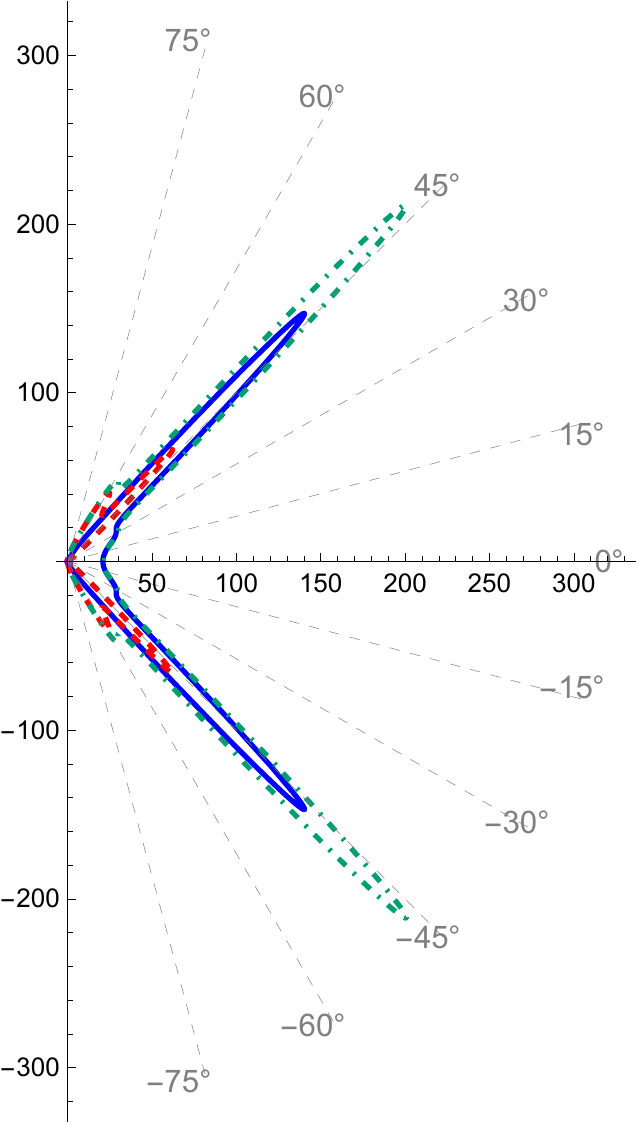}
        \caption{$\mu=100$ meV}
    \end{subfigure}

   \caption{Polar plots of the angular dependence of the transmission function $\mathcal{T}(E,\phi)$ (blue, solid), the reflection function $\mathcal{R}(E,\phi)$ (red, dashed), and the flux-balance term $1+\Gamma(E,\phi)$ (green, dotdashed), evaluated at $E=\mu$, across a complex barrier of length $L=100\,\mathrm{nm}$ with $U=0$ and $W=10\,\mathrm{meV}$. The thick purple curve marks the unit circle.}
    \label{fig:polar_vs_EF}
\end{figure}

\begin{figure}[t]
    \centering
    \begin{subfigure}{0.32\linewidth}
        \centering
        \includegraphics[width=\linewidth]{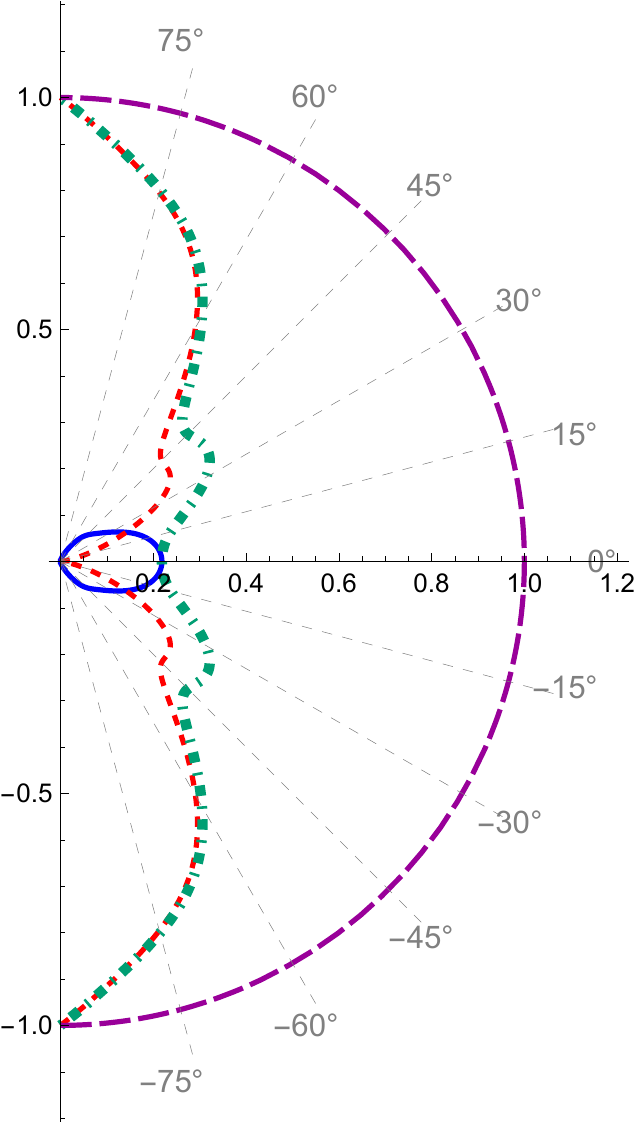}
        \caption{$W=-5$ meV}
    \end{subfigure}
    \begin{subfigure}{0.32\linewidth}
        \centering
        \includegraphics[width=\linewidth]{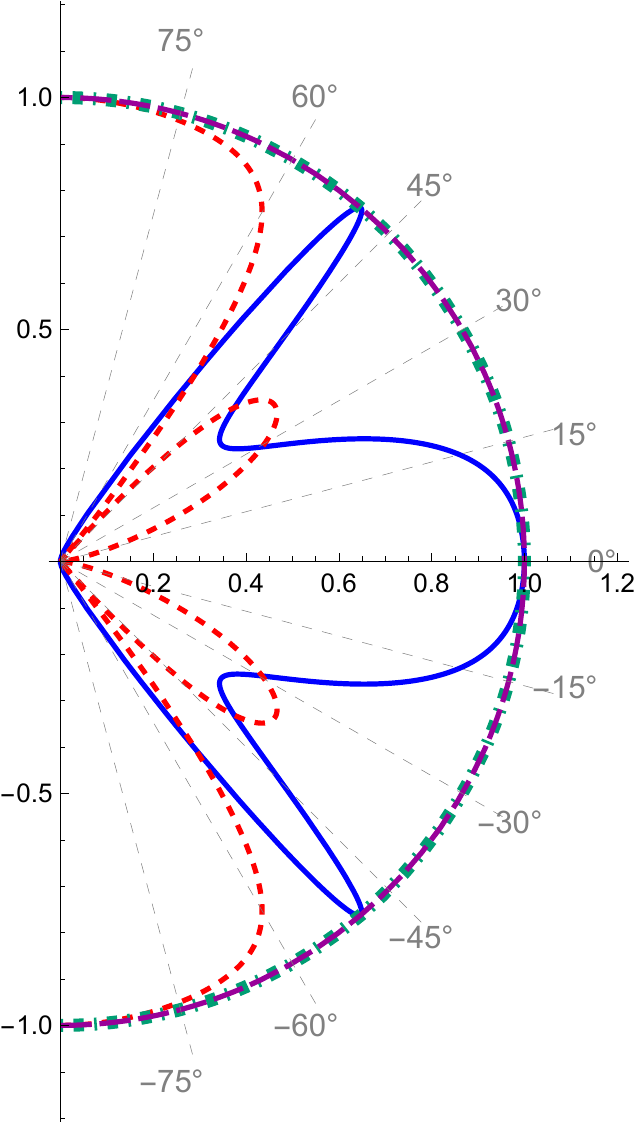}
        \caption{$W=0$ meV}
    \end{subfigure}
    \begin{subfigure}{0.32\linewidth}
        \centering
        \includegraphics[width=\linewidth]{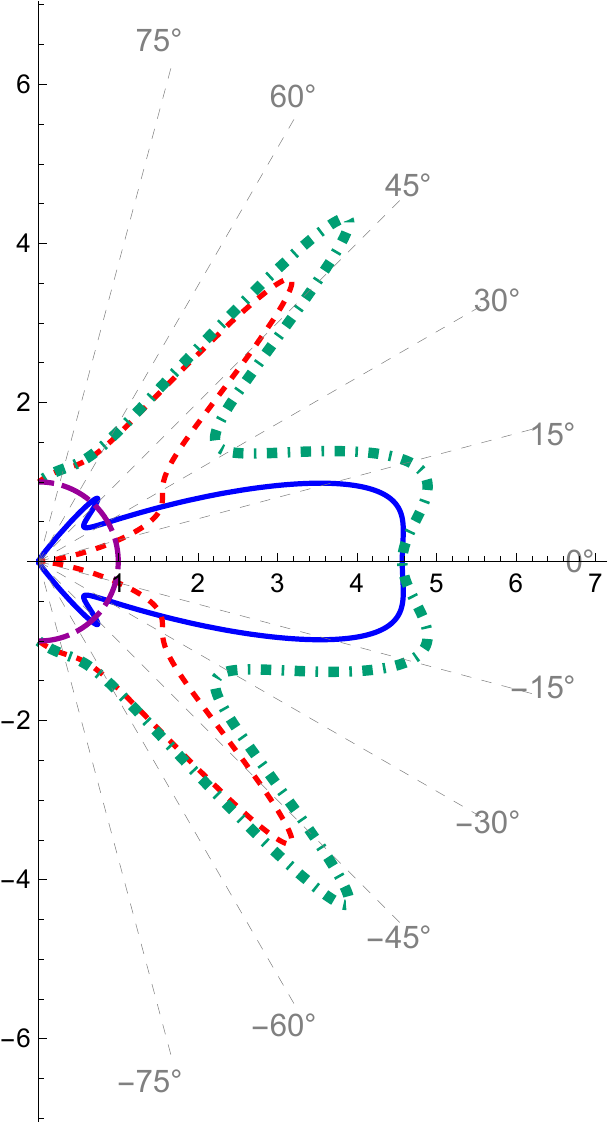}
        \caption{$W=5$ meV}
    \end{subfigure}

    \caption{Polar plots of the angular dependence of the transmission function $\mathcal{T}(E,\phi)$ (blue, solid), the reflection function $\mathcal{R}(E,\phi)$ (red, dashed), and the flux-balance term $1+\Gamma(E,\phi)$ (green, dotdashed), evaluated at $E=\mu$, for a barrier of length $L=100\,\mathrm{nm}$ with $U=200\,\mathrm{meV}$ and $\mu=80\,\mathrm{meV}$. The purple curve marks the unit circle.}
    \label{fig:polar_vs_W}
\end{figure}
FIG.~\ref{fig:polar_vs_EF} shows the angular dependence of the transmission and reflection functions defined in Eqs.~\eqref{eq:T_def} and \eqref{eq:R_def}, together with the gain-loss factor defined in Eq.~\eqref{eq:Gamma}. For all three values of $\mu$, the curves are symmetric under $\phi\to-\phi$, but their magnitude grows strongly with chemical potential. At $\mu=20\,\mathrm{meV}$, the transmission is dominated by a broad lobe around normal incidence, while reflection stays smaller and $1+\Gamma$ already lies above the unit circle. As $\mu$ increases to $60$ and $100\,\mathrm{meV}$, the response is amplified and becomes more anisotropic. The broad forward lobe gives way to two sharper lobes at finite angles. This already shows the main non-Hermitian effect: the imaginary part of the barrier does not only change the total flux, it also redistributes it in angle. The same plots make clear that the Hermitian identity $\mathcal{R}+\mathcal{T}=1$ does not hold here. The correct balance is the generalized relation in Eq.~\eqref{eq:1-R}, where the excess or deficit is measured by $\Gamma(E,\phi)$.

FIG.~\ref{fig:polar_vs_W} shows more clearly what the sign of $W$ does. For $W=0$, the green curve collapses onto the unit circle, so $\Gamma=0$ and $\mathcal{R}+\mathcal{T}=1$. In this limit, our amplitudes reduce, up to notation, to the usual rectangular-barrier result in graphene \cite{Geim_Klein}. The comparison with the literature must therefore be understood at the Hermitian level, since those works consider real electrostatic barriers or $p$-$n$ junctions rather than a complex barrier. In particular, \cite{Geim_Klein} studies chiral tunneling through a single barrier, \cite{Pereira2007} analyzes resonant tunneling in graphene double barriers, and \cite{Shekhar2025} studies locally periodic and superperiodic real barriers. Within that common Hermitian setting, all of them show the same basic trend seen in panel (b): perfect transmission at normal incidence and narrow resonant lobes at finite angles. In \cite{Pereira2007}, these features are linked to resonant hole states inside the barrier, while \cite{Shekhar2025} shows in polar plots that the peaks sharpen as $|\phi|$ grows and explains their multiplicity in terms of transfer-matrix resonances. The experimental works \cite{Huard2007,Sutar2012} do not use the same model either, but they support the same picture of strong angular selectivity in graphene junction transport. Once $W<0$, the whole pattern contracts and $1+\Gamma$ stays mostly below the unit circle, which signals loss and suppresses both the broad forward lobe and the finite-angle resonances. For $W>0$, the opposite occurs: the same selected channels are amplified, $1+\Gamma$ moves above unity, and both $\mathcal{T}$ and $\mathcal{R}$ can exceed one.

Taken together, FIGS.~\ref{fig:polar_vs_EF} and \ref{fig:polar_vs_W} show a simple general trend. The imaginary part of the barrier does not create a completely new angular pattern. It reshapes the usual Hermitian resonances. As $|W|$ grows, the response becomes stronger, the dominant lobes become narrower, and the outgoing flux is concentrated into selected incidence windows. At the same time, the non-Hermitian barrier removes the Hermitian constraint set by Klein tunneling at normal incidence. In the real-barrier problem, flux is conserved and chirality suppresses backscattering at $\phi=0$, which fixes the response to perfect transmission. Here, the term $iW$ makes the barrier region a source or sink of density, as follows from the modified continuity equation and the generalized flux balance. For that reason, the scattering coefficients are no longer constrained by the standard unitary balance, and the response near normal incidence can be either suppressed or enhanced. Since the essential non-Hermitian features already appear for small $|W|$, and larger values mainly sharpen them, in the rest of this section we focus on weak imaginary parts to analyze the transport response more clearly.

\subsection{Non-Hermitian flux balance}

\begin{figure}[!ht]
    \centering
    \begin{subfigure}{\linewidth}
        \centering
        \includegraphics[width=0.8\linewidth]{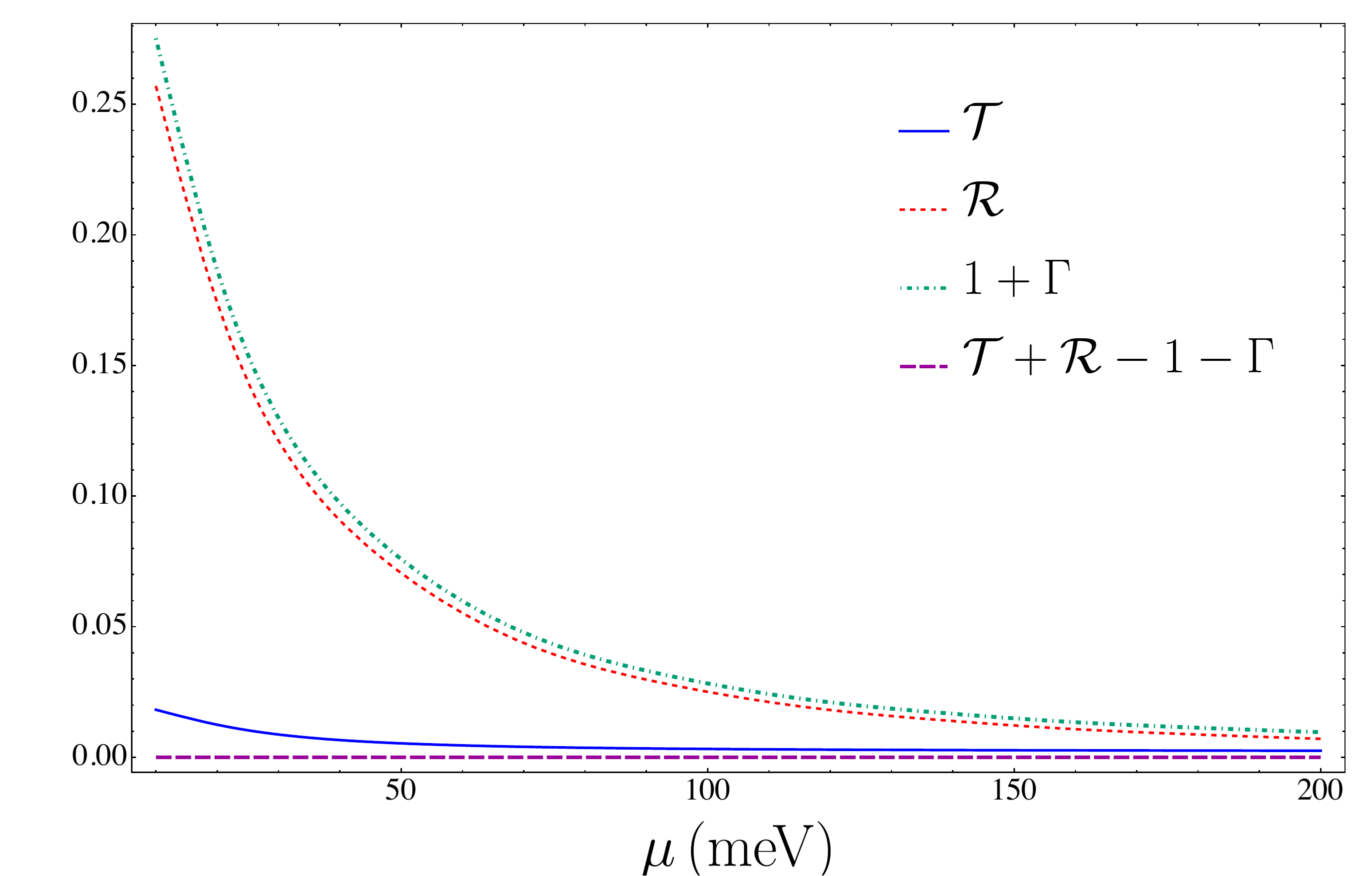}
        \caption{$W=-10$ meV}
    \end{subfigure}
    \begin{subfigure}{\linewidth}
        \centering
        \includegraphics[width=0.8\linewidth]{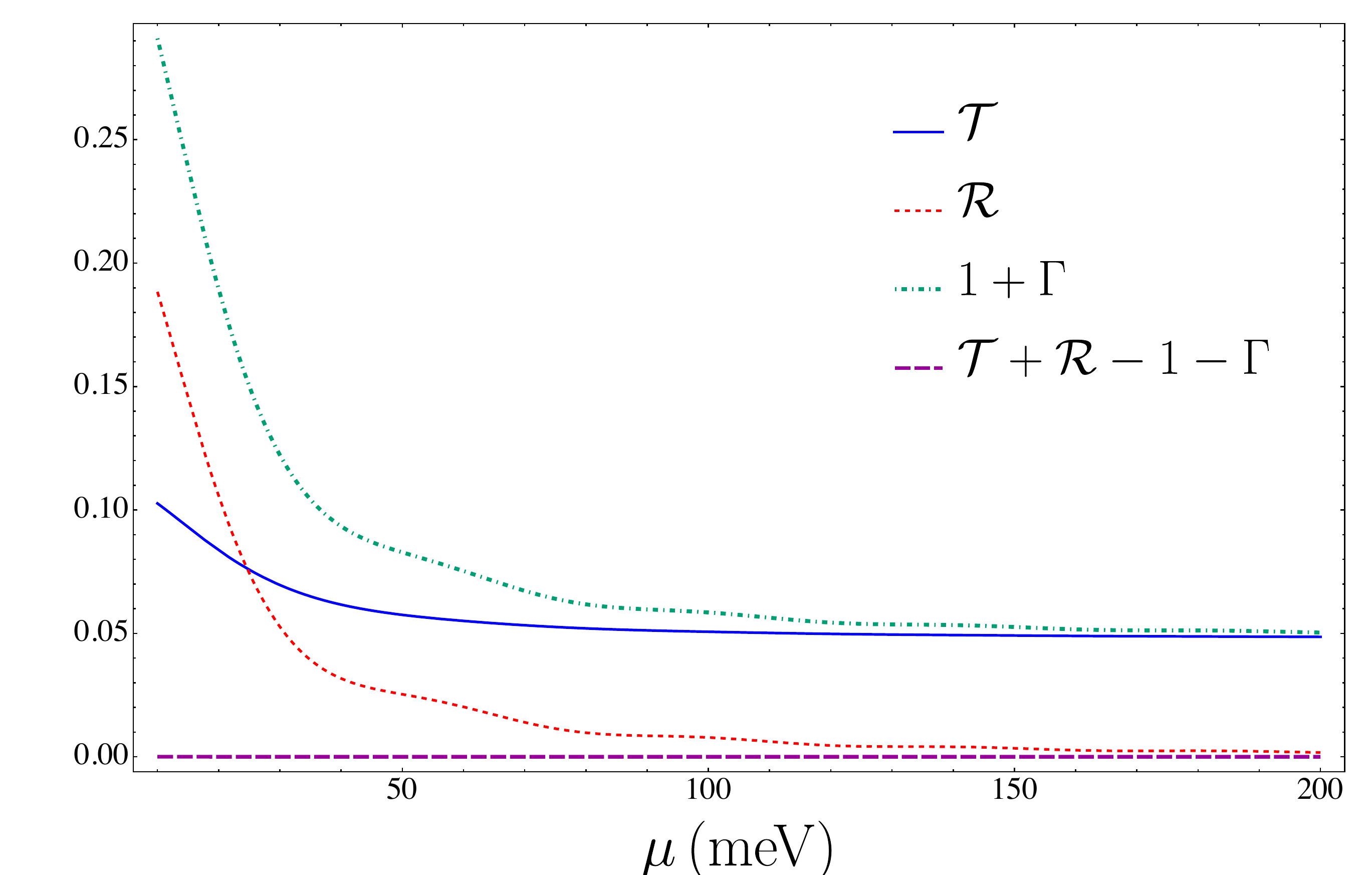}
        \caption{$W=-5$ meV}
    \end{subfigure}
    \begin{subfigure}{\linewidth}
        \centering
        \includegraphics[width=0.8\linewidth]{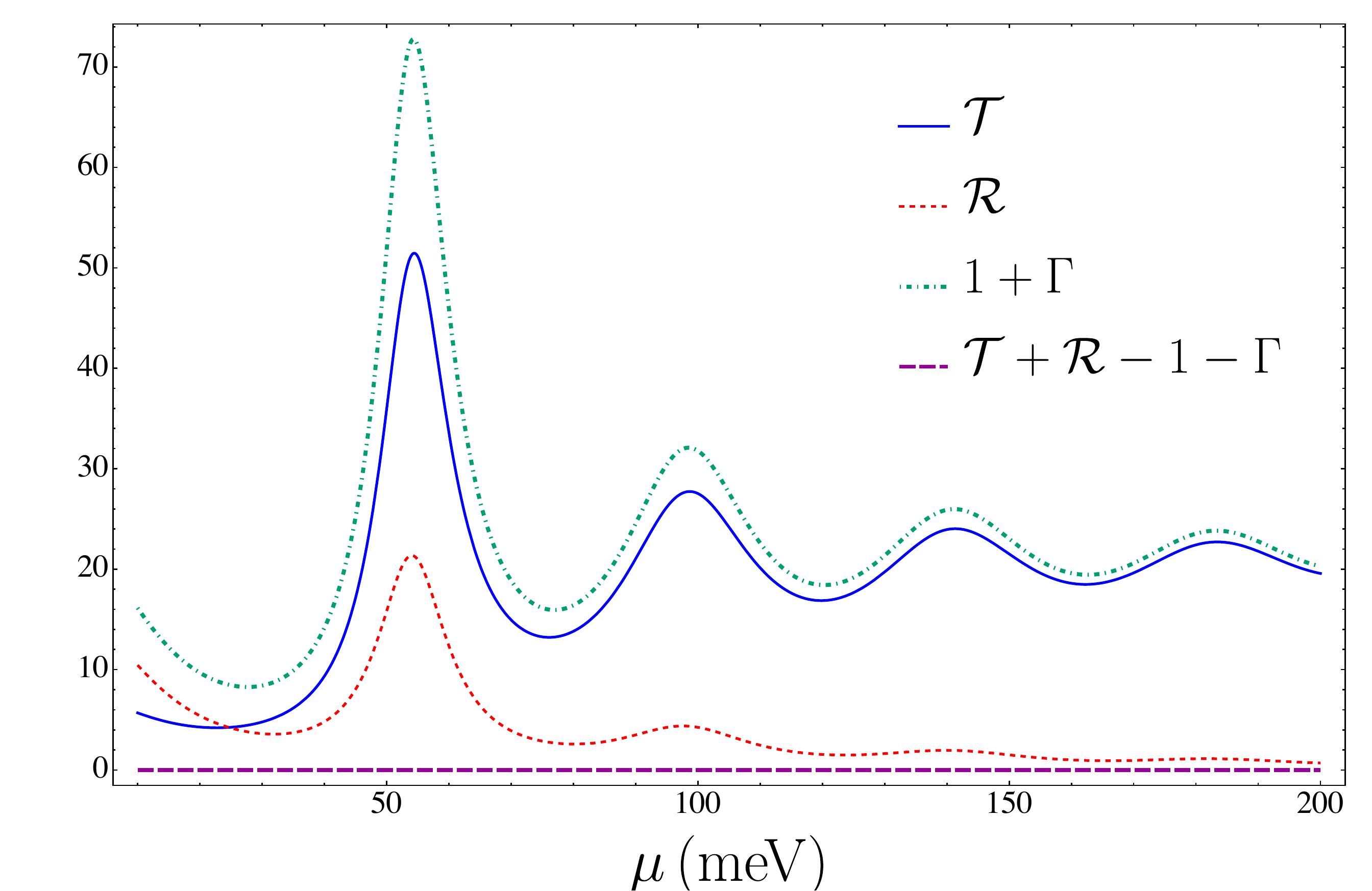}
        \caption{$W=5$ meV}
    \end{subfigure}
    \begin{subfigure}{\linewidth}
        \centering
        \includegraphics[width=0.8\linewidth]{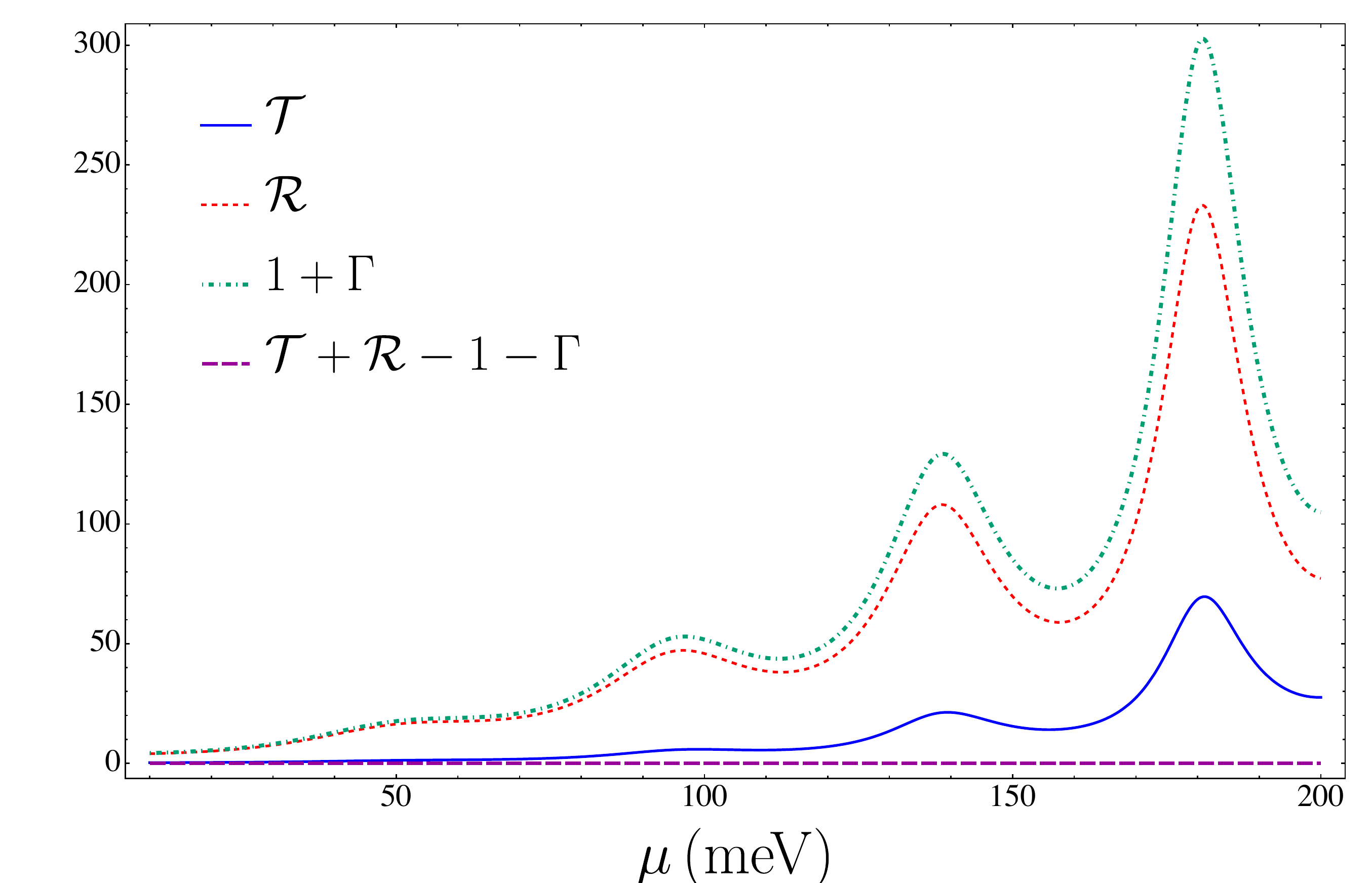}
        \caption{$W=10$ meV}
    \end{subfigure}
    \caption{Chemical-potential dependence of the dimensionless quantities $\mathcal{T}(E,\phi)$, $\mathcal{R}(E,\phi)$, $1+\Gamma(E,\phi)$, and the residual $\mathcal{T}(E,\phi)+\mathcal{R}(E,\phi)-1-\Gamma(E,\phi)$, evaluated at $E=\mu$, for $L=100\,\mathrm{nm}$, $U=0$, and $\phi=\pi/3$.}
    \label{fig:cons_vs_EF}
\end{figure}

\begin{figure}[t]
    \centering
    \begin{subfigure}{\linewidth}
        \centering
        \includegraphics[width=0.8\linewidth]{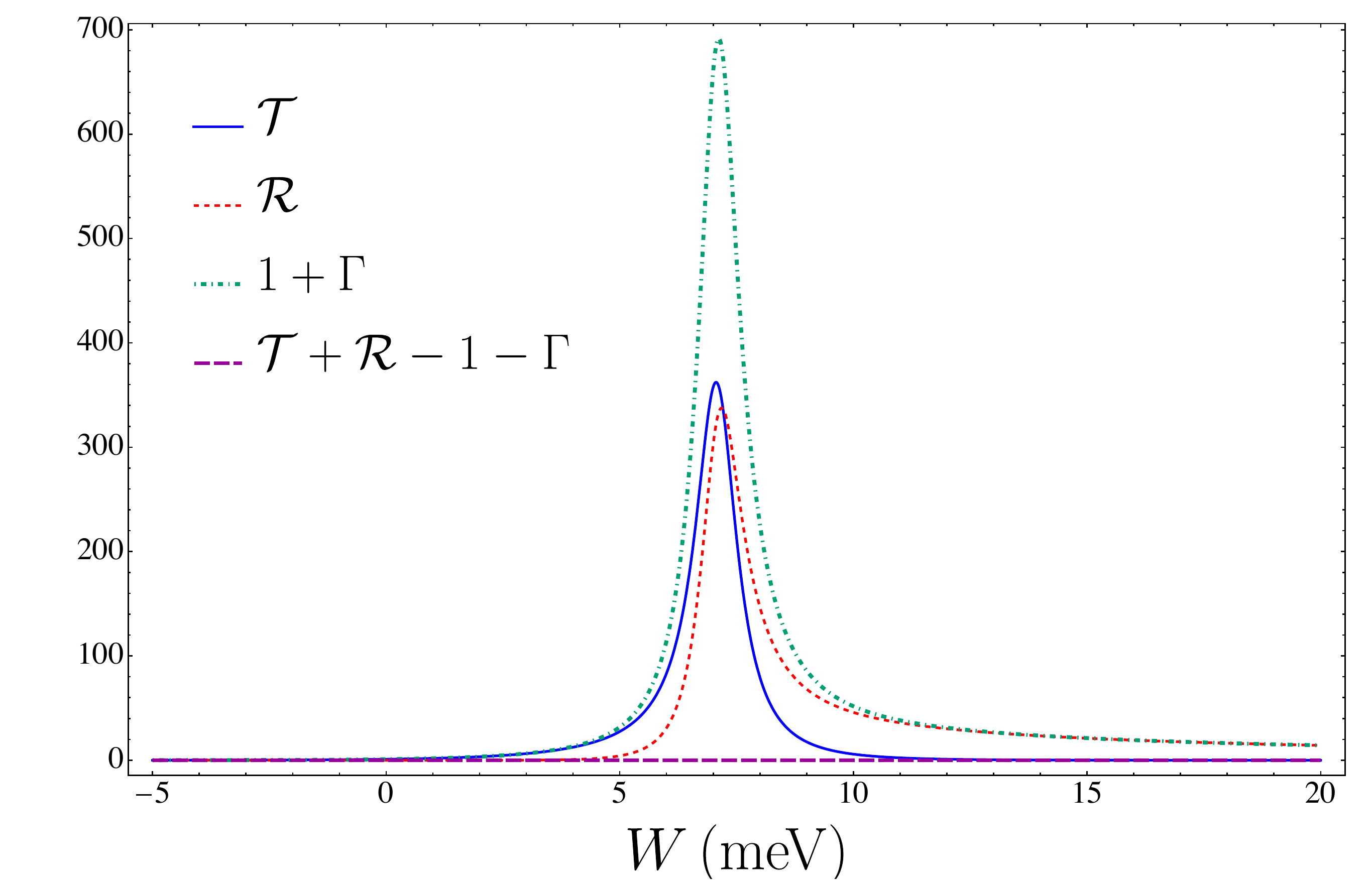}
        \caption{$\mu=100$ meV}
    \end{subfigure}
    \begin{subfigure}{\linewidth}
        \centering
        \includegraphics[width=0.8\linewidth]{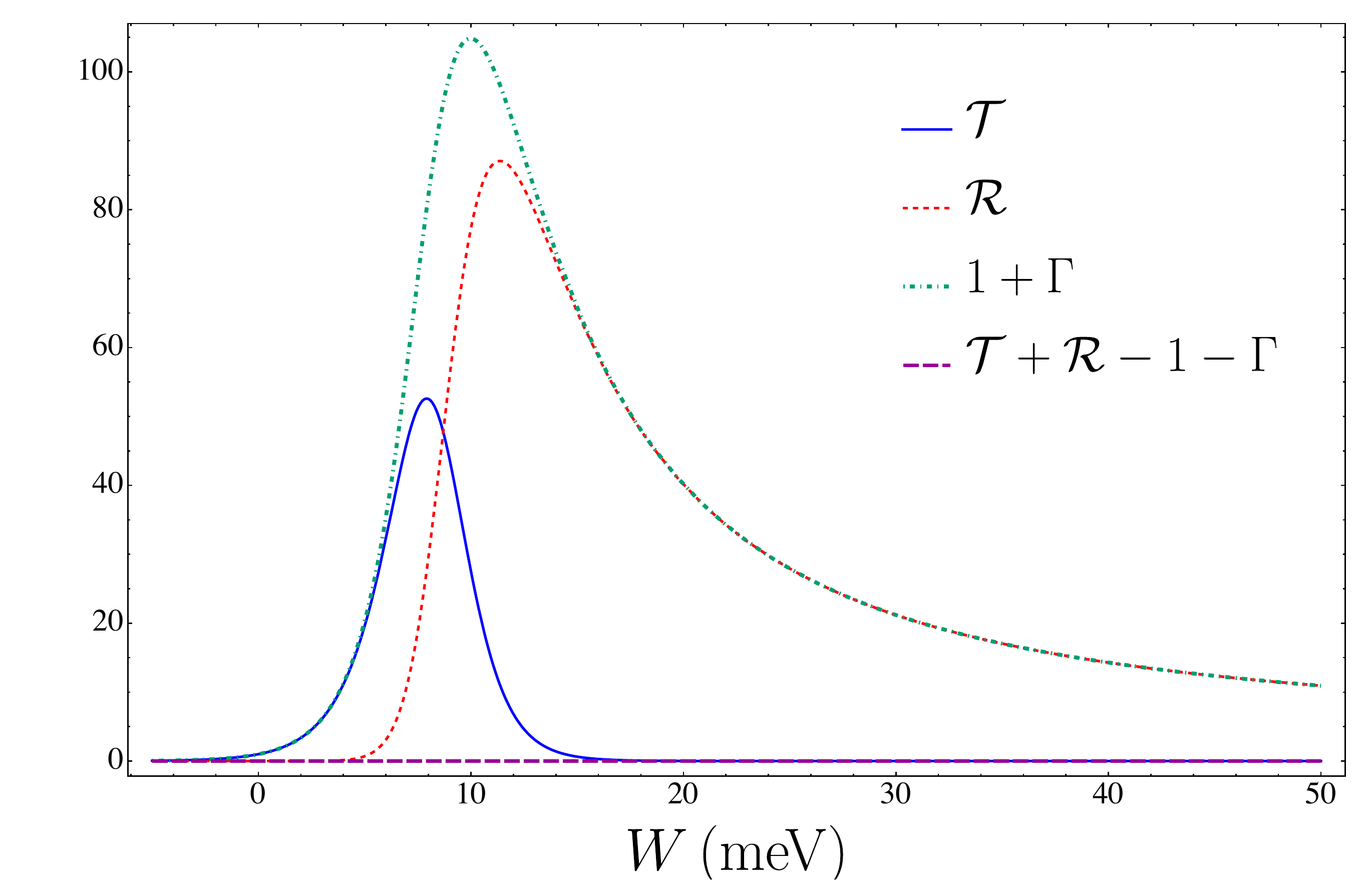}
        \caption{$\mu=200$ meV}
    \end{subfigure}
    \caption{Dimensionless flux-balance quantities as functions of $W$, evaluated at $E=\mu$, for $L=100\,\mathrm{nm}$, $U=0$, and $\phi=\pi/3$.}
    \label{fig:cons_vs_W}
\end{figure}

\begin{figure}
    \centering
    \begin{subfigure}{\linewidth}
        \centering
        \includegraphics[width=0.8\linewidth]{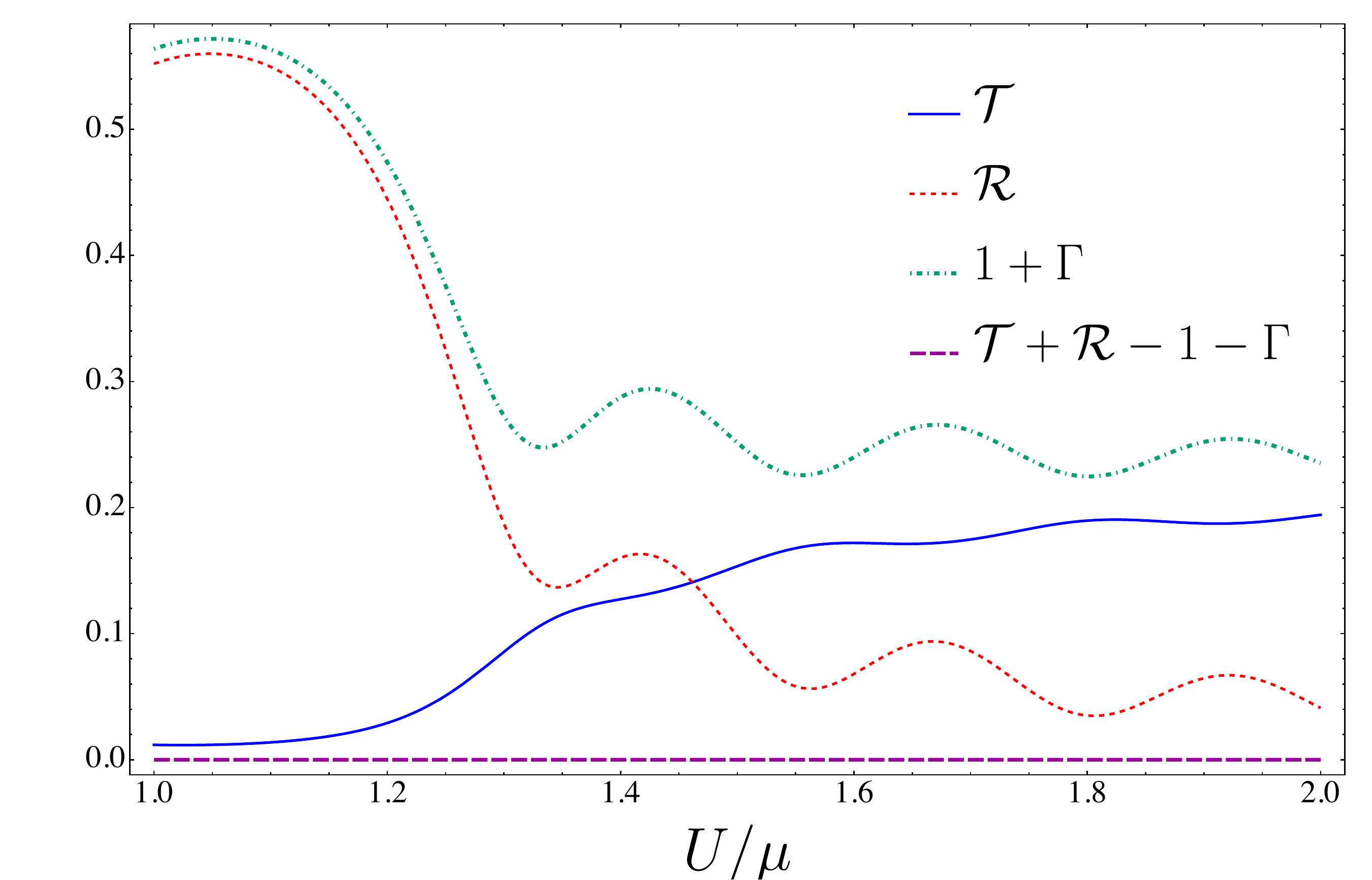}
        \caption{$W=-5$ meV}
    \end{subfigure}
    \begin{subfigure}{\linewidth}
        \centering
        \includegraphics[width=0.8\linewidth]{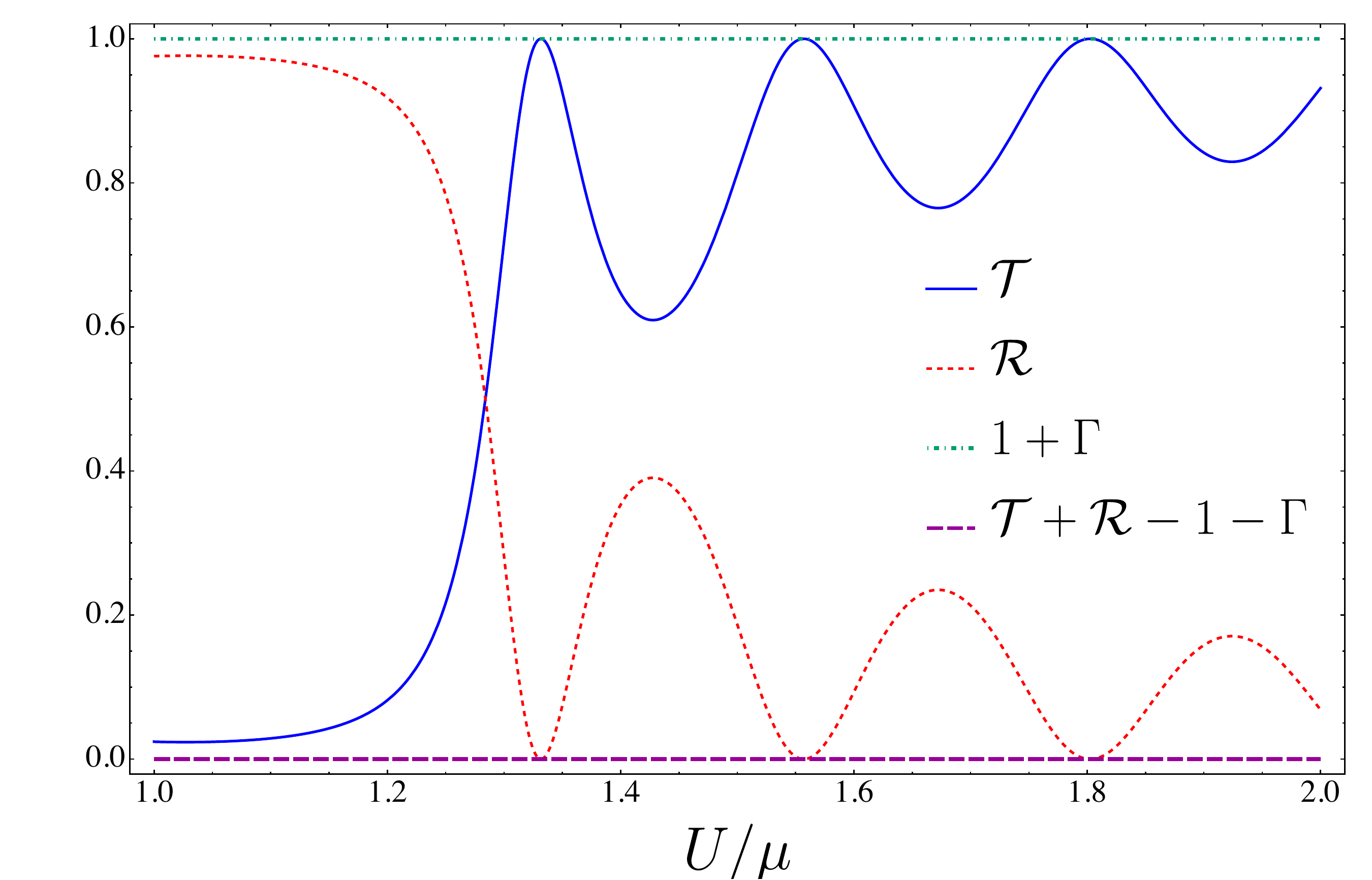}
        \caption{$W=0$ meV}
    \end{subfigure}
    \begin{subfigure}{\linewidth}
        \centering
        \includegraphics[width=0.8\linewidth]{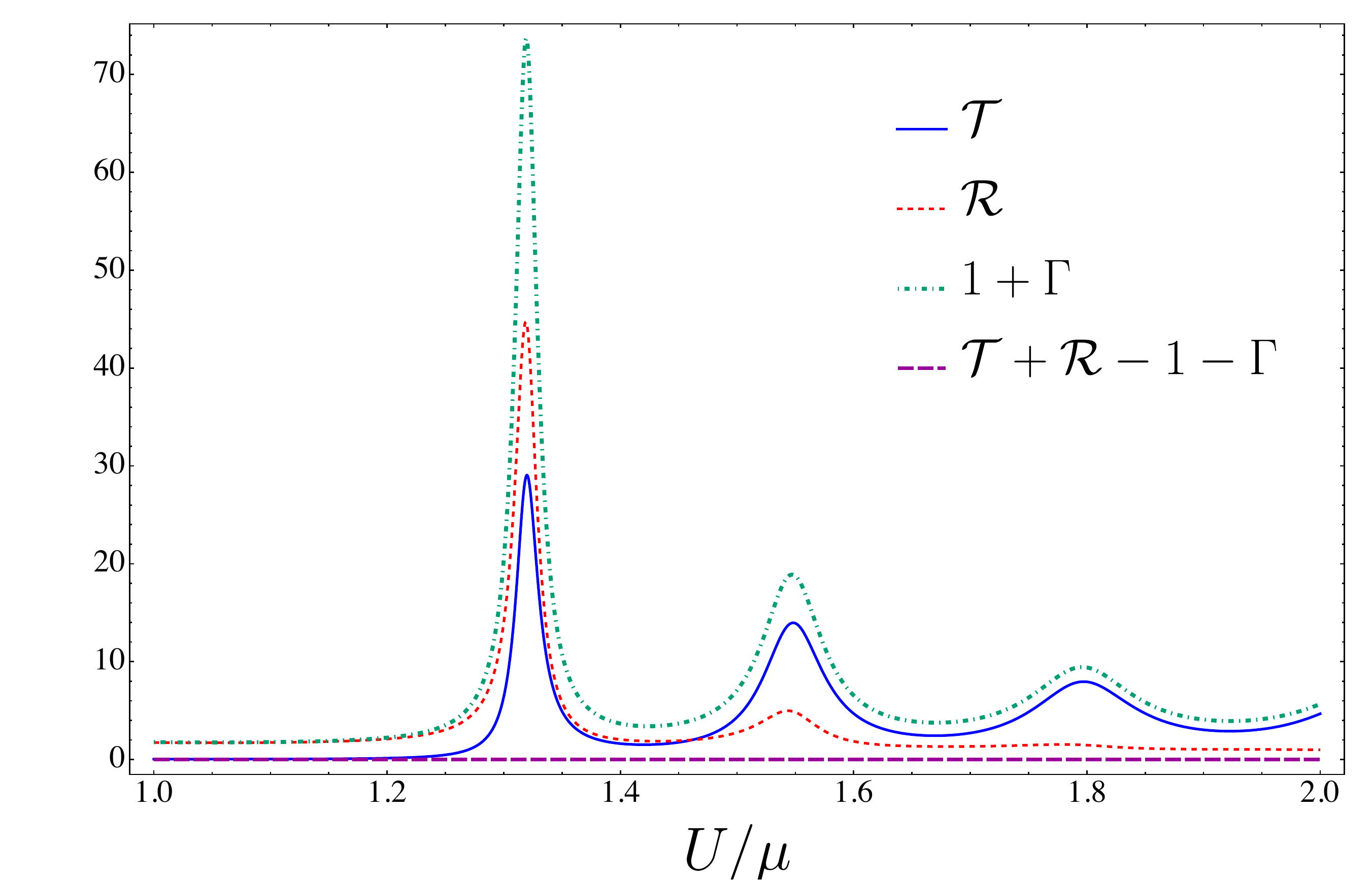}
        \caption{$W=5$ meV}
    \end{subfigure}

    \caption{Dimensionless flux-balance quantities as functions of $U/\mu$, evaluated at $E=\mu$, for $\mu=80\,\mathrm{meV}$, $L=100\,\mathrm{nm}$, and $\phi=\pi/15$.}
    \label{fig:cons_vs_U}
\end{figure}

The flux-balance plots confirm that the nonunitary scattering problem is still governed by an exact balance law. In FIG.~\ref{fig:cons_vs_EF}, the purple residual stays at zero within the scale of the plot in all panels, which confirms the generalized relation in Eq.~\eqref{eq:1-R}. For $W<0$, both $\mathcal{T}$ and $\mathcal{R}$ stay below unity and decay smoothly with $\mu$, while $1+\Gamma<1$, as expected for a lossy barrier. For $W>0$, the response becomes strongly resonant: $\mathcal{T}$ and $\mathcal{R}$ develop a sequence of peaks in $\mu$, $1+\Gamma$ follows their sum, and the resonances become higher and sharper when $W$ increases from $5$ to $10\,\mathrm{meV}$. These peaks are the same resonant channels already visible in the polar plots of FIGS.~\ref{fig:polar_vs_EF} and \ref{fig:polar_vs_W}, now seen in the energy domain. For fixed $L$, $\phi$, and barrier parameters, both $\mathcal{T}$ and $\mathcal{R}$ are controlled by the complex longitudinal wave vector inside the barrier, $q=q_R+i q_I$, through factors of the form $\exp(\pm iqL)$. As $\mu$ changes, the real part $q_R$ changes the accumulated phase and the system repeatedly approaches constructive interference conditions. At the Hermitian level, this is the same mechanism behind Fabry-P\'erot oscillations and resonant tunneling in graphene barriers and junction cavities \cite{Pereira2007,RamezaniMasir2010,YoungKim2009,Grushina2013,Shekhar2025}. In particular, \cite{Pereira2007} links the resonant structure to resonant electron states in the wells or hole states in the barriers, while \cite{RamezaniMasir2010} explicitly interprets the peaks as Fabry-P\'erot resonances and notes that the corresponding sharp resonant states can be viewed as quasibound states. The experiments of \cite{YoungKim2009,Grushina2013} support the same interpretation by showing that the oscillatory conductance comes from phase-coherent cavity interference between two interfaces. In our non-Hermitian case, the imaginary part $q_I$ does not remove this interference mechanism. It controls how strongly the same channels are attenuated or amplified. Loss damps the internal round trips and washes out the resonances, while gain enhances the same channels and makes the peaks larger and sharper.

FIG.~\ref{fig:cons_vs_W} shows the same physics from another angle. For fixed $\mu$ and $\phi$, the response is not monotonic in $W$ but reaches a maximum at a finite gain strength. Small positive $W$ strengthens the internal multiple scattering and increases all outgoing fluxes. When $W$ enters a resonant window, phase matching and amplification work together and produce a sharp enhancement of $\mathcal{T}$, $\mathcal{R}$, and $1+\Gamma$. Beyond that point, however, the growing imaginary part of $q$ detunes the resonance and the cavity buildup becomes less effective, so the peak collapses even though the barrier still provides gain. This is why the strongest response occurs at a finite $W$ rather than growing without bound. For negative $W$, the opposite happens: loss suppresses the internal buildup and the three quantities remain small. The residual stays at zero in both panels, again confirming the exact flux-balance relation. Increasing $\mu$ shifts the main enhancement to larger $W$ and makes it less singular, which is consistent with the weaker relative effect of the non-Hermitian term when $W/\mu$ is smaller.

The role of the real part of the barrier is shown in FIG.~\ref{fig:cons_vs_U}. Here $U$ mainly sets the phase-matching condition, while $W$ decides whether the selected channels are damped or amplified. In the Hermitian case, $W=0$, the green curve collapses onto unity and the purple residual stays at zero, so $\mathcal{R}+\mathcal{T}=1$ as expected. The oscillatory dependence on $U/\mu$ is also the standard one: near normal incidence the transmission stays large because backscattering is suppressed, while the finite barrier length creates Fabry-P\'erot-like oscillations between transmission maxima and reflection minima. This is the same Hermitian behavior discussed for rectangular graphene barriers and related resonant cavities in Refs.~\cite{Geim_Klein,Lejarreta2013,RamezaniMasir2010,YoungKim2009,Shekhar2025}. The crossover near $U\simeq\mu$ is especially important because the longitudinal wave vector inside the barrier becomes small there and the propagation changes from electron-like to hole-like, which makes the response especially sensitive to the barrier height. Once $W\neq0$, the same phase-selected structure stays in place, but it is no longer unitary. For $W<0$, loss smooths the Hermitian oscillations and absorbs flux inside the barrier. For $W>0$, gain strongly enhances the same resonance positions, so the peaks become narrower and larger. Therefore, the real barrier sets where the resonances occur, and the imaginary part decides how strongly they appear.

\subsection{Zero-temperature conductance behavior}

\begin{figure}[t]
    \centering
    \begin{subfigure}{\linewidth}
        \centering
        \includegraphics[width=0.8\linewidth]{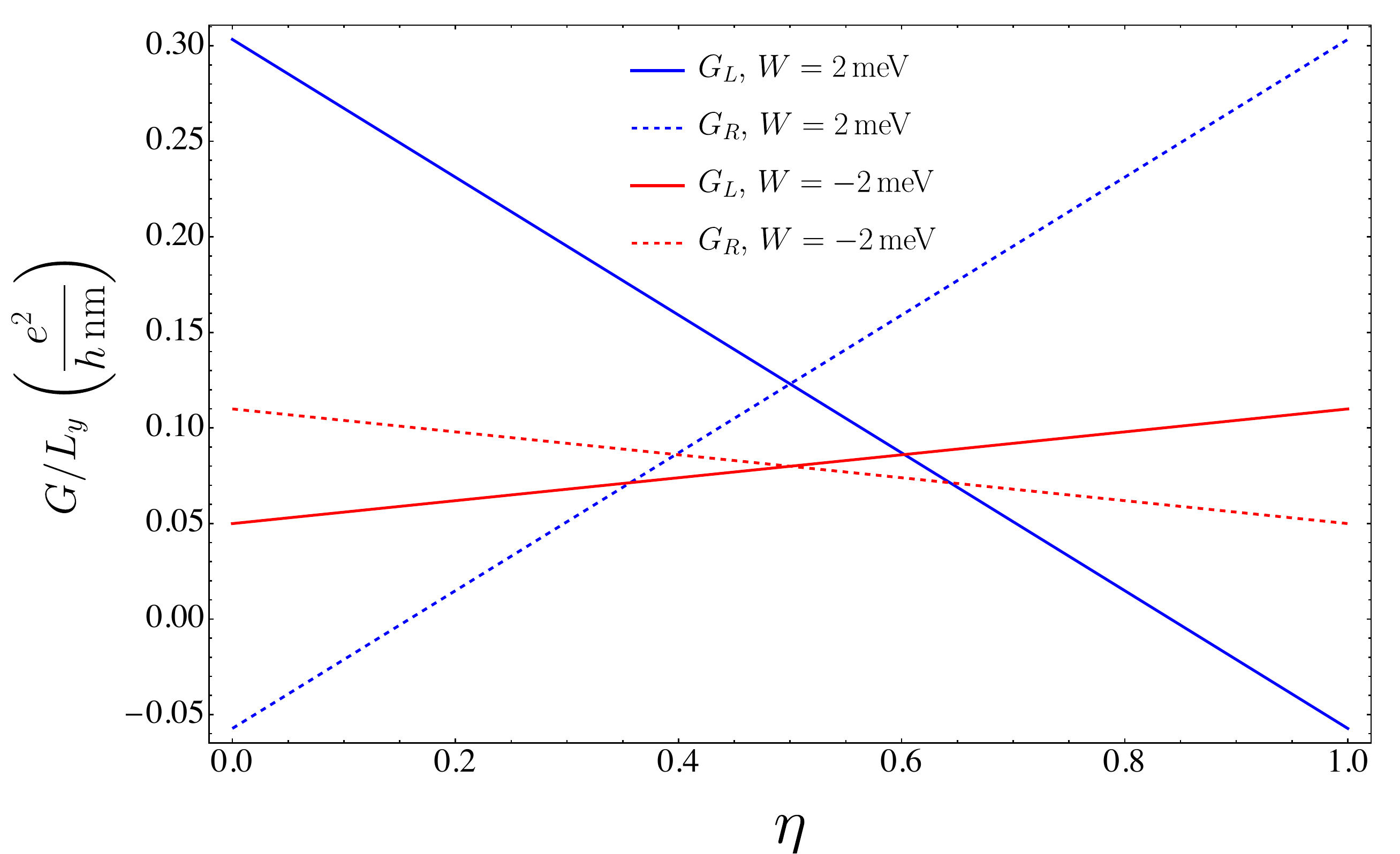}
        \caption{}
    \end{subfigure}

    \vspace{0.4cm}

    \begin{subfigure}{\linewidth}
        \centering
        \includegraphics[width=0.8\linewidth]{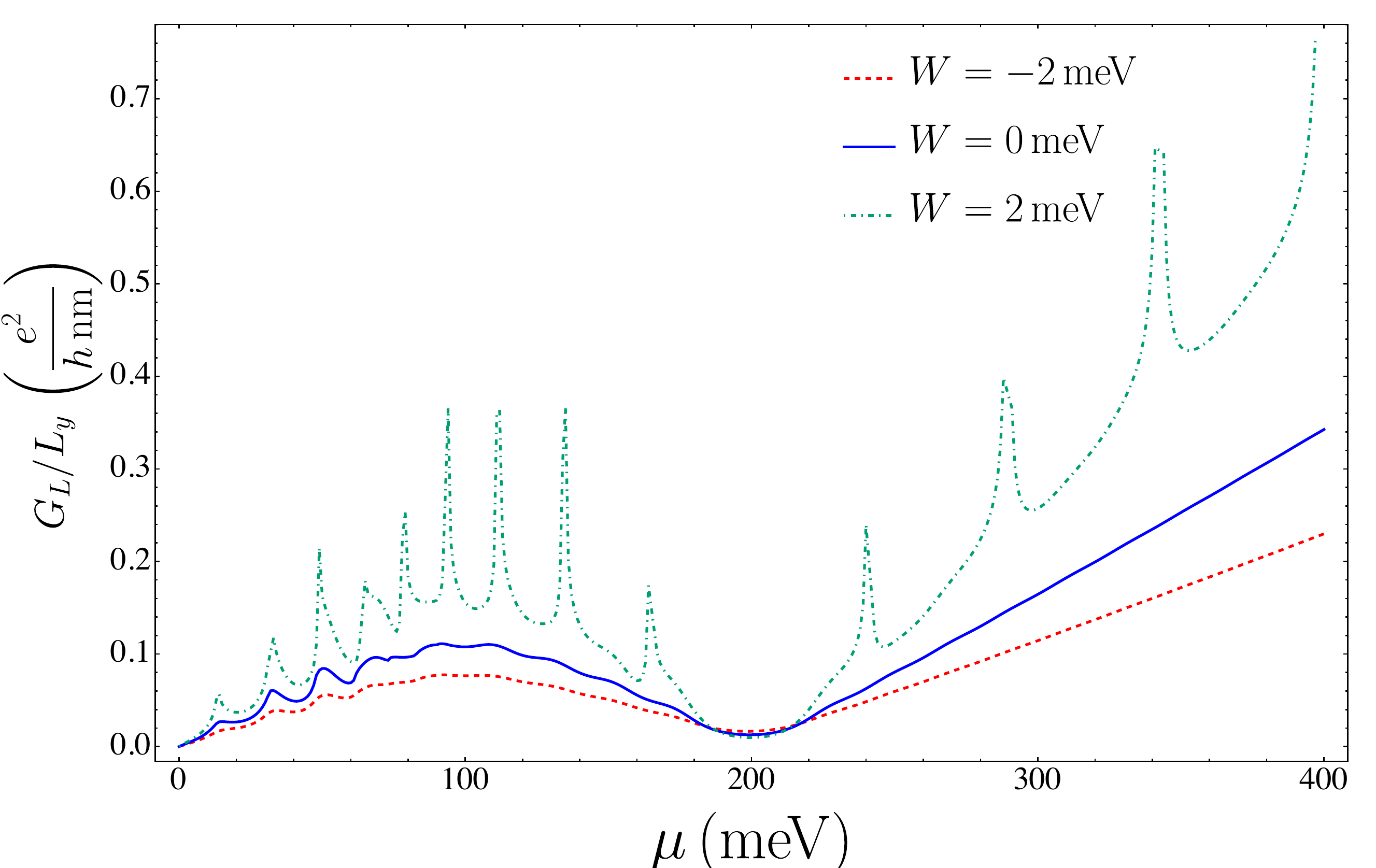}
        \caption{}
    \end{subfigure}
    \begin{subfigure}{\linewidth}
        \centering
        \includegraphics[width=0.8\linewidth]{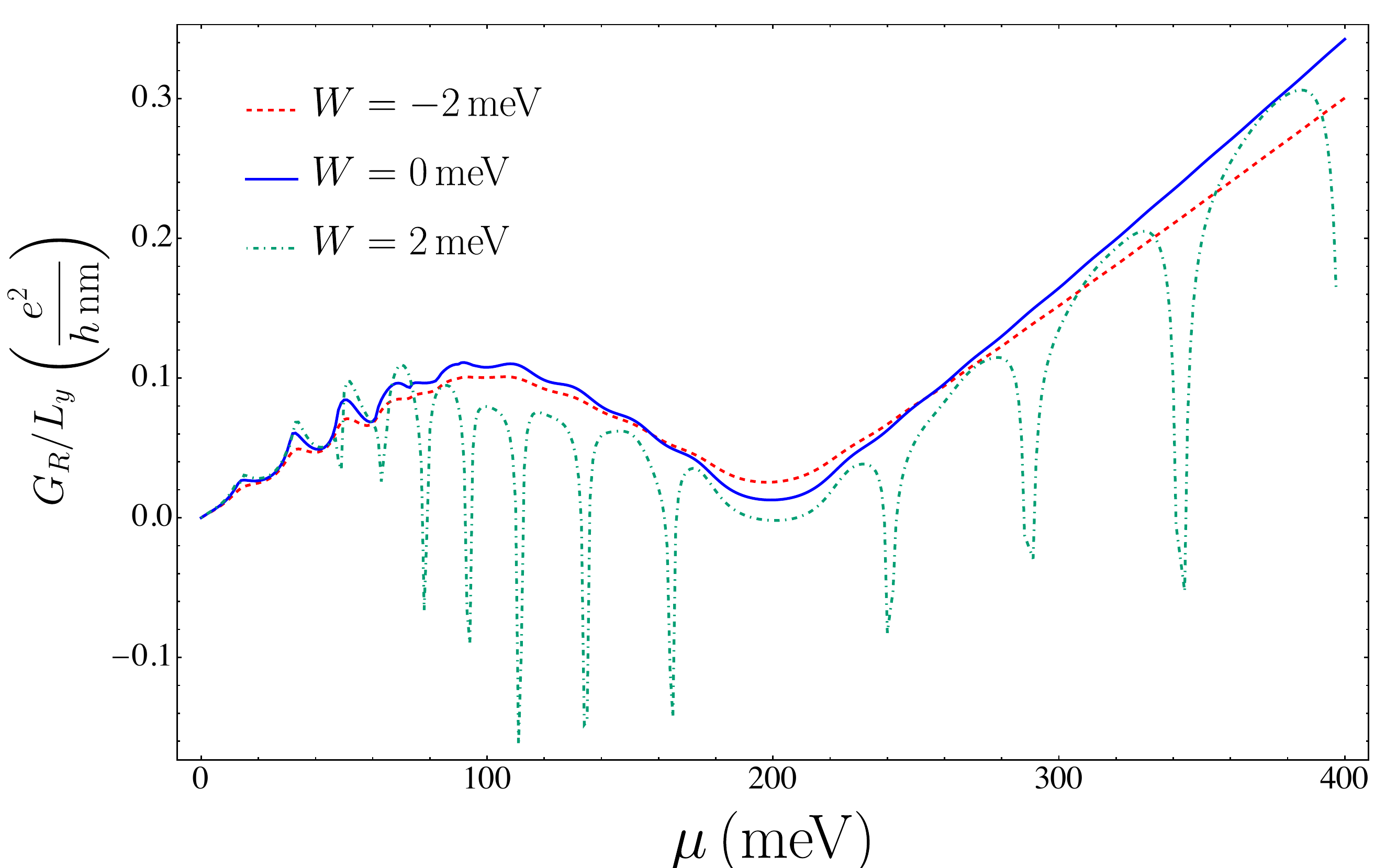}
        \caption{}
    \end{subfigure}

    \caption{Zero-temperature electrical conductances for the non-Hermitian graphene barrier. In all panels, we take $U=200\,\mathrm{meV}$ and $L=100\,\mathrm{nm}$. Panel (a) shows $G_L/L_y$ and $G_R/L_y$ as functions of the bias-asymmetry parameter $\eta$ at fixed $\mu=80\,\mathrm{meV}$ and $W=\pm2\,\mathrm{meV}$. Panels (b) and (c) show $G_L/L_y$ and $G_R/L_y$, respectively, as functions of $\mu$ for $\eta=1/3$ and $W=-2,0,2\,\mathrm{meV}$.}
    \label{fig:conductance_eta_mu}
\end{figure}

\begin{figure*}[t]
    \centering
    \begin{subfigure}{0.45\textwidth}
        \centering
        \includegraphics[width=\linewidth]{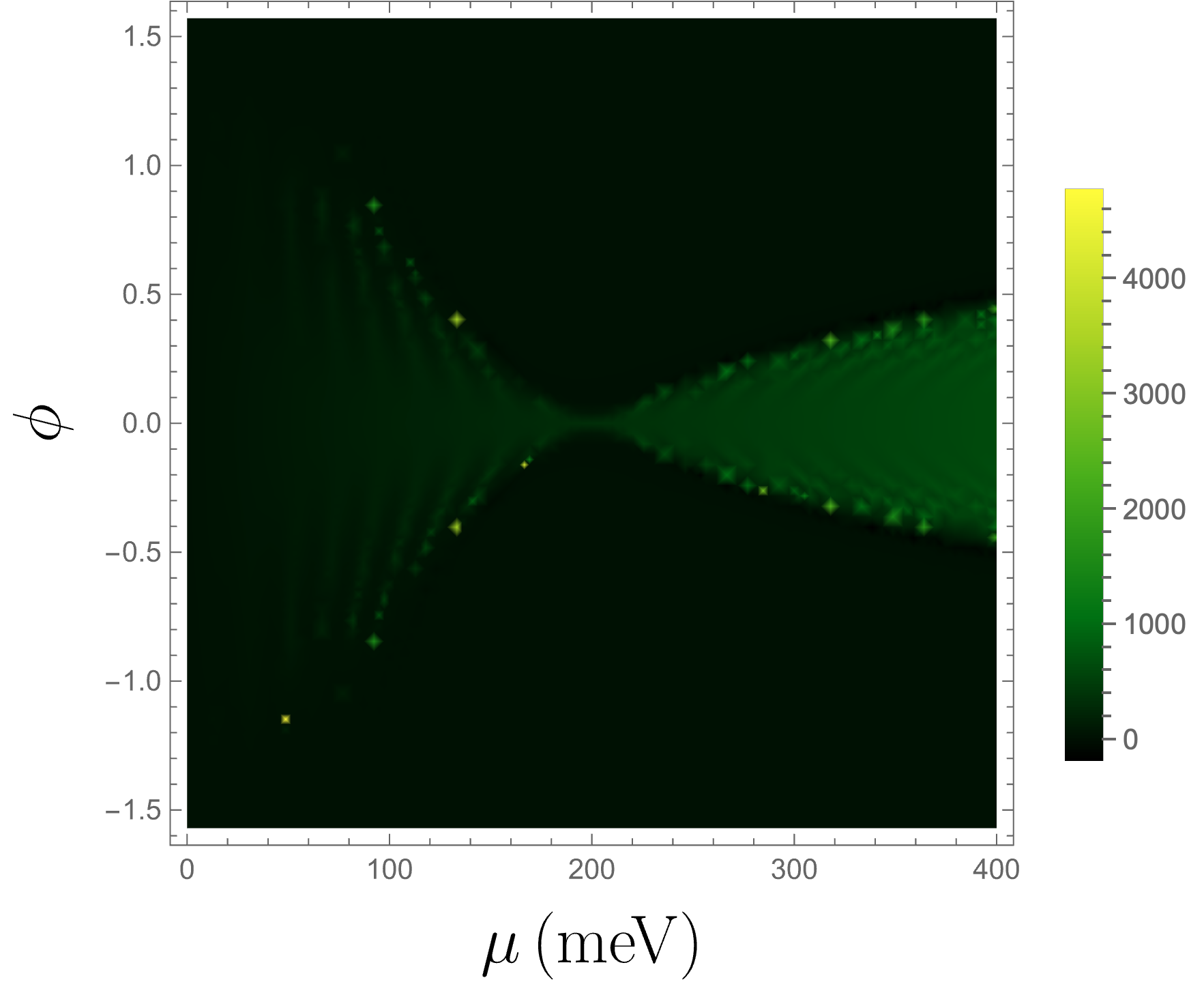}
        \caption{$W=2$ meV}
    \end{subfigure}
    \begin{subfigure}{0.45\textwidth}
        \centering
        \includegraphics[width=\linewidth]{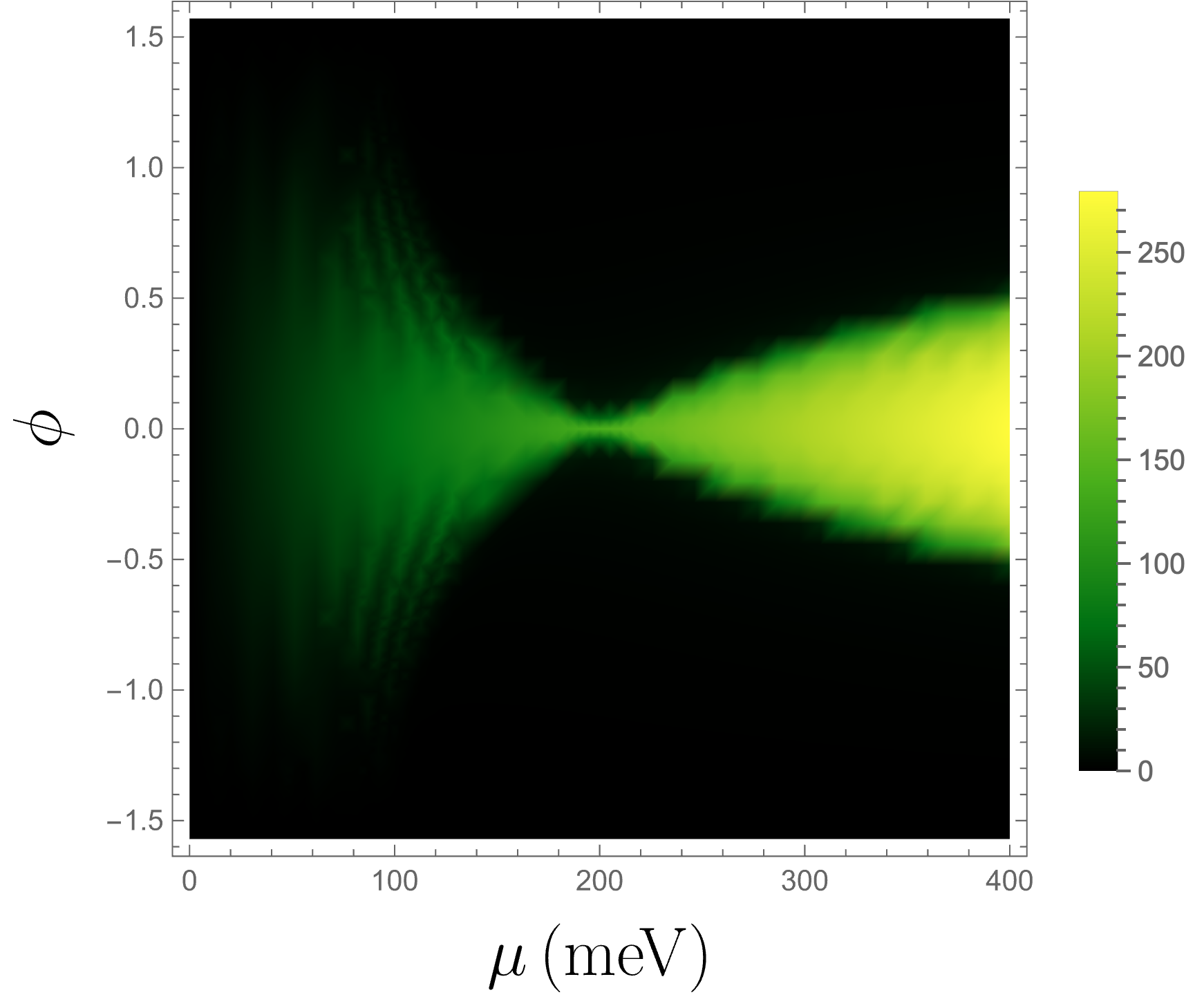}
        \caption{$W=-2$ meV}
    \end{subfigure}

    \vspace{0.4cm}

    \begin{subfigure}{0.45\textwidth}
        \centering
        \includegraphics[width=\linewidth]{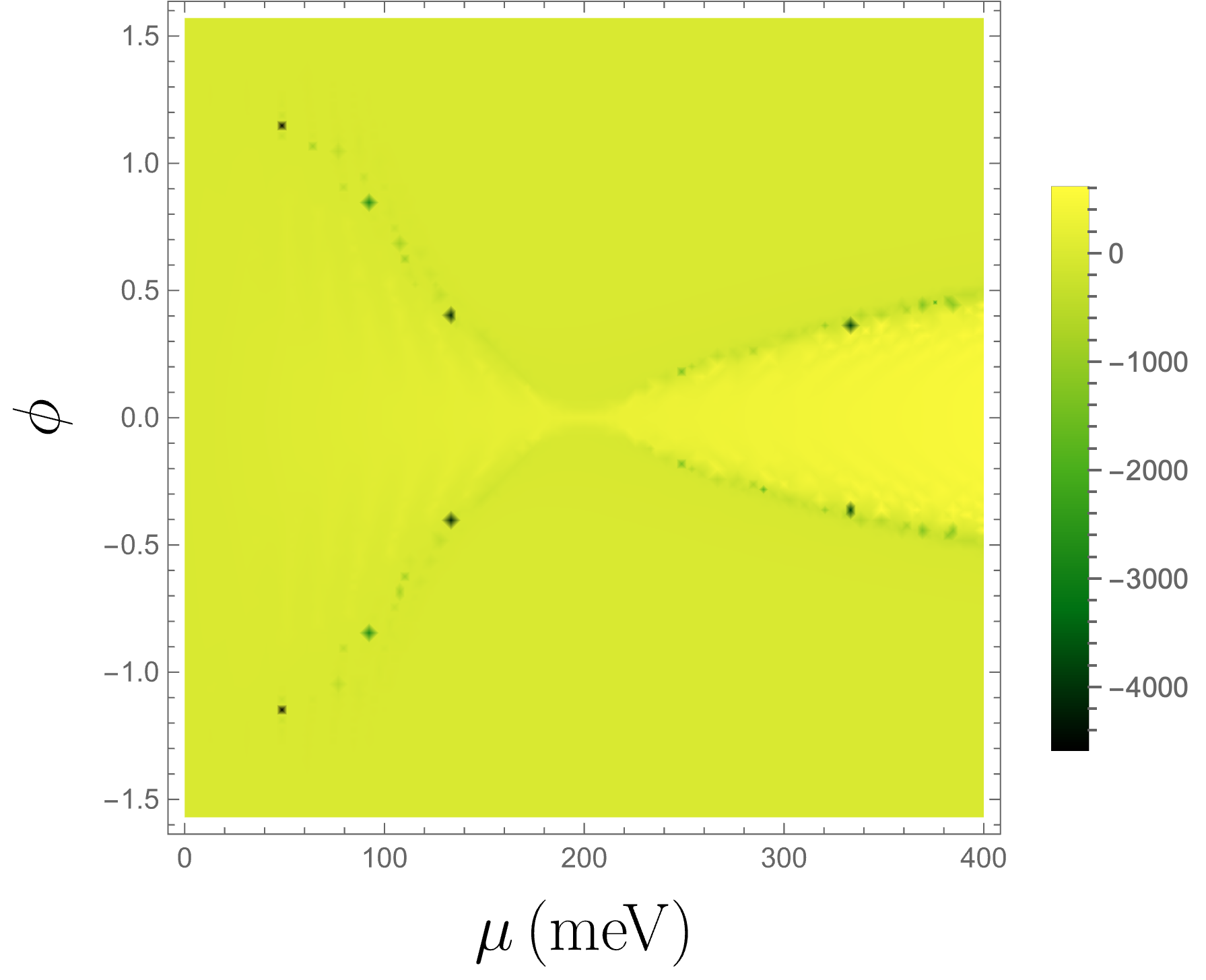}
        \caption{$W=2$ meV}
    \end{subfigure}
    \begin{subfigure}{0.45\textwidth}
        \centering
        \includegraphics[width=\linewidth]{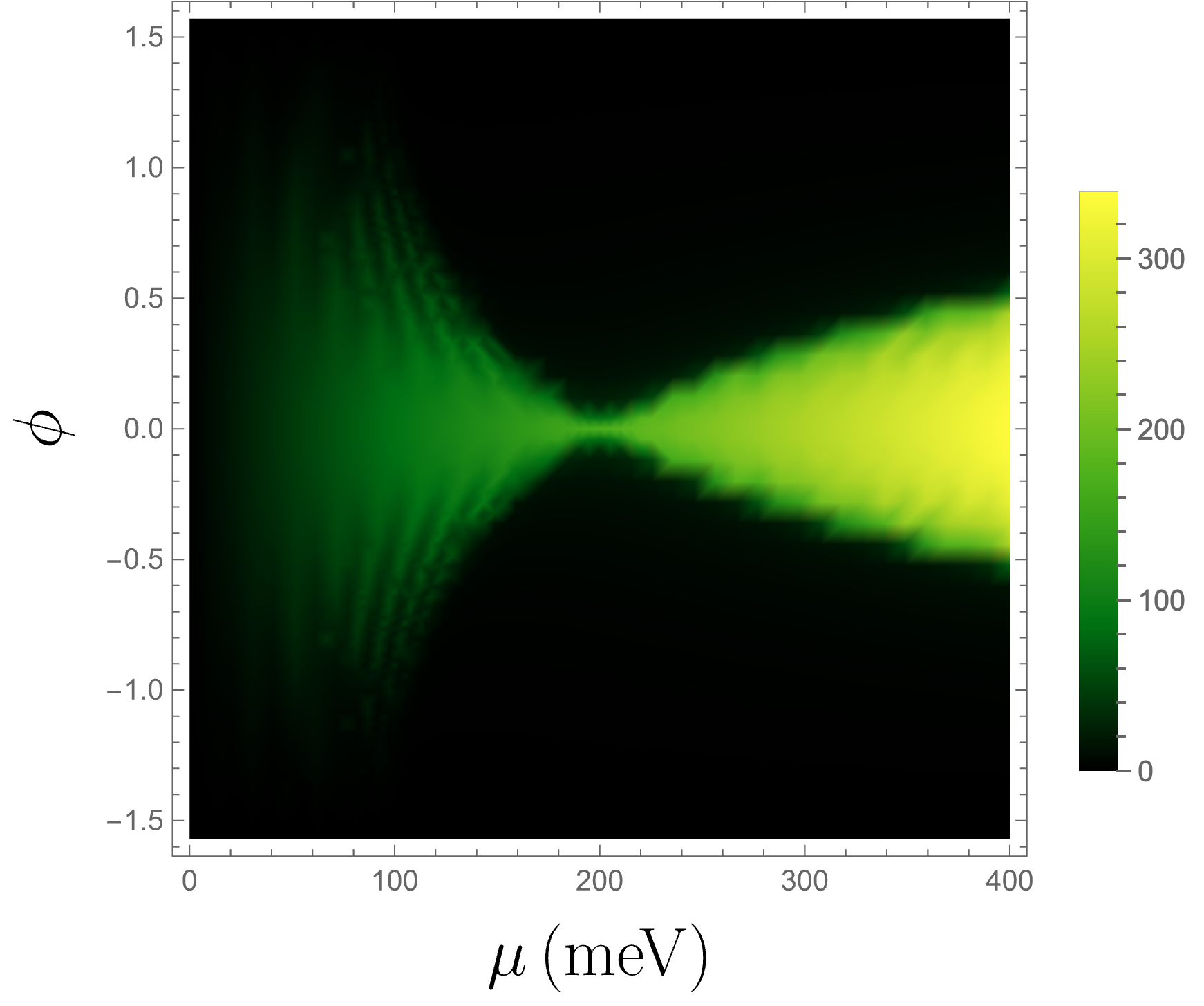}
        \caption{$W=-2$ meV}
    \end{subfigure}

    \caption{Angular integrands entering the zero-temperature conductances. Panels (a) and (b) show the integrand of $G_L$ for $W=2$ and $-2\,\mathrm{meV}$, respectively, while panels (c) and (d) show the integrand of $G_R$ for the same values of $W$. In all panels, we take $U=200\,\mathrm{meV}$, $L=100\,\mathrm{nm}$, and $\eta=1/3$. The horizontal axis is the chemical potential $\mu$, and the vertical axis is the incidence angle $\phi$.}
    \label{fig:integrands_GL_GR}
\end{figure*}

FIG.~\ref{fig:conductance_eta_mu} summarizes the zero-temperature conductances defined in Eq.~\eqref{eq:G_R/L_T0}, with the bias partition introduced in Eqs.~\eqref{eq:bias_partition}-\eqref{eq:eta_xi}. Panel (a) shows the main non-Hermitian effect very clearly. Since at $T=0$ the scattering coefficients are evaluated at $E=\mu$ and do not depend on $\eta$, Eq.~\eqref{eq:G_R/L_T0} predicts a linear dependence on the bias-asymmetry parameter. In the Hermitian limit, $\mathcal{R}+\mathcal{T}=1$, so the $\eta$ dependence cancels and one gets $G_L=G_R$ for any partition of the same voltage drop. For $W\neq0$, by contrast, $\mathcal{R}+\mathcal{T}=1+\Gamma$, and the two lead-resolved conductances differ by a term proportional to $(2\eta-1)\Gamma$. This is why the slopes reverse when the sign of $W$ changes, why the asymmetry is much stronger for $W=2\,\mathrm{meV}$ than for $W=-2\,\mathrm{meV}$, and why both curves meet at the symmetric choice $\eta=1/2$. For $W=2\,\mathrm{meV}$ the effect is strong enough that one of the lead-resolved conductances becomes negative near the extreme partitions $\eta\to0$ or $\eta\to1$.

This bias dependence gives a direct operational signature of the loss of gauge invariance in the effective two-terminal conductance. Following \cite{Wei2025}, $G_{\alpha\beta}$ is the linear-response conductance matrix element that measures how the current in lead $\alpha$ changes when the voltage at lead $\beta$ is varied, so that $I_\alpha=\sum_\beta G_{\alpha\beta}v_\beta$. Gauge invariance requires that a uniform shift of all lead voltages leave the current unchanged, which is equivalent to $\sum_\beta G_{\alpha\beta}=0$. In the non-Hermitian case of \cite{Wei2025}, the complex potential generates an additional source-sink contribution that can be represented as an effective fictitious probe $F$. The quantity $G_{\alpha F}$ is then the conductance between the physical lead $\alpha$ and that fictitious probe, and their result $\sum_\beta G_{\alpha\beta}=-G_{\alpha F}$ shows that gauge invariance is lost whenever $G_{\alpha F}\neq0$. Physically, this means that the current depends not only on the total voltage drop, but also on how that drop is distributed among the leads. This is exactly the behavior seen in panel (a). From this viewpoint, the present model also has a practical motivation. If in an experiment one finds that the measured conductance depends on the bias partition, but the missing channels responsible for that effect are not known microscopically, then a complex potential can be used as a simple effective model for them. In that reading, the imaginary part of the barrier mimics unresolved gain-loss channels or an effective probe. Of course, this is only a simplified reduced model, not a substitute for a full microscopic open-system treatment. Still, it provides a simple way to analyze and parameterize the observed gauge-invariance breaking. This interpretation is consistent with recent non-Hermitian transport work, where the usual Landauer-B\"uttiker expression acquires extra source-sink terms and gauge invariance is no longer automatic unless those terms vanish or are redistributed in a more complete description \cite{Yan2024,Wei2025,Yang2026}.

Panels (b) and (c) show the same physics as a function of chemical potential. For $W=0$, the left and right conductances collapse onto the same curve, as they should in the Hermitian case, and the result is independent of $\eta$. The overall behavior agrees with the standard graphene barrier problem discussed in Refs.~\cite{Geim_Klein,Lejarreta2013,Krstajic2011,Fuentevilla2015,Pereira2007,RamezaniMasir2010,Shekhar2025}. In particular, for $\mu<U$ the conductance oscillates because the finite barrier acts as a resonant cavity and the transmission is modulated by Fabry-P\'erot-like interference. As $\mu$ approaches $U$, the longitudinal wave vector inside the barrier becomes small and the conductance develops a pronounced minimum. This is the same mechanism identified in \cite{Krstajic2011,Fuentevilla2015}, where the dip is traced to the near-vanishing of the real wave vector in the barrier region. Once $\mu>U$, propagation inside the barrier becomes electron-like again, the transmission recovers, and the conductance grows approximately with $\mu$, in agreement with \cite{Fuentevilla2015}. The sharp structures that remain below the barrier are therefore not a new non-Hermitian effect by themselves. They are the familiar Hermitian resonance pattern of a finite graphene barrier, now reshaped by the imaginary part.

This reshaping becomes very clear in FIG.~\ref{fig:integrands_GL_GR}, which helps explain why the $W>0$ conductance curves in FIG.~\ref{fig:conductance_eta_mu} are so sharp. For $W=-2\,\mathrm{meV}$, the angular integrands are relatively smooth and spread over a broad angular sector, so the angular integral averages the response in a regular way. For $W=2\,\mathrm{meV}$, by contrast, the integrands develop strong oscillations in both angle and chemical potential, together with narrow high-intensity ridges and sharp resonant spots. These structures are especially pronounced below and around $\mu\simeq U$, where the scattering amplitudes are most sensitive to the barrier region. This is fully consistent with the angular lobes in FIGS.~\ref{fig:polar_vs_EF} and \ref{fig:polar_vs_W}, and with the resonant peaks in the flux-balance plots of FIGS.~\ref{fig:cons_vs_EF} and \ref{fig:cons_vs_W}. In short, the spiky conductance for positive $W$ does not come from a smooth overall enhancement. It comes from the angular integral over a highly structured resonant pattern. In the same way, negative $W$ damps those channels and produces a much smoother conductance. The same change of scale also explains why the numerical integration becomes harder in the gain regime: once the integrand is dominated by very narrow angular peaks, an accurate calculation requires a much finer angular resolution than in the lossy case. Finally, these curves should not be confused with the pseudodiffusive minimum-conductivity problem of undoped graphene between heavily doped leads studied in \cite{Tworzydlo2006,Katsnelson2006EPJB}; see also the finite-temperature ballistic extension in \cite{MullerBrauningerTrauzettel2009}. Our setup is different: here the central region is a finite rectangular barrier with $U=200\,\mathrm{meV}$, and the relevant minimum is the barrier-induced dip near $\mu\simeq U$, not the universal conductivity at the Dirac point.

\subsection{Finite-temperature thermoelectric transport}

\begin{figure*}
    \centering
    \begin{subfigure}{0.45\textwidth}
        \centering
        \includegraphics[width=\linewidth]{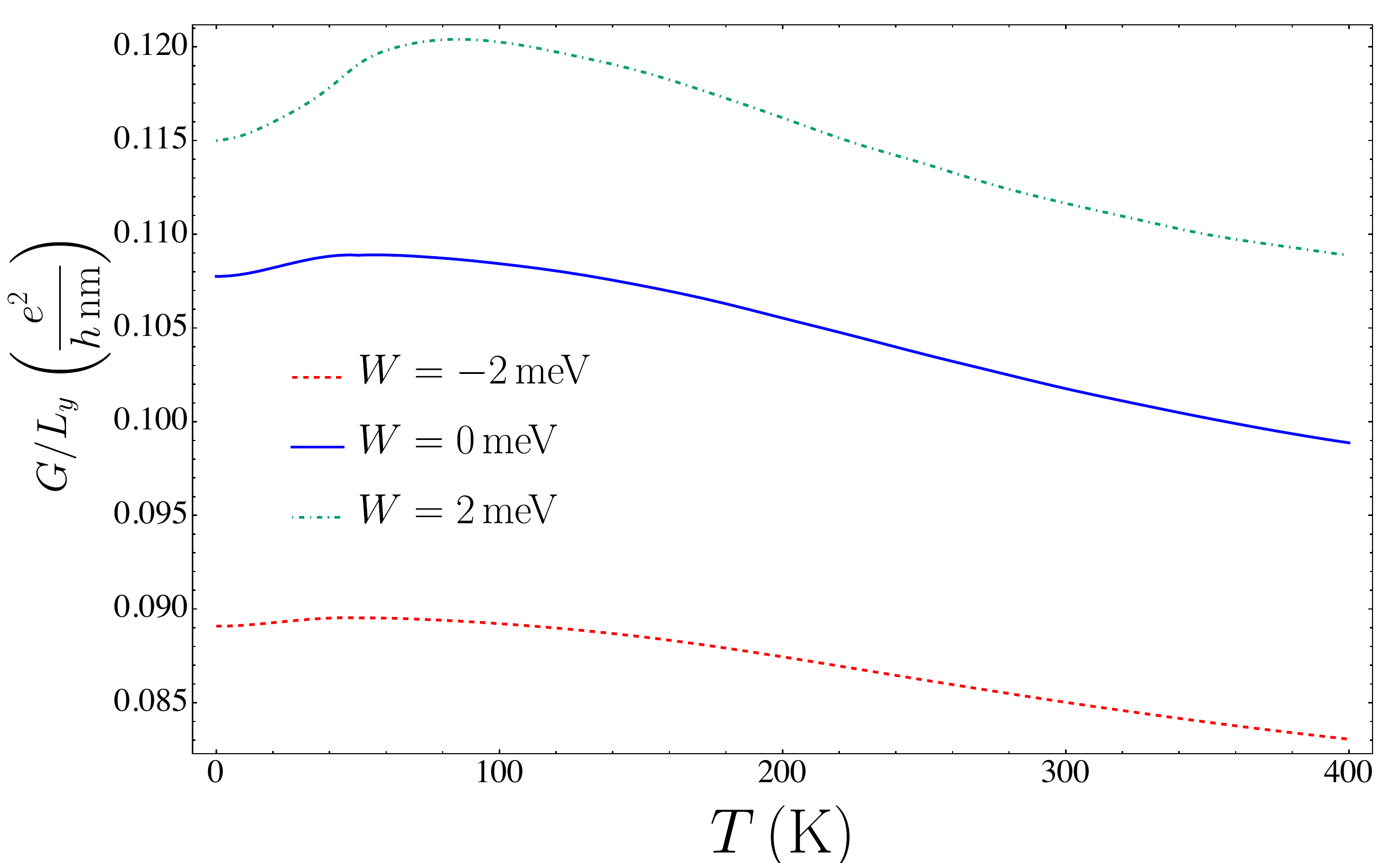}
        \caption{Electrical conductance $G/L_y$}
    \end{subfigure}
    \begin{subfigure}{0.45\textwidth}
        \centering
        \includegraphics[width=\linewidth]{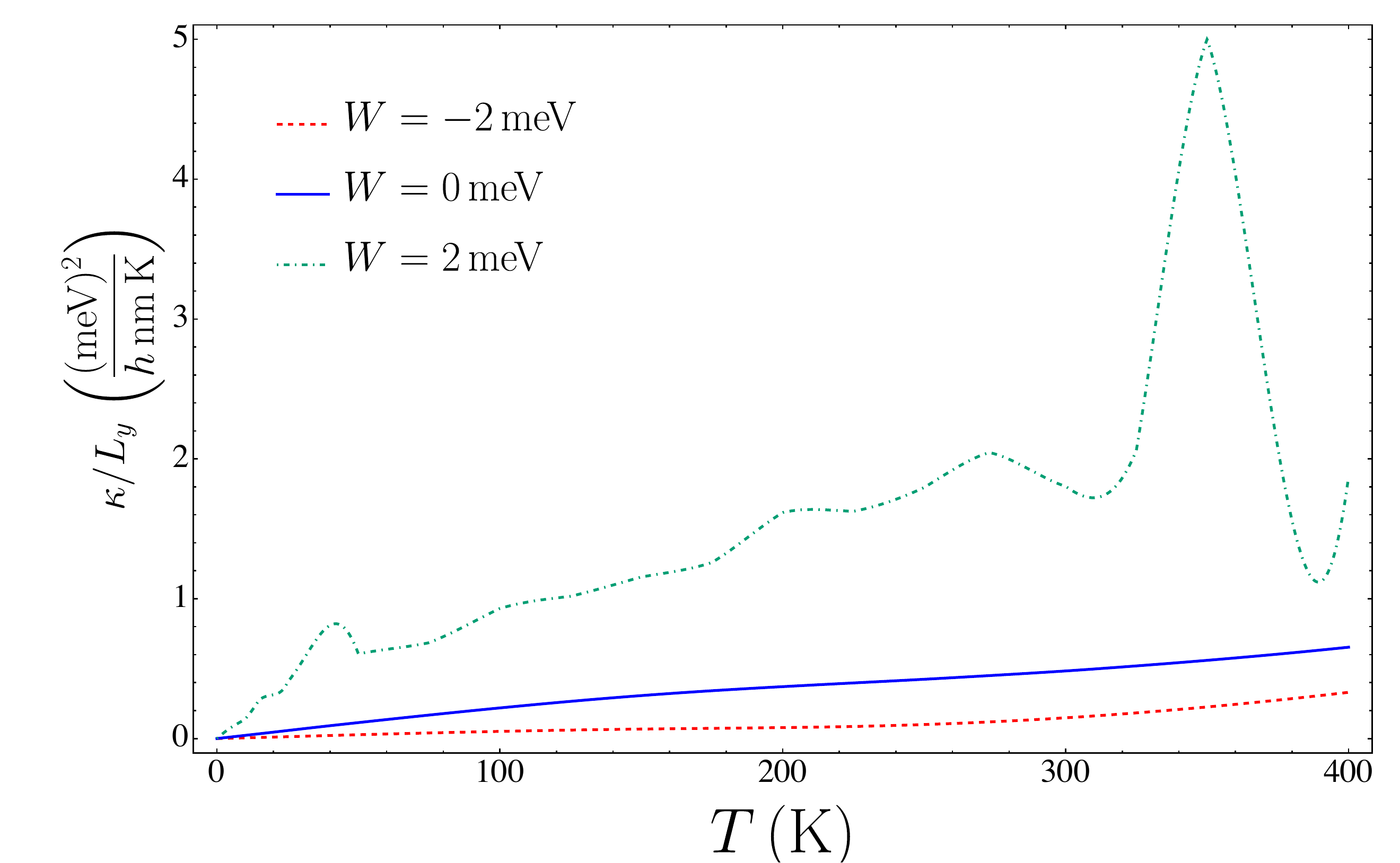}
        \caption{Thermal conductance $\kappa/L_y$}
    \end{subfigure}

    \vspace{0.4cm}

    \begin{subfigure}{0.45\textwidth}
        \centering
        \includegraphics[width=\linewidth]{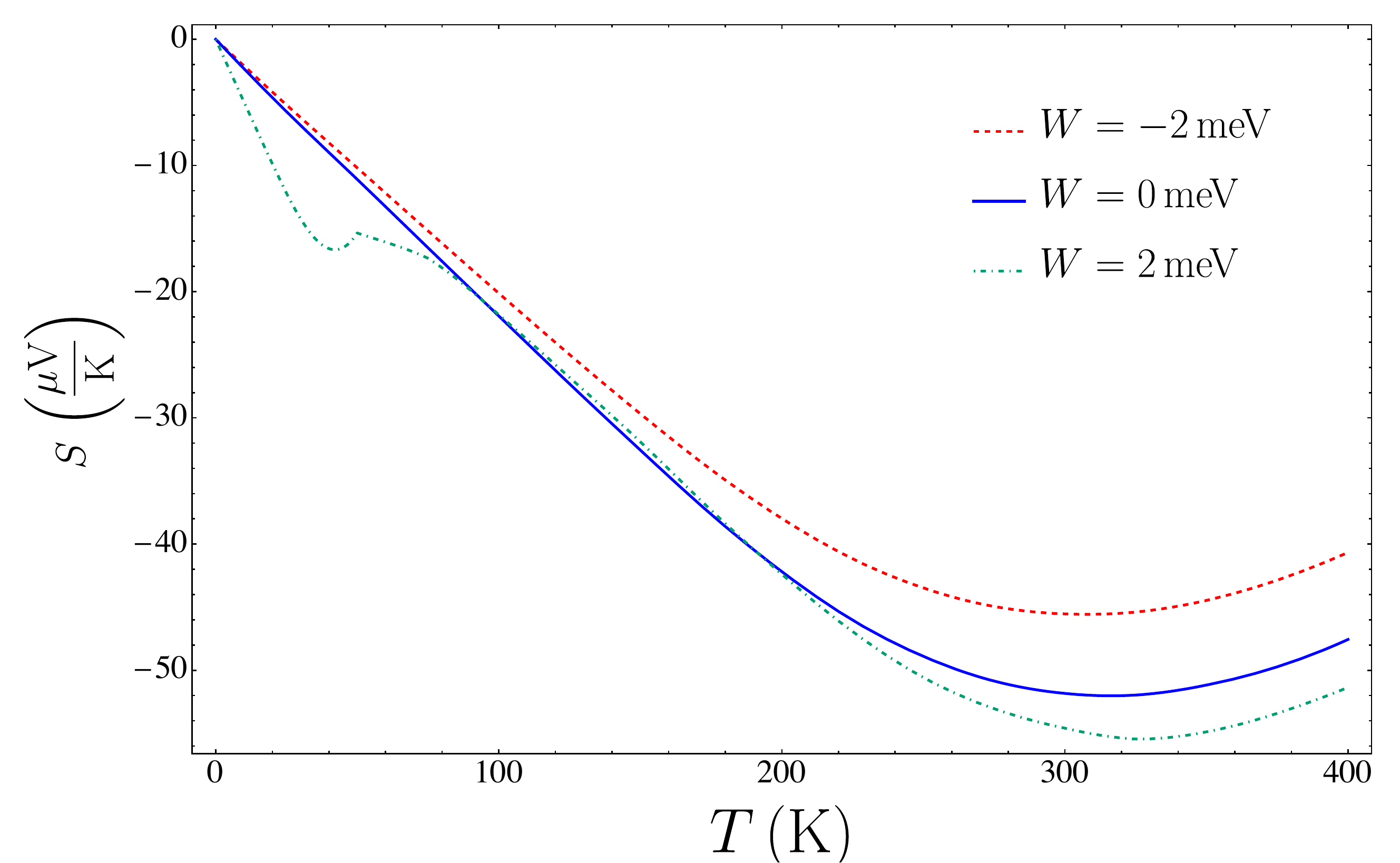}
        \caption{Seebeck coefficient $S$}
    \end{subfigure}
    \begin{subfigure}{0.45\textwidth}
        \centering
        \includegraphics[width=\linewidth]{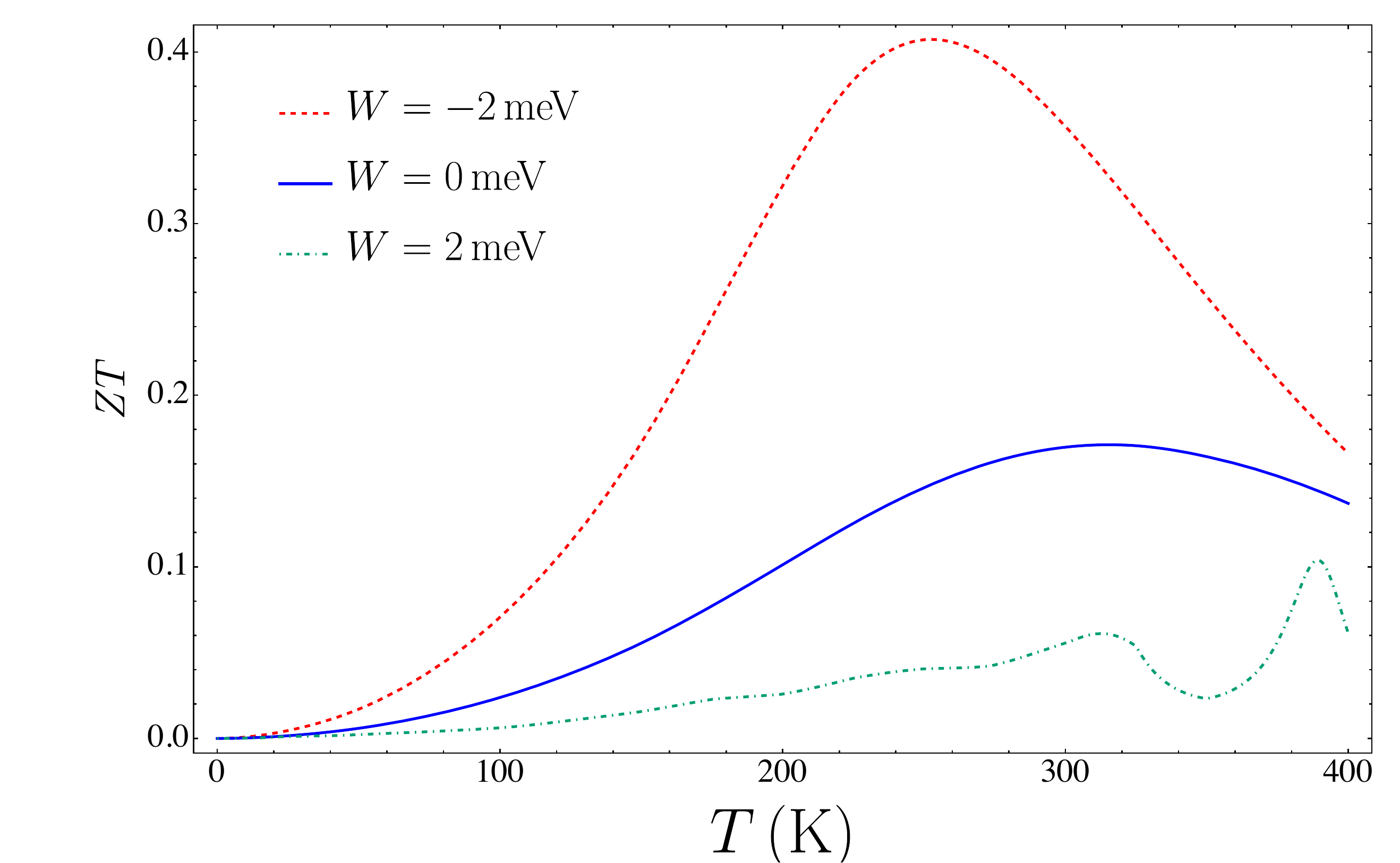}
        \caption{Figure of merit $ZT$}
    \end{subfigure}

    \caption{Temperature dependence of the electrical conductance per unit width $G/L_y$, the thermal conductance per unit width $\kappa/L_y$, the Seebeck coefficient $S$, and the thermoelectric figure of merit $ZT$ for a non-Hermitian graphene barrier. The barrier parameters are $L=100~\mathrm{nm}$, $\mu=200~\mathrm{meV}$, and $U=280~\mathrm{meV}$. In each panel, the different curves correspond to the values of the imaginary potential $W$ indicated in the plots.}
    \label{fig:G_kappa_S_ZT_vs_T}
\end{figure*}

We now turn to the finite-temperature transport coefficients shown in FIG.~\ref{fig:G_kappa_S_ZT_vs_T}. These quantities were obtained numerically from the exact linear-response expressions for the electrical conductance and heat coefficients, Eqs.~\eqref{eq:G_R/L} to \eqref{eq:M_R/L}, together with the slope definitions of the Seebeck coefficient, thermal conductance, and figure of merit in Eqs.~\eqref{eq:seebeck_ratio}, \eqref{eq:kappa_MLK}, and \eqref{eq:ZT_def}. In this part we set $\eta=\xi=1/2$ in Eqs.~\eqref{eq:bias_partition}-\eqref{eq:eta_xi}, so that the voltage and temperature drops are shared symmetrically and the discussion refers to transport through the whole device, without the left-right imbalance that appears for asymmetric partitions. We also choose $\mu=200\,\mathrm{meV}$ and $U=280\,\mathrm{meV}$. This keeps the reservoirs in an electron-dominated regime, well away from the charge-neutrality point, which is numerically more stable and physically closer to metallic transport, while the barrier remains in the $n$-$p$-$n$ configuration. Since the non-Hermitian barrier produces finite zero-bias offsets, the finite-temperature coefficients must be understood as linear-response slopes around the reference state rather than as naive ratios through the origin, exactly as discussed in the formalism above and in recent open-system extensions of Landauer transport \cite{Yang2026}. The same non-Hermitian structure also affects reciprocity: because Eq.~\eqref{eq:M_R/L} contains an additional term proportional to $1-\mathcal{R}-\mathcal{T}$, the usual Onsager relation does not hold in general and is recovered only in the Hermitian limit. At low temperature, the numerical curves are also consistent with the Sommerfeld expansion: below about $10\,\mathrm{K}$, the electrical conductance shows the expected leading $T^2$ behavior, while the thermal conductance grows linearly with $T$.

The plotted trends show that the imaginary part of the barrier provides a clear thermoelectric control parameter. Positive $W$ enhances the electrical conductance over the whole temperature range, while negative $W$ suppresses it. At $T=300\,\mathrm{K}$, and taking a representative transverse width $L_y=1\,\mu\mathrm{m}$, the plotted values of $G/L_y$ correspond to total conductances $G\simeq 4.33\,\mathrm{mS}$, $3.94\,\mathrm{mS}$, and $3.29\,\mathrm{mS}$ for $W=2$, $0$, and $-2\,\mathrm{meV}$, respectively. The same ordering appears even more strongly in the thermal conductance: restoring the absolute scale from the $300\,\mathrm{K}$ values gives $\kappa\simeq 69.88\,\mathrm{nW/K}$, $18.72\,\mathrm{nW/K}$, and $5.75\,\mathrm{nW/K}$ for $W=2$, $0$, and $-2\,\mathrm{meV}$, respectively. The Seebeck coefficient stays negative in all cases, with $S\simeq-54.6\,\mu\mathrm{V/K}$, $-51.8\,\mu\mathrm{V/K}$, and $-45.5\,\mu\mathrm{V/K}$ at $300\,\mathrm{K}$, which is fully consistent with electron-dominated transport. In other words, the non-Hermitian term changes the magnitude of the thermopower, but not its sign, so the dominant carriers remain electrons throughout this regime. This sign trend and the overall scale, namely tens of $\mu\mathrm{V/K}$ together with subunitary $ZT$, are consistent with recent graphene thermoelectric analyses away from neutrality \cite{CanasBonillaMartinRuiz2025,CanasBonillaPerezPedrazaMartinRuiz2026}. The nonmonotonic behavior of $G(T)$ is also easy to understand. Since $\mu<U$, the thermal window around the Fermi level first samples states closer to the top of the barrier, where transmission is larger than exactly at $E=\mu$, so the conductance initially increases. At higher temperature, however, the same thermal broadening averages over a much wider energy interval, including less transmitting subbarrier states and the oscillatory structure around the barrier region, so the conductance decreases again. Because this maximum appears for all three values of $W$, its origin is mainly set by the real barrier scale $U-\mu$, and is therefore already present in the Hermitian problem.

The most important point is that the best electrical transport and the best thermoelectric efficiency do not occur for the same sign of $W$. Gain, $W>0$, gives the largest $G$, the largest $\kappa$, and also the largest $|S|$. This means that it amplifies the same energy-selective channels already seen in the angular plots, in the flux-balance resonances, and in the sharp zero-temperature conductance peaks. This agrees with the general idea, emphasized in graphene thermoelectric barrier structures, that a sharper energy dependence in the transmission tends to enhance the Seebeck response \cite{Miniya2022,Mishra2017JAP,Mishra2020SPMI}. In the present case, the same mechanism also explains why the $W>0$ curves are less regular and need not be monotonic: as temperature increases, the thermal window scans across gain-enhanced resonant channels, and this affects the heat coefficient more strongly than the charge coefficient because $\kappa$ carries the extra $(E-\mu)^2$ weight. Therefore, once a resonance enters the thermally active window, the thermal conductance can increase very rapidly, which is exactly what happens for positive $W$. However, in our case the increase in thermal conductance is much stronger than the gain in $S^2G$, so the figure of merit is reduced. At $300\,\mathrm{K}$ one finds $ZT\simeq0.0553$ for $W=2\,\mathrm{meV}$, compared with $ZT\simeq0.1697$ in the Hermitian case and $ZT\simeq0.3566$ for $W=-2\,\mathrm{meV}$. Therefore, if the goal is to maximize charge and heat throughput, positive $W$ is preferable, whereas if the goal is thermoelectric performance, negative $W$ is clearly better because it suppresses $\kappa$ much more efficiently than it suppresses $S$. The Hermitian case remains intermediate and more balanced. Read together with the previous figures, this means that non-Hermiticity gives a genuine tuning mechanism: by changing the sign and magnitude of $W$, one can move continuously between a high-conductance regime and a higher-efficiency thermoelectric regime without changing the real barrier geometry.

It is important to stress that the present thermoelectric analysis includes only the electronic contribution to transport. In particular, the thermal conductance discussed here corresponds to electronic heat flow through the non-Hermitian graphene barrier, while the lattice contribution associated with graphene phonons has not been included. For that reason, these results should not be interpreted as the complete thermoelectric coefficients of the device, especially for the figure of merit $ZT$, whose quantitative value requires the total thermal conductance, including both electronic and lattice parts. This point is especially important in graphene, where the phononic thermal channel is known to be large. A fully realistic assessment of thermoelectric performance therefore requires extending the present treatment to include lattice thermal transport and its interplay with the electronic non-Hermitian scattering, which we leave for future work.

\section{Conclusions} \label{conclusions}

We studied thermoelectric transport in monolayer graphene across a finite complex barrier $V(x)=U+iW$. From the exact Dirac-Weyl scattering problem, we obtained the full scattering amplitudes and showed that the imaginary part of the barrier renders the problem nonunitary. The usual Hermitian relation $\mathcal{R}+\mathcal{T}=1$ is then replaced by the generalized flux-balance law $\mathcal{R}+\mathcal{T}=1+\Gamma$, where $\Gamma$ quantifies the net gain or loss inside the barrier. In the Hermitian limit, the standard graphene barrier behavior is recovered, including perfect transmission at normal incidence and the usual Fabry-P\'erot-type resonances. For $W\neq0$, however, the same resonant channels are selectively damped or amplified: negative $W$ smooths the response, while positive $W$ sharpens the angular lobes and enhances the resonant peaks.

At the transport level, the non-Hermitian barrier produces effects that are absent in the usual Hermitian $n$-$p$-$n$ junction. The left and right lead currents are no longer equivalent, the zero-temperature conductances become explicitly dependent on the bias partition, and the naive two-terminal response no longer satisfies gauge invariance. In the same way, the finite-temperature coefficients are built around finite zero-bias offsets, and the usual Onsager relation is not generally recovered because the mixed heat coefficient acquires an additional contribution proportional to $1-\mathcal{R}-\mathcal{T}$. These are all direct consequences of the fact that the barrier acts as an effective source or sink of flux. One practical conclusion is that the complex barrier extends the range of transport behaviors available in graphene: unlike the Hermitian junction, where the response is restricted by unitary scattering and the usual Klein-tunneling constraint, here one can tune angular selectivity, lead asymmetry, and the conductance profile through the imaginary part of the barrier.

Our finite-temperature analysis shows the same degree of tunability. Positive $W$ increases the electrical and thermal conductances and also enhances the magnitude of the Seebeck coefficient, whereas negative $W$ suppresses the thermal conductance more efficiently and yields the largest $ZT$ in the parameter range studied here. Therefore, the best electrical performance and the best thermoelectric efficiency do not occur for the same sign of $W$. This is the second main conclusion of the paper: the complex barrier provides a simple way to move between a high-conductance regime and a more favorable thermoelectric regime without changing the real barrier geometry. In this sense, it extends the usual Hermitian $n$-$p$-$n$ setup by allowing control over transport features that are otherwise fixed.

A further point is that this effective description may also be useful beyond the specific barrier problem studied here. If an experiment shows a measurable dependence of the conductance on the bias partition, but the extra channels or probes responsible for that effect are not known or not under control, then a complex potential can be used as a simple effective model for them. In that reading, the imaginary part plays the role of an effective source-sink channel that captures missing or unresolved degrees of freedom within a reduced description. This is, of course, a simplified model with clear limitations, but it provides a practical way to parameterize measurable bias effects when a full microscopic open-system treatment is not available.

Finally, our thermoelectric analysis includes only the electronic contribution to transport. In particular, the thermal conductance and the resulting $ZT$ do not yet include the lattice thermal channel of graphene. The present values should therefore be understood as the electronic part of the response, not as the complete device performance. Even with this limitation, the results establish a clear proof of principle: a finite complex barrier in graphene is sufficient to produce transport effects beyond the Hermitian case and to provide a simple tunable platform for modeling and controlling them.

\

\section*{Acknowledgments}
D.A.B. was supported by the DGAPA-UNAM Posdoctoral Program. J.A.C. gratefully acknowledges the support of SECIHTI through the program \textit{Becas Nacionales para estudios de Posgrado}, under grant number 4018746. J.C.P.-P was supported by the SECIHTI under the program ``Estancias Posdoctorales por México'' with CVU
number 671687.  A.M.-R. acknowledges financial support by UNAM-PAPIIT project No. IG100224, UNAM-PAPIME project No. PE109226, by SECIHTI project No. CBF-2025-I-1862 and by the Marcos Moshinsky Foundation.

\bibliography{prb_refs.bib}

@book{Nazarov,
  author    = {Nazarov, Yuli V. and Blanter, Yaroslav M.},
  title     = {Quantum Transport: Introduction to Nanoscience},
  publisher = {Cambridge University Press},
  address   = {Cambridge, United Kingdom},
  year      = {2009},
  isbn      = {978-0-521-83246-5}
}

@article{CastroNeto2009,
  author  = {Castro Neto, A. H. and Guinea, F. and Peres, N. M. R. and Novoselov, K. S. and Geim, A. K.},
  title   = {The electronic properties of graphene},
  journal = {Rev. Mod. Phys.},
  volume  = {81},
  pages   = {109--162},
  year    = {2009},
  doi     = {10.1103/RevModPhys.81.109}
}

@article{DasSarma2011,
  author  = {Das Sarma, S. and Adam, S. and Hwang, E. H. and Rossi, E.},
  title   = {Electronic transport in two-dimensional graphene},
  journal = {Rev. Mod. Phys.},
  volume  = {83},
  pages   = {407--470},
  year    = {2011},
  doi     = {10.1103/RevModPhys.83.407}
}

@book{Katsnelson2012,
  author    = {Katsnelson, Mikhail I.},
  title     = {Graphene: Carbon in Two Dimensions},
  publisher = {Cambridge University Press},
  address   = {Cambridge, United Kingdom},
  year      = {2012},
  isbn      = {978-0-521-19540-9}
}

@book{Datta1995,
  author    = {Datta, Supriyo},
  title     = {Electronic Transport in Mesoscopic Systems},
  publisher = {Cambridge University Press},
  address   = {Cambridge, United Kingdom},
  year      = {1995},
  isbn      = {978-0-521-59943-6}
}

@article{Landauer1970,
  author  = {Landauer, R.},
  title   = {Electrical resistance of disordered one-dimensional lattices},
  journal = {Philos. Mag.},
  volume  = {21},
  pages   = {863--867},
  year    = {1970},
  doi     = {10.1080/14786437008238472}
}

@article{Buttiker1986,
  author  = {B{\"u}ttiker, M.},
  title   = {Four-terminal phase-coherent conductance},
  journal = {Phys. Rev. Lett.},
  volume  = {57},
  pages   = {1761--1764},
  year    = {1986},
  doi     = {10.1103/PhysRevLett.57.1761}
}

@article{Bender2007,
  author  = {Bender, C. M.},
  title   = {Making sense of non-Hermitian Hamiltonians},
  journal = {Rep. Prog. Phys.},
  volume  = {70},
  pages   = {947--1018},
  year    = {2007},
  doi     = {10.1088/0034-4885/70/6/R03}
}

@book{Moiseyev2011,
  author    = {Moiseyev, Nimrod},
  title     = {Non-Hermitian Quantum Mechanics},
  publisher = {Cambridge University Press},
  address   = {Cambridge, United Kingdom},
  year      = {2011},
  isbn      = {978-0-521-19560-7}
}

@article{Rotter2009,
  author  = {Rotter, Ingrid},
  title   = {A non-Hermitian Hamilton operator and the physics of open quantum systems},
  journal = {J. Phys. A: Math. Theor.},
  volume  = {42},
  pages   = {153001},
  year    = {2009},
  doi     = {10.1088/1751-8113/42/15/153001}
}

@article{ElGanainy2018,
  author  = {El-Ganainy, R. and Makris, K. G. and Khajavikhan, M. and Musslimani, Z. H. and Rotter, S. and Christodoulides, D. N.},
  title   = {Non-Hermitian physics and {PT} symmetry},
  journal = {Nat. Phys.},
  volume  = {14},
  pages   = {11--19},
  year    = {2018},
  doi     = {10.1038/nphys4323}
}

@article{Schomerus2013,
  author  = {Schomerus, Henning},
  title   = {From scattering theory to complex wave dynamics in non-Hermitian {PT}-symmetric resonators},
  journal = {Philos. Trans. R. Soc. A},
  volume  = {371},
  pages   = {20120194},
  year    = {2013},
  doi     = {10.1098/rsta.2012.0194}
}

@article{FoaTorres2020,
  author  = {Foa Torres, Luis E. F.},
  title   = {Perspective on topological states of non-Hermitian lattices},
  journal = {J. Phys. Mater.},
  volume  = {3},
  pages   = {014002},
  year    = {2020},
  doi     = {10.1088/2515-7639/ab4092}
}

@article{Tzortzakakis2021,
  author  = {Tzortzakakis, A. F. and Makris, K. G. and Szameit, A. and Economou, E. N.},
  title   = {Transport and spectral features in non-Hermitian open systems},
  journal = {Phys. Rev. Research},
  volume  = {3},
  pages   = {013208},
  year    = {2021},
  doi     = {10.1103/PhysRevResearch.3.013208}
}

@article{Muga2004,
  author  = {Muga, J. G. and Palao, J. P. and Navarro, B. and Egusquiza, I. L.},
  title   = {Complex absorbing potentials},
  journal = {Phys. Rep.},
  volume  = {395},
  pages   = {357--426},
  year    = {2004},
  doi     = {10.1016/j.physrep.2004.03.002}
}

@article{Bagarello2016,
  author  = {Bagarello, F.},
  title   = {P{T}-symmetric graphene under a magnetic field},
  journal = {Proc. R. Soc. A},
  volume  = {472},
  number  = {2193},
  pages   = {20160365},
  year    = {2016},
  doi     = {10.1098/rspa.2016.0365}
}

@article{Fernandez2022,
  author  = {Fern{\'a}ndez, David J. and Garc{\'\i}a-Mu{\~n}oz, Juan D.},
  title   = {Graphene in complex magnetic fields},
  journal = {Eur. Phys. J. Plus},
  volume  = {137},
  pages   = {1013},
  year    = {2022},
  doi     = {10.1140/epjp/s13360-022-03221-5}
}

@article{Sarisaman2018,
  author  = {Sarisaman, M. and Tas, M.},
  title   = {Unidirectional invisibility and {PT} symmetry with graphene},
  journal = {Phys. Rev. B},
  volume  = {97},
  pages   = {045409},
  year    = {2018},
  doi     = {10.1103/PhysRevB.97.045409}
}

@article{Chaturvedi2025,
  author  = {Chaturvedi, R. and K{\"o}nye, V. and Hankiewicz, E. M. and van den Brink, J. and Fulga, I. C.},
  title   = {Non-Hermitian topology of transport in the quantum Hall phases in graphene},
  journal = {Phys. Rev. B},
  volume  = {111},
  pages   = {245424},
  year    = {2025},
  doi     = {10.1103/j5n3-dbwr}
}

@article{Ozer2024,
  author  = {{\"O}zer, B. and Ochkan, K. and Chaturvedi, R. and Maltsev, E. and K{\"o}nye, V. and Giraud, R. and Veyrat, A. and Hankiewicz, E. M. and Watanabe, K. and Taniguchi, T. and B{\"u}chner, B. and van den Brink, J. and Fulga, I. C. and Dufouleur, J. and Veyrat, L.},
  title   = {Non-Hermitian topology in the quantum Hall effect of graphene},
  journal = {arXiv},
  pages   = {arXiv:2410.14329},
  year    = {2024},
  doi     = {10.48550/arXiv.2410.14329}
}

@article{Geim_Klein,
  author  = {Katsnelson, M. I. and Novoselov, K. S. and Geim, A. K.},
  title   = {Chiral tunnelling and the Klein paradox in graphene},
  journal = {Nat. Phys.},
  volume  = {2},
  number  = {9},
  pages   = {620--625},
  year    = {2006},
  doi     = {10.1038/nphys384}
}

@article{Imry1986,
  title = {Multichannel Landauer formula for thermoelectric transport with application to thermopower near the mobility edge},
  author = {Sivan, U. and Imry, Y.},
  journal = {Phys. Rev. B},
  volume = {33},
  issue = {1},
  pages = {551--558},
  numpages = {0},
  year = {1986},
  month = {Jan},
  publisher = {American Physical Society},
  doi = {10.1103/PhysRevB.33.551},
}

@article{Butcher1990,
  author  = {Butcher, P. N.},
  title   = {Thermal and electrical transport formalism for electronic microstructures with many terminals},
  journal = {Journal of Physics: Condensed Matter},
  volume  = {2},
  number  = {22},
  pages   = {4869--4878},
  year    = {1990},
  doi     = {10.1088/0953-8984/2/22/008}
}

@article{Kloeckner2017_PRB96_205405,
  title     = {Thermal conductance of metallic atomic-size contacts: Phonon transport and Wiedemann-Franz law},
  author    = {Kl{\"o}ckner, J. C. and Matt, M. and Nielaba, P. and Pauly, F. and Cuevas, J. C.},
  journal   = {Phys. Rev. B},
  volume    = {96},
  number    = {20},
  pages     = {205405},
  year      = {2017},
  month     = nov,
  publisher = {American Physical Society},
  doi       = {10.1103/PhysRevB.96.205405},
  url       = {https://doi.org/10.1103/PhysRevB.96.205405}
}

@article{Landauer1981,
  author  = {Landauer, R.},
  title   = {Conductance determined by transmission: probes and quantised constriction resistance},
  journal = {J. Phys.: Condens. Matter},
  volume  = {1},
  pages   = {8099--8110},
  year    = {1989},
  doi     = {10.1088/0953-8984/1/43/011}
}

@article{Buttiker1985,
  title   = {Generalized many-channel conductance formula with application to small rings},
  author  = {B{\"u}ttiker, M. and Imry, Y. and Landauer, R. and Pinhas, S.},
  journal = {Phys. Rev. B},
  volume  = {31},
  pages   = {6207--6215},
  year    = {1985},
  doi     = {10.1103/PhysRevB.31.6207}
}

@book{Cuevas2010,
  author    = {Juan Carlos Cuevas and Elke Scheer},
  title     = {Molecular Electronics: An Introduction to Theory and Experiment},
  publisher = {World Scientific Publishing},
  address   = {Singapore},
  year      = {2010},
  volume    = {1},
  doi       = {10.1142/7434},
  isbn      = {978-981-4282-58-1}
}

@article{Yang2026,
  title = {Extended Landauer-B\"uttiker Formula for Current through Open Quantum Systems with Gain or Loss},
  author = {Yang, Chao and Wang, Yucheng},
  journal = {Phys. Rev. Lett.},
  volume = {136},
  issue = {5},
  pages = {056304},
  numpages = {11},
  year = {2026},
  month = {Feb},
  publisher = {American Physical Society},
  doi = {10.1103/trty-58r2},
  url = {https://link.aps.org/doi/10.1103/trty-58r2}
}

@article{Kawabata105165137,
  author  = {Kawabata, Kohei and Shiozaki, Ken and Ryu, Shinsei},
  title   = {Many-body topology of non-Hermitian systems},
  journal = {Phys. Rev. B},
  volume  = {105},
  pages   = {165137},
  year    = {2022},
  doi     = {10.1103/PhysRevB.105.165137}
}

@article{GhaemiDizicheh2023,
  author  = {Ghaemi-Dizicheh, Hamed and Kottos, Tsampikos and Shapiro, Boris},
  title   = {Transport effects in non-Hermitian nonreciprocal systems},
  journal = {Phys. Rev. B},
  volume  = {107},
  pages   = {125155},
  year    = {2023},
  doi     = {10.1103/PhysRevB.107.125155}
}

@article{Yan2024,
  author  = {Yan, Qing and Li, Hailong and Sun, Qing-Feng and Xie, X. C.},
  title   = {Transport theory in non-Hermitian systems},
  journal = {Phys. Rev. B},
  volume  = {110},
  pages   = {045138},
  year    = {2024},
  doi     = {10.1103/PhysRevB.110.045138}
}

@article{Wei2025,
  author  = {Wei, Miaomiao and Wang, Bin and Wang, Jian},
  title   = {Gauge invariant quantum transport theory for non-Hermitian systems},
  journal = {Phys. Rev. B},
  volume  = {111},
  pages   = {075135},
  year    = {2025},
  doi     = {10.1103/PhysRevB.111.075135}
}

@article{MunozSotoGarrido2017,
  author  = {Mu{\~n}oz, Enrique and Soto-Garrido, Rodrigo},
  title   = {Analytic approach to magneto-strain tuning of electronic transport through a graphene nanobubble: perspectives for a strain sensor},
  journal = {J. Phys.: Condens. Matter},
  volume  = {29},
  number  = {44},
  pages   = {445302},
  year    = {2017},
  doi     = {10.1088/1361-648X/aa89bc}
}

@article{SotoGarridoMunoz2018,
  author  = {Soto-Garrido, Rodrigo and Mu{\~n}oz, Enrique},
  title   = {Electronic transport in torsional strained Weyl semimetals},
  journal = {J. Phys.: Condens. Matter},
  volume  = {30},
  number  = {19},
  pages   = {195302},
  year    = {2018},
  doi     = {10.1088/1361-648X/aaba07}
}

@article{MunozSotoGarrido2019,
  author  = {Mu{\~n}oz, Enrique and Soto-Garrido, Rodrigo},
  title   = {Thermoelectric transport in torsional strained Weyl semimetals},
  journal = {J. Appl. Phys.},
  volume  = {125},
  pages   = {082507},
  year    = {2019},
  doi     = {10.1063/1.5051966}
}

@article{BonillaMunozSotoGarrido2021,
  author  = {Bonilla, Daniel and Mu{\~n}oz, Enrique and Soto-Garrido, Rodrigo},
  title   = {Thermo-Magneto-Electric Transport through a Torsion Dislocation in a Type I Weyl Semimetal},
  journal = {Nanomaterials},
  volume  = {11},
  number  = {11},
  pages   = {2972},
  year    = {2021},
  doi     = {10.3390/nano11112972}
}

@article{BonillaMunoz2022,
  author  = {Bonilla, Daniel and Mu{\~n}oz, Enrique},
  title   = {Electronic Transport in Weyl Semimetals with a Uniform Concentration of Torsional Dislocations},
  journal = {Nanomaterials},
  volume  = {12},
  number  = {20},
  pages   = {3711},
  year    = {2022},
  doi     = {10.3390/nano12203711}
}

@article{BonillaCastanoYepesMartinRuizMunoz2023,
  author  = {Bonilla, Daniel A. and Casta{\~n}o-Yepes, Jorge David and Mart{\'i}n-Ruiz, A. and Mu{\~n}oz, Enrique},
  title   = {Electromagnetic coupling and transport in a topological insulator--graphene heterostructure},
  journal = {Phys. Rev. B},
  volume  = {107},
  pages   = {245103},
  year    = {2023},
  doi     = {10.1103/PhysRevB.107.245103}
}

@article{BonillaMunoz2024,
  author  = {Bonilla, Daniel A. and Mu{\~n}oz, Enrique},
  title   = {Thermoelectric transport in Weyl semimetals under a uniform concentration of torsional dislocations},
  journal = {Nanoscale Adv.},
  volume  = {6},
  pages   = {2701--2712},
  year    = {2024},
  doi     = {10.1039/d4na00056k}
}

@article{CanasBonillaMartinRuiz2025,
  author  = {Ca{\~n}as, Juan A. and Bonilla, Daniel A. and Mart{\'i}n-Ruiz, A.},
  title   = {Thermoelectric transport in graphene under strain fields modeled by Dirac oscillators},
  journal = {Phys. Rev. B},
  volume  = {112},
  pages   = {104206},
  year    = {2025},
  doi     = {10.1103/3r17-kfy7}
}

@article{CanasBonillaPerezPedrazaMartinRuiz2026,
  author  = {Ca{\~n}as, Juan A. and Bonilla, Daniel A. and P{\'e}rez-Pedraza, J. C. and Mart{\'i}n-Ruiz, A.},
  title   = {Charge and energy transport in graphene with smooth finite-range disorder},
  journal = {Physica B: Condens. Matter},
  volume  = {729},
  pages   = {418431},
  year    = {2026},
  doi     = {10.1016/j.physb.2026.418431}
}

@article{Pereira2007,
  author  = {Pereira, J. Milton, Jr. and Vasilopoulos, P. and Peeters, F. M.},
  title   = {Graphene-based resonant-tunneling structures},
  journal = {Appl. Phys. Lett.},
  volume  = {90},
  pages   = {132122},
  year    = {2007},
  doi     = {10.1063/1.2717092}
}

@article{RamezaniMasir2010,
  author  = {Ramezani Masir, M. and Vasilopoulos, P. and Peeters, F. M.},
  title   = {Fabry-P{\'e}rot resonances in graphene microstructures: Influence of a magnetic field},
  journal = {Phys. Rev. B},
  volume  = {82},
  pages   = {115417},
  year    = {2010},
  doi     = {10.1103/PhysRevB.82.115417}
}

@article{Krstajic2011,
  author  = {Krstajic, P. M. and Vasilopoulos, P.},
  title   = {Ballistic transport through graphene nanostructures of velocity and potential barriers},
  journal = {J. Phys.: Condens. Matter},
  volume  = {23},
  pages   = {135302},
  year    = {2011},
  doi     = {10.1088/0953-8984/23/13/135302}
}

@article{Lejarreta2013,
  author  = {Lejarreta, J. D. and Fuentevilla, C. H. and Diez, E. and Cervero, Jose M.},
  title   = {An exact transmission coefficient with one and two barriers in graphene},
  journal = {J. Phys. A: Math. Theor.},
  volume  = {46},
  pages   = {155304},
  year    = {2013},
  doi     = {10.1088/1751-8113/46/15/155304}
}

@article{Fuentevilla2015,
  author  = {Fuentevilla, C. H. and Lejarreta, J. D. and Cobaleda, C. and Diez, E.},
  title   = {Angle dependent conductivity in graphene FET transistors},
  journal = {Solid-State Electron.},
  volume  = {104},
  pages   = {47--52},
  year    = {2015},
  doi     = {10.1016/j.sse.2014.11.007}
}

@article{Shekhar2025,
  author  = {Shekhar, Sudhanshu and Mandal, Bhabani Prasad and Dutta, Anirban},
  title   = {Relativistic particles in a superperiodic potential: Exploring graphene and fractal systems},
  journal = {Phys. Rev. B},
  volume  = {112},
  pages   = {165106},
  year    = {2025},
  doi     = {10.1103/s82p-5sxv}
}

@article{Mishra2020SPMI,
  author  = {Mishra, Shakti Kumar and Kumar, Amar and Kaushik, Chetan Prakash and Dikshit, Biswaranjan},
  title   = {An efficient tunable thermoelectric device based on n and p doped graphene superlattice heterostructures},
  journal = {Superlattices Microstruct.},
  volume  = {142},
  pages   = {106520},
  year    = {2020},
  doi     = {10.1016/j.spmi.2020.106520}
}

@article{Mishra2017JAP,
  author  = {Mishra, Shakti Kumar and Kumar, Amar and Kaushik, Chetan Prakash and Dikshit, Biswaranjan},
  title   = {Resonant tunneling effect in graphene superlattice heterostructures by tuning the electric potential of defect layer},
  journal = {J. Appl. Phys.},
  volume  = {121},
  pages   = {184301},
  year    = {2017},
  doi     = {10.1063/1.4983151}
}

@article{Grushina2013,
  author  = {Grushina, Anya L. and Ki, Dong-Keun and Morpurgo, Alberto F.},
  title   = {A ballistic pn junction in suspended graphene with split bottom gates},
  journal = {Appl. Phys. Lett.},
  volume  = {102},
  pages   = {223102},
  year    = {2013},
  doi     = {10.1063/1.4807888}
}

@article{Katsnelson2006EPJB,
  author  = {Katsnelson, M. I.},
  title   = {Zitterbewegung, chirality, and minimal conductivity in graphene},
  journal = {Eur. Phys. J. B},
  volume  = {51},
  pages   = {157--160},
  year    = {2006},
  doi     = {10.1140/epjb/e2006-00203-1}
}

@article{YoungKim2009,
  author  = {Young, Andrea F. and Kim, Philip},
  title   = {Quantum interference and Klein tunnelling in graphene heterojunctions},
  journal = {Nat. Phys.},
  volume  = {5},
  pages   = {222--226},
  year    = {2009},
  doi     = {10.1038/nphys1198}
}

@article{Tworzydlo2006,
  author  = {Tworzydlo, J. and Trauzettel, B. and Titov, M. and Rycerz, A. and Beenakker, C. W. J.},
  title   = {Sub-Poissonian Shot Noise in Graphene},
  journal = {Phys. Rev. Lett.},
  volume  = {96},
  pages   = {246802},
  year    = {2006},
  doi     = {10.1103/PhysRevLett.96.246802}
}

@article{MullerBrauningerTrauzettel2009,
  author  = {M{\"u}ller, Markus and Br{\"a}uninger, Matthias and Trauzettel, Bj{\"o}rn},
  title   = {Temperature Dependence of the Conductivity of Ballistic Graphene},
  journal = {Phys. Rev. Lett.},
  volume  = {103},
  pages   = {196801},
  year    = {2009},
  doi     = {10.1103/PhysRevLett.103.196801}
}

@article{Huard2007,
  author  = {Huard, B. and Sulpizio, J. A. and Stander, N. and Todd, K. and Yang, B. and Goldhaber-Gordon, D.},
  title   = {Transport Measurements Across a Tunable Potential Barrier in Graphene},
  journal = {Phys. Rev. Lett.},
  volume  = {98},
  pages   = {236803},
  year    = {2007},
  doi     = {10.1103/PhysRevLett.98.236803}
}

@article{Miniya2022,
  author  = {Miniya, M. and Oubram, O. and Reynaud-Morales, A. G. and Rodr{\'\i}guez-Vargas, I. and Gaggero-Sager, L. M.},
  title   = {Thermoelectric effects in self-similar multibarrier structure based on monolayer graphene},
  journal = {Fractals},
  volume  = {30},
  number  = {3},
  pages   = {2250068},
  year    = {2022},
  doi     = {10.1142/S0218348X22500682}
}

@article{Sutar2012,
  author  = {Sutar, S. and Comfort, E. S. and Liu, J. and Taniguchi, T. and Watanabe, K. and Lee, J. U.},
  title   = {Angle-Dependent Carrier Transmission in Graphene p-n Junctions},
  journal = {Nano Lett.},
  volume  = {12},
  pages   = {4460--4464},
  year    = {2012},
  doi     = {10.1021/nl3011897}
}

@article{nanolett3c04703,
author = {García, Santiago Galván y and Betancur-Ocampo, Yonatan and Sánchez-Ochoa, Francisco and Stegmann, Thomas},
title = {Atomically Thin Current Pathways in Graphene through Kekulé-O Engineering},
journal = {Nano Letters},
volume = {24},
number = {7},
pages = {2322-2327},
year = {2024},
doi = {10.1021/acs.nanolett.3c04703}
}

@article{Celardo2009,
  title = {{Superradiance transition in one-dimensional nanostructures: An effective non-Hermitian Hamiltonian formalism}},
  author = {Celardo, G. L. and Kaplan, L.},
  journal = {Phys. Rev. B},
  volume = {79},
  issue = {15},
  pages = {155108},
  numpages = {9},
  year = {2009},
  month = {Apr},
  publisher = {American Physical Society},
  doi = {10.1103/PhysRevB.79.155108},
  url = {https://link.aps.org/doi/10.1103/PhysRevB.79.155108}
}

@article{Giusteri2015,
  title = {{Non-Hermitian Hamiltonian approach to quantum transport in disordered networks with sinks: Validity and effectiveness}},
  author = {Giusteri, Giulio G. and Mattiotti, Francesco and Celardo, G. Luca},
  journal = {Phys. Rev. B},
  volume = {91},
  issue = {9},
  pages = {094301},
  numpages = {18},
  year = {2015},
  month = {Mar},
  publisher = {American Physical Society},
  doi = {10.1103/PhysRevB.91.094301},
  url = {https://link.aps.org/doi/10.1103/PhysRevB.91.094301}
}

@article{Cook2011,
  title = {{Calculation of electron transport in multiterminal systems using complex absorbing potentials}},
  author = {Cook, Brandon G. and Dignard, Peter and Varga, K\'alm\'an},
  journal = {Phys. Rev. B},
  volume = {83},
  issue = {20},
  pages = {205105},
  numpages = {12},
  year = {2011},
  month = {May},
  publisher = {American Physical Society},
  doi = {10.1103/PhysRevB.83.205105},
  url = {https://link.aps.org/doi/10.1103/PhysRevB.83.205105}
}

@article{Xie2014,
    author = {Xie, Hang and Kwok, Yanho and Jiang, Feng and Zheng, Xiao and Chen, GuanHua},
    title = {{Complex absorbing potential based Lorentzian fitting scheme and time dependent quantum transport}},
    journal = {The Journal of Chemical Physics},
    volume = {141},
    number = {16},
    pages = {164122},
    year = {2014},
    month = {10},
    issn = {0021-9606},
    doi = {10.1063/1.4898729},
    url = {https://doi.org/10.1063/1.4898729},
}

@article{Geng2023,
  title = {{Nonreciprocal charge and spin transport induced by non-Hermitian skin effect in mesoscopic heterojunctions}},
  author = {Geng, H. and Wei, J. Y. and Zou, M. H. and Sheng, L. and Chen, Wei and Xing, D. Y.},
  journal = {Phys. Rev. B},
  volume = {107},
  issue = {3},
  pages = {035306},
  numpages = {10},
  year = {2023},
  month = {Jan},
  publisher = {American Physical Society},
  doi = {10.1103/PhysRevB.107.035306},
  url = {https://link.aps.org/doi/10.1103/PhysRevB.107.035306}
}

@article{Wu2022,
	author = {Wu, Chao and Fan, Annan and Liang, Shi-Dong},
	da = {2022/12/09},
	doi = {10.1007/s43673-022-00065-0},
	id = {Wu2022},
	isbn = {2309-4710},
	journal = {AAPPS Bulletin},
	number = {1},
	pages = {39},
	title = {{Complex Berry curvature and complex energy band structures in non-Hermitian graphene model}},
	ty = {JOUR},
	url = {https://doi.org/10.1007/s43673-022-00065-0},
	volume = {32},
	year = {2022}}

\end{document}